\documentclass[fleqn,usenatbib]{mnras}

\usepackage{newtxtext,newtxmath}
\usepackage[T1]{fontenc}

\DeclareRobustCommand{\VAN}[3]{#2}
\let\VANthebibliography\thebibliography
\def\thebibliography{\DeclareRobustCommand{\VAN}[3]{##3}\VANthebibliography}

\usepackage{geometry} % Margins
\usepackage{amsmath,amsfonts} % Maths

\usepackage[dvipsnames]{xcolor}  % Colors
\usepackage{graphicx, float, subfig, overpic, comment} %Figures
\usepackage{stmaryrd} %Normes rares
\usepackage{faktor} % Quocients
\usepackage{mathrsfs,mathtools} % Caligrafiques
\usepackage{titlesec} %Titols capitols
\usepackage[none]{hyphenat} % Evitar hyphenation
\usepackage[font=small,labelfont=bf,tableposition=top]{caption} % Captions

\usepackage{stfloats}

\usepackage{cancel}
\usepackage{url}

\usepackage{hyperref} % Hyperlinks
\hypersetup{
	colorlinks=true,
	linkcolor=blue,
	citecolor=green,
}

\newcommand{\R}{\mathbb{R}}
 \newcommand{\HH}{\mathcal{H}}

\title[Co-orbital motion classification through ML]{Asteroids co-orbital motion classification based on Machine Learning }

\author[]{
Giulia Ciacci$^{1}$\thanks{E-mail: giuliaciacci8@gmail.com}
Andrea Barucci$^{1}$\thanks{E-mail: a.barucci@ifac.cnr.it}
Sara Di Ruzza$^{2}$\thanks{E-mail:  sara.diruzza@unipa.it}
and Elisa Maria Alessi$^{3}$\thanks{E-mail: elisamaria.alessi@cnr.it} 
\\
$^{1}$IFAC-CNR, Istituto di Fisica Applicata ``Nello Carrara'', Consiglio Nazionale delle Ricerche, via Madonna del Piano 10, 50019 Sesto Fiorentino (FI), Italy\\
$^{2}$Dipartimento di Matematica e Informatica, Universit\`a di Palermo, Via Archirafi 34, 90123 Palermo, Italy\\ 
$^{3}$IMATI-CNR, Istituto di Matematica Applicata e Tecnologie informatiche ``E. Magenes'', Consiglio Nazionale delle Ricerche, Via Alfonso Corti 12, 20133 Milano, Italy
}

\date{Accepted XXX. Received YYY; in original form ZZZ}

\pubyear{2015}

\begin{document}
\label{firstpage}
\pagerange{\pageref{firstpage}--\pageref{lastpage}}
\maketitle

% Abstract of the paper
\begin{abstract}
%This is a simple template for authors to write new MNRAS papers.
%The abstract should briefly describe the aims, methods, and main results of the paper.
%It should be a single paragraph not more than 250 words (200 words for Letters).
%No references should appear in the abstract.

In this work, we explore how to classify asteroids in co-orbital motion with a given planet using Machine Learning. We consider four different kinds of motion in mean motion resonance with the planet, nominally \textit{Tadpole}, \textit{Horseshoe} and \textit{Quasi-satellite}, building 3 datasets defined as Real (taking the ephemerides of real asteroids from the JPL Horizons system), Ideal and Perturbed (both simulated, obtained by propagating initial conditions considering two different dynamical systems) for training and testing the Machine Learning algorithms in different conditions. 

The time series of the variable $\theta$ (angle related to the resonance) are studied with a data analysis pipeline defined {\it ad hoc} for the problem and composed by: data creation and annotation, time series features extraction thanks to the  \textit{tsfresh} package (potentially followed by selection and standardization) and the application of Machine Learning algorithms for Dimensionality Reduction and Classification. Such approach, based on features extracted from the time series, allows to work with a smaller number of data with respect to Deep Learning algorithms, also allowing to define a ranking of the importance of the features. Physical Interpretability of the features is another key point of this approach. In addition, we introduce the SHapley Additive exPlanations for Explainability technique.

Different training and test sets are used, in order to understand the power and the limits of our approach. The results show how the algorithms are able to identify and classify correctly the time series, with a high degree of performance. %These results suggest that the proposed approach to time series classification of co-orbital motion can be exploited in the context of celestial mechanics. 

\end{abstract}

% Select between one and six entries from the list of approved keywords.
% Don't make up new ones.

%%%%%%%%%%%%%%%%%%%%%%%%%%%%%%%%%%%%%%%%%%%%%%%%%%

%%%%%%%%%%%%%%%%% BODY OF PAPER %%%%%%%%%%%%%%%%%%

\section{Introduction}

%\subsection{State of art of Machine Learning applied to Astronomy and Celestial Mechanics}

In the last decades, the use of Artificial Intelligence (AI) for data analysis has significantly increased in scientific applications, in particular thanks to its sub-field known as Machine Learning (ML), where an algorithm is said to improve its performance on a specific task by experience \citep[e.g.,][]{hastie2009elements,jordan2015machine}.  More recently, many authors started to use such methods in astronomy and solar system science \citep[e.g.,][]{ball2010data,ivezic2014statistics}. 
Although well-known and broadly applied in several contexts, we recall here the general concepts of AI and ML, for the sake of completeness. With AI we mean methods by which a computer makes decisions or discoveries that would usually require human intelligence, while with ML we mean automated processes that learn by examples in order to classify, predict, discover or generate new data. Part of ML is the class of algorithms known as {\it Deep Learning} (DL) which is based on artificial neural networks \citep[e.g.,][]{lecun2015deep,goodfellow2016deep}. ML and DL are the key of the  success of AI nowadays. There are three classes of ML algorithms (see, for example, \citet{hastie_etal2009} for more details): {\it supervised learning}, where a labeled dataset is used to help to train and tune the algorithm, with the goal to create a map that links inputs to outputs; {\it unsupervised learning}, where no labels are provided and the goal is to discover hidden patterns allowing the data to speak for itself; {\it reinforcement learning}, where an agent learns by interacting with an environment and modifying its behavior to maximize its reward.
It is important to keep in mind that this line between classes can occasionally become hazy and fluid because numerous applications frequently combine them in inventive and unique ways (e.g. self-supervised learning, see \citet{liu2021self}).

These approaches are firmly established in astronomy and an important survey of the state of art can be found in \citet{fluke&jacobs2020}, who analyse the published articles in the last years. They highglight applications in many sub-fields of astronomy where ML could be used for several activities, as classification, regression, clustering, forecasting, generation of data, discovering, development of new scientific insights. \citet{fluke&jacobs2020} also classify the different fields of astronomy where ML is used as \lq\lq emerging\rq\rq, \lq\lq progressing\rq\rq \, and \lq\lq established\rq\rq, depending on the progress of its use. 

The first approach in astronomy to Principal Component Analysis (PCA), an algorithm devoted to Dimensionality Reduction, which is nowadays a standard technique, was introduced in the 1980s for morphological classification of spiral galaxies \citep[e.g.,][]{whitmore1984}, in the 1990s for quasar detection \citep[e.g.,][]{francis_etal1992} and spectral classification \citep[e.g.,][]{singh_etal1998}, while more recent applications with ML have been done for discovering extrasolar planets \citep[e.g.,][]{pearson_etal2018,Shallue_etal2018}, for studying gravitationally lensed systems \citep[e.g.,][]{jacobs_etal2017,lanusse_etal2018,Pourrahmani_etal2018} and for  discovering and classifying transient objects \citep[e.g.,][]{Connor_etal2018,farah_etal2018}. For a complete and detailed bibliography about all the ML applications in the astronomical fields we suggest a careful reading of \citet{fluke&jacobs2020}. 

The analysis of motion of the solar system bodies is considered one \lq\lq progressing\rq\rq \, field of application of ML. Several authors in the last years studied problems related to solar system objects as, for example, applications to TransNeptunian objects \citep[e.g.,][]{chen_etal2018}, or detection and classification of asteroids through taxonomies of spectrophotometry, as studied in \citet{Erasmus_etal2017,Erasmus_etal2018}. 

One \lq\lq emerging\rq\rq \, field concerns asteroid dynamics \citep[e.g.,][]{carruba_etal2022}. Indeed, the numerical propagation of asteroids' orbits, based on continuous improved information, implies a large volume of data, that requires fast and novel methods to be analyzed. % this field requires numerical integration of asteroid orbits and with the increasing volume of new and improved information, fast and novel methods can be useful to handle and analyse huge amount of data. 
For example, in \citet{Smirnov&Markov2017}, the authors use ML methods to identify three-body mean motion resonance asteroids in the main belt without requiring numerical integration. They use proper elements which are quasi-integral of motion that are stable for a long time \citep[e.g.,][]{knezevic_milani1994,knezevic_etal2002}, and use four different supervised ML methods as reported in  \citet{hastie_etal2009}. The authors compare their results with the ones of the previous paper by \citet{Smirnov_etal2013} remarking that, with the new approach, the identification of the objects trapped in  mean motion resonance is very good and the procedure requires few seconds, while the numerical integration requires days and weeks. Very recently, \citet{SMIRNOV2023} provides a new open-source package for identifying objects trapped in mean motion resonances (MMR). The main objective they have is to distinguish resonant and non-resonant orbits, but they do not aim at distinguishing different classes of 1:1 MMR, like we will do here.

Other new works comparing results from ML algorithms with previous known asteroid classifications are, for example,  \citet{smullen&volk2020}, where the authors classify objects of the Kuiper belt into four classes based on their dynamics; \citet{carruba_etal2019}, where hierarchical clustering algorithms for supervised learning are applied to identify 6 new families and 13 new clustering of asteroids; \citet{carruba_etal2020}, where ML classification algorithms are used to identify new families of asteroids based on the orbital distribution in the parameters $(a,e,\sin(i))$ (where $a,e,i$ are, respectively, the semi-major axis, the eccentricity and the inclination of the asteroid orbit) of previous known family objects.

Some other very interesting and recent works explore the use of ML to classify regular or chaotic motions. For example, \citet{kamath_2022} studies and classifies orbits in Poincar\'e maps: the major challenge of this problem is solved by creating high-quality training sets with few mislabeled orbits and converting the coordinates of the points into features that are discriminating, despite the apparent similarities between orbits of different classes. \citet{celletti_etal2022} use DL methods, such as convolutional neural networks (CNNs), to show how it is possible to classify different types of motion, starting from time series, without any prior knowledge of the dynamics. Indeed, the identification of a  motion usually requires a knowledge and the solution of the differential equations governing the dynamical system. Instead using CNNs trained on one dynamical model, the type of motion could be predicted, for example, from observational data. 

All these examples show how ML algorithms are increasingly used in astronomy, as well as in dynamical systems and in particular in celestial mechanics.

In this paper, leveraging on the recent work \citet{DiRuzza2023}, we focus on asteroids that are in co-orbital motion (1:1 Mean Motion Resonance) with a planet of the solar system. We apply ML methods to classify the various types of co-orbital motion that can arise in the planar case, through features derived from time series corresponding to the evolution of a specific variable -- the angle $\theta$, that we will define in the following.

The current paper is organized as follows. In Section \ref{coorbital}, we recall the averaged problem of circular restricted three-body problem for the co-orbital motion in the planar case and how the approximation can be applied to classify co-orbital objects in the solar system. In Section \ref{Sec_data}, it is explained how the training and testing data are generated. In Section \ref{pipeline}, the whole algorithmic pipeline is detailed, while in Section \ref{sec:results} the results are given together with a critical analysis on the procedure. In Sections \ref{sec:future} and \ref{sec:conclusions} a possible future direction is proposed and the conclusions are drawn. 

\section{Co-planar co-orbital asteroids in the solar system}
\label{coorbital}

%Let us recall, in this section, the setting and the results of the paper .

%\subsubsection{General setting}
The main idea considered by \citet{DiRuzza2023} was to show how an integrable approximation of the restricted three-body problem can be applied to describe the dynamics of real natural objects and the goal was to provide a general catalogue of co-orbital objects in the solar system in the co-planar case and a tool to visualize them. 

We recall here the general setting and main features that will be important for the present work. More details can be found in \citet{pousse&alessi2022} and \citet{DiRuzza2023}.
The theoretical model is the Planar Circular Restricted Three-Body Problem (PCR3BP) where a massless body is interacting by gravitational attraction with two massive bodies. The Hamiltonian describing the motion of the massless body can be written as
\begin{equation}\label{ham_R3BP}
 \mathcal{H}\left(\mathbf{r}, \dot{\mathbf{r}}, \lambda_p\right)
	=	\frac{\|\dot{\mathbf{r}}\|^{2}}{2}
	-	\frac{\mu}{\|\mathbf{r}\|}
	-	\frac{(\mu + \mu_p) \, \varepsilon}{\left\|\mathbf{r}-\mathbf{r}_p\left(\lambda_p\right)\right\|}+	(\mu + \mu_p) \, \varepsilon \, 
	\mathbf{r} \cdot\mathbf{r}_p\left(\lambda_p\right) \,,
\end{equation} 
where $\mathbf{r}, \dot{\mathbf{r}} \in  {\R}^2  $ are, respectively, the heliocentric position and velocity vectors of the massless body (the asteroid); $\mu, \mu_p$ are the mass parameters of the massive primary body (the Sun) and of the massive secondary body (the planet), respectively; 
\begin{equation*}
    \varepsilon := \frac{\mu_p}{\mu + \mu_p}
\end{equation*} 
is a dimensionless parameter characterizing the mass ratio of the Sun-planet system; the heliocentric vector $\mathbf{r}_p\left(\lambda_p\right)$ denotes the position of the planet, for a given value of the mean longitude $\lambda_p$, which follows the solution of the two-body problem for the Sun-planet system. Usually, the Hamiltonian~\eqref{ham_R3BP} is analyzed in the synodic reference frame rotating with the planet. It is well-known that the problem admits 5 equilibrium points, called Lagrangian points and denoted by $L_j$ for $j=1, \ldots, 5$. If $\varepsilon$ is small enough, we could rewrite the Hamiltonian~\eqref{ham_R3BP} as 
\begin{equation*}
\mathcal{H}\left(\mathbf{r}, \dot{\mathbf{r}}, \lambda_p\right) = \mathcal{H}_{\mathrm{K}} \left(\mathbf{r}, \dot{\mathbf{r}} \right)
	+(\mu + \mu_p)\, \varepsilon \,    \mathcal{H}_{\mathrm{P}}\left(\mathbf{r}, \lambda_p\right)\,,
\end{equation*} 
where $\HH_{\rm K} $ is the unperturbed Kepler motion of the massless body (around the Sun) and $\HH_{\rm P} $ is the perturbation depending on the gravitational influence of the planet and, then, we consider the averaged problem with respect to the fast angle $\lambda_p$ obtaining the new Hamiltonian
\begin{equation*}
 \overline \HH = \HH_{\rm K}+\overline \HH_{\rm P} \,,
\end{equation*}
where $\overline \HH_{\rm P}$ is the average over the period of revolution of the planet with respect to the fast angle  $\lambda_p$. %We refer to \citet{pousse&alessi2022} for the details on the domain of validity of the averaged problem.

We assume that the particle and the secondary are in a $1:1$ Mean Motion Resonance (MMR), that is, their orbits have the same value of semi-major axis. Within this approximation, the problem can be studied by means of the action-angle variables $(\theta, u)$, defined as follows:
%In the three-body problem, if the ratio of periods of revolution of the second and the third bodies has the form $p/q$ with $p$ and $q$ integers, the two bodies are in $p:q$ mean motion resonance. If the periods are the same, the two bodies are in a $1:1$ mean motion resonance called also co-orbital resonance. In this case the semi-major axes of the two bodies are the same.	In this context, can be useful introducing new symplectic variables to study the problem. Let us define the action-angle variable $(\theta, u)$, where
\begin{equation*}
    \theta := \lambda - \lambda_p
\end{equation*} 
is the resonant angle (being $\lambda$ the mean longitude of the asteroid) % that characterizes the commensurability between asteroid and planet 
and 
\begin{equation*}
    u := \sqrt{\frac{a}{a_p}} - 1
\end{equation*} 
is its conjugated action whose modulus measures the distance to the exact Mean Motion Resonance, with $a$ and $a_p$ being the semi-major axis of the asteroid and of the planet orbit, respectively; the exact $1:1$ MMR is obtained for $(\dot{\theta},u)=(0,0)$.

\begin{figure*}
\centering
\includegraphics[scale = 0.15]{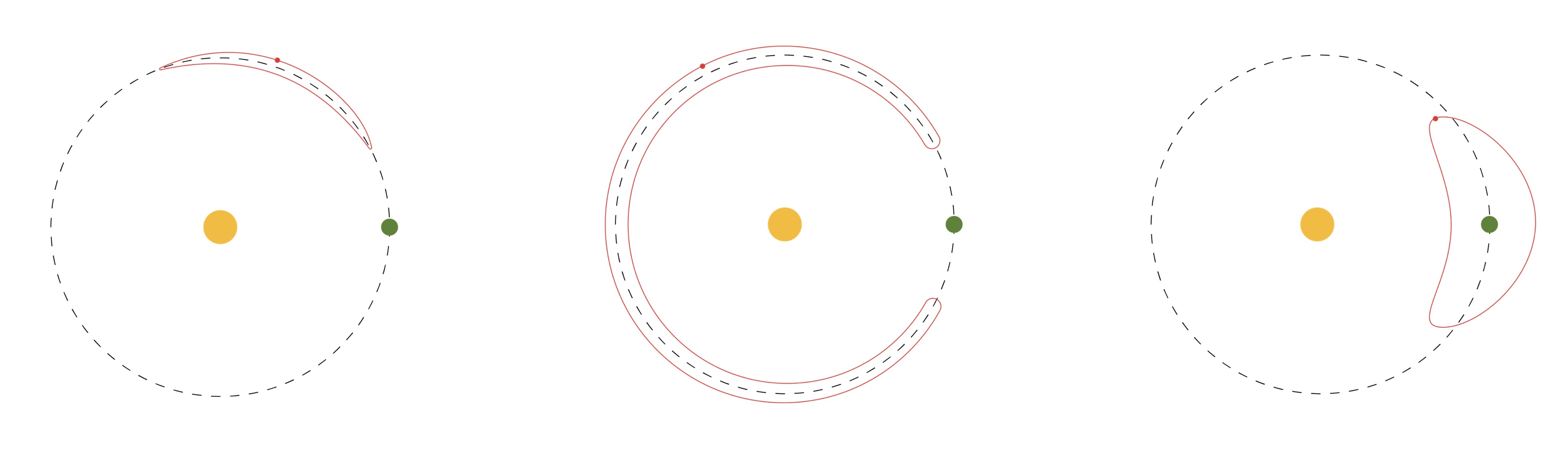}
 \caption{In red, a sketch of the tadpole motion (left), horseshoe motion (center), quasi-satellite motion (right), in the synodic reference system. The yellow circle represents the Sun and the green one the planet. }
    \label{fig:QSHSTP}
\end{figure*}

In this system, the quantity
\begin{equation*}
    \Gamma  = \sqrt{a_p}\left(1 - \sqrt{1 - e^2}\right) \, 
\end{equation*}
is a first integral of the problem, being $e$ the eccentricity of the asteroid orbit. For different values of $\Gamma \in [0:\sqrt{a_p}]$, the phase portrait in resonant variables $(\theta,u)$ allows to understand the whole co-orbital motion structure. In the planar circular case we can have three types of co-orbital motion, depicted in Fig.~\ref{fig:QSHSTP} in the synodic reference system. The tadpole (TP) motion (on the left) stemming from $L_j$ with $j=4,5$ is such that $\theta$ experiences a periodic oscillation around a given $\theta_j(\Gamma)$ satisfying ${23.9^\circ< (-1)^j\theta_j(\Gamma)< 180^\circ}$; the horseshoe (HS) motion (in the middle), stemming from $L_3$ is such that $\theta$ oscillates around $180^\circ$ with a large amplitude that decreases as long as $\Gamma$ increases; the quasi-satellite (QS) regime (on the right) is such that $\theta$ librates around zero for $\Gamma>0$.

\begin{figure}
\centering
\includegraphics[width=1.\columnwidth]{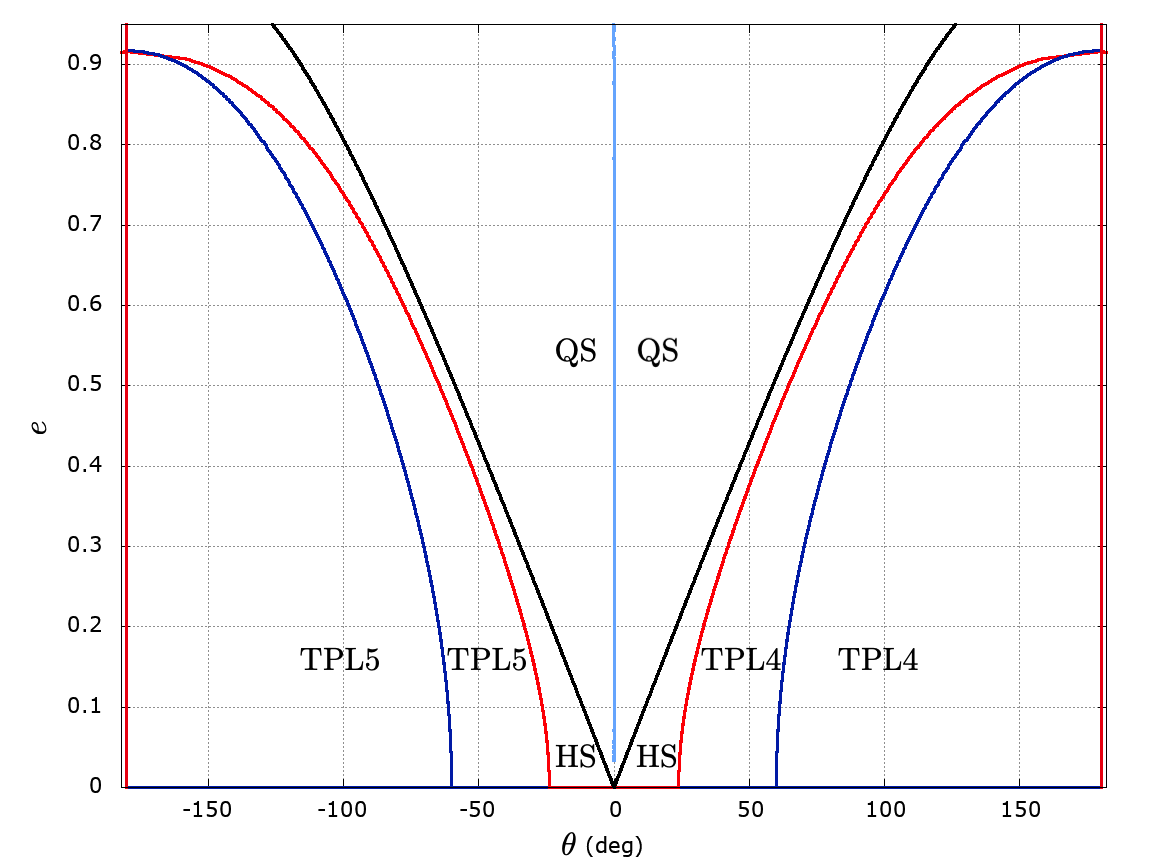}
\caption{The $(\theta,e)$-map of the co-orbital motion defined by the section $u = 0$. The black and red thick curves stand, respectively, for the singularity of collision and the crossing of the separatrices that originate from $L_3$ (thick red curve). They divide the map in three regions. 
The QS domain is between the dark curves; the HS region, split in two parts, is between the separatrix (red curve) and the the dark curve; the TP regions are inside the separatrices (respectively, TPL4 for positive values of the angle $\theta$ and TPL5 for negative values of the angle $\theta$).}
\label{fig:Map}
\end{figure}

In the given phase space, the co-orbital trajectories are solutions located in the neighborhood of $u=0$ and such that $\theta$ oscillates around the given value. The crossing with the section $u=0$, that corresponds to $a=a_p$, provides a way to understand the global evolution of the dynamics at varying $\Gamma$, or equivalently, the eccentricity $e$ of the asteroid's orbit. In this way it is possible to derive a $(\theta,e)$-map, represented in Fig.~\ref{fig:Map}, that allows to classify the different domains of co-orbital motion. We remark that, in first approximation, this map is invariant with respect to the mass parameter $\varepsilon$, so it has the same features for all the planets.

\begin{figure*}
    \centering
     \includegraphics[scale = 0.19]{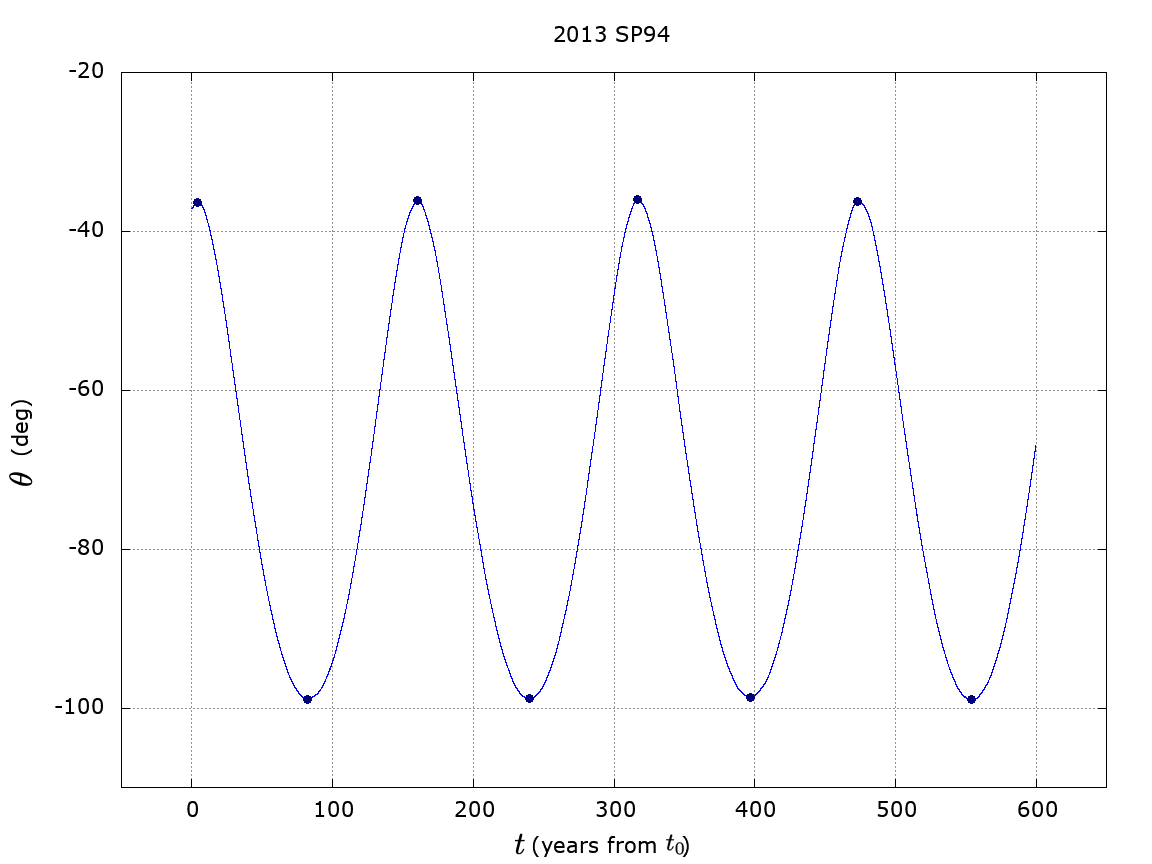}
      \includegraphics[scale = 0.19]{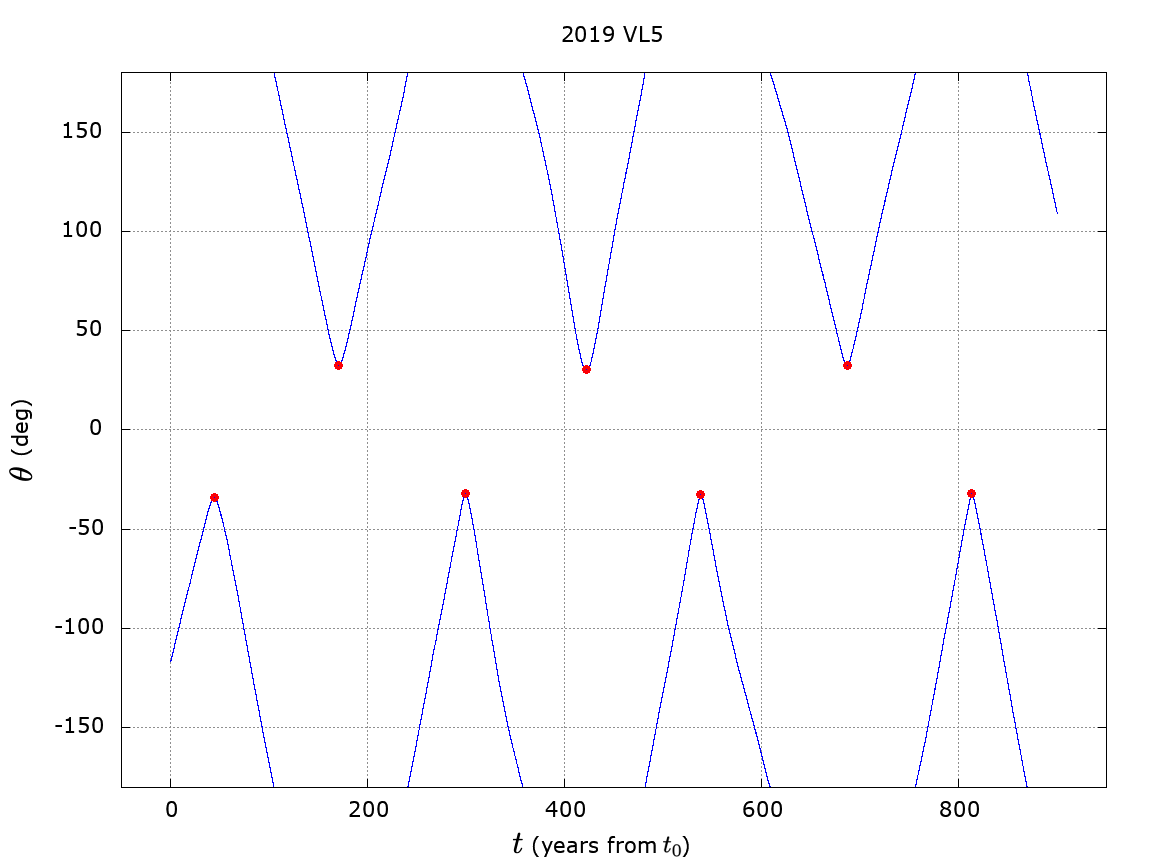}
            \includegraphics[scale = 0.19]{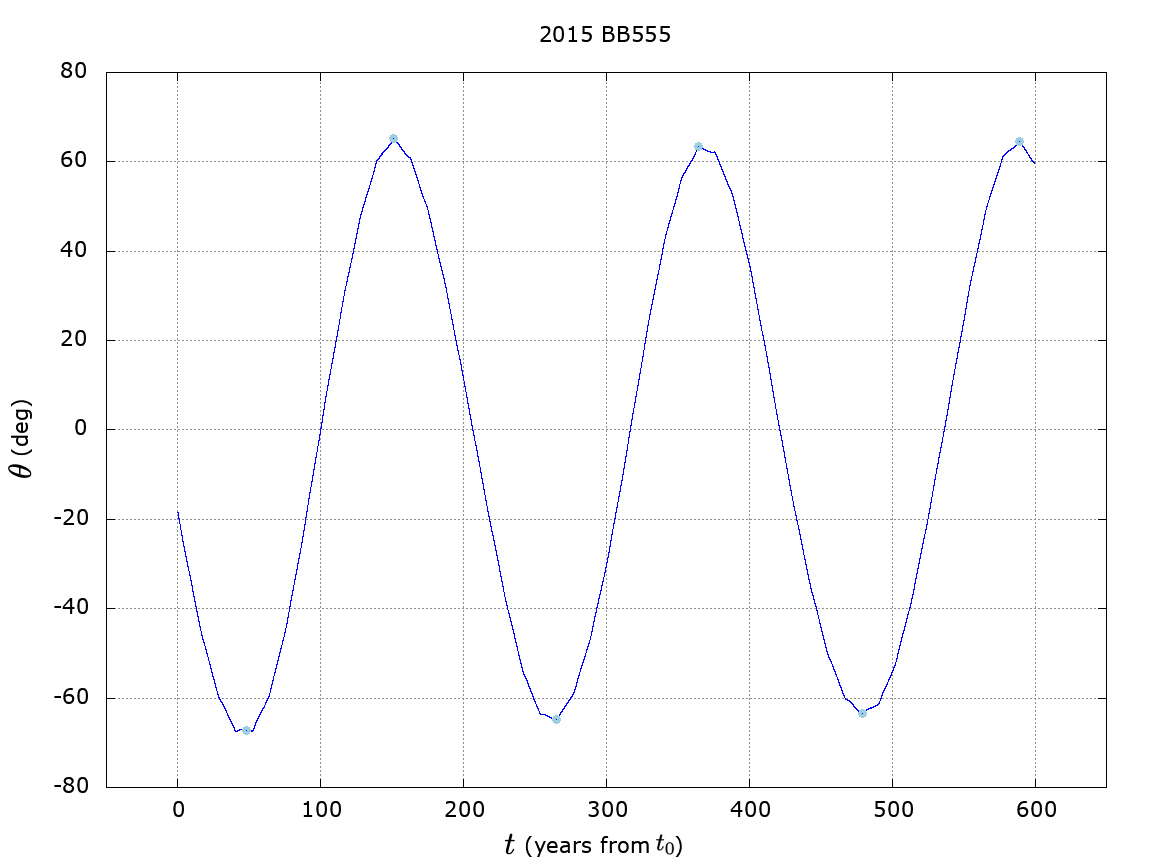}\\
                    \includegraphics[scale = 0.19]{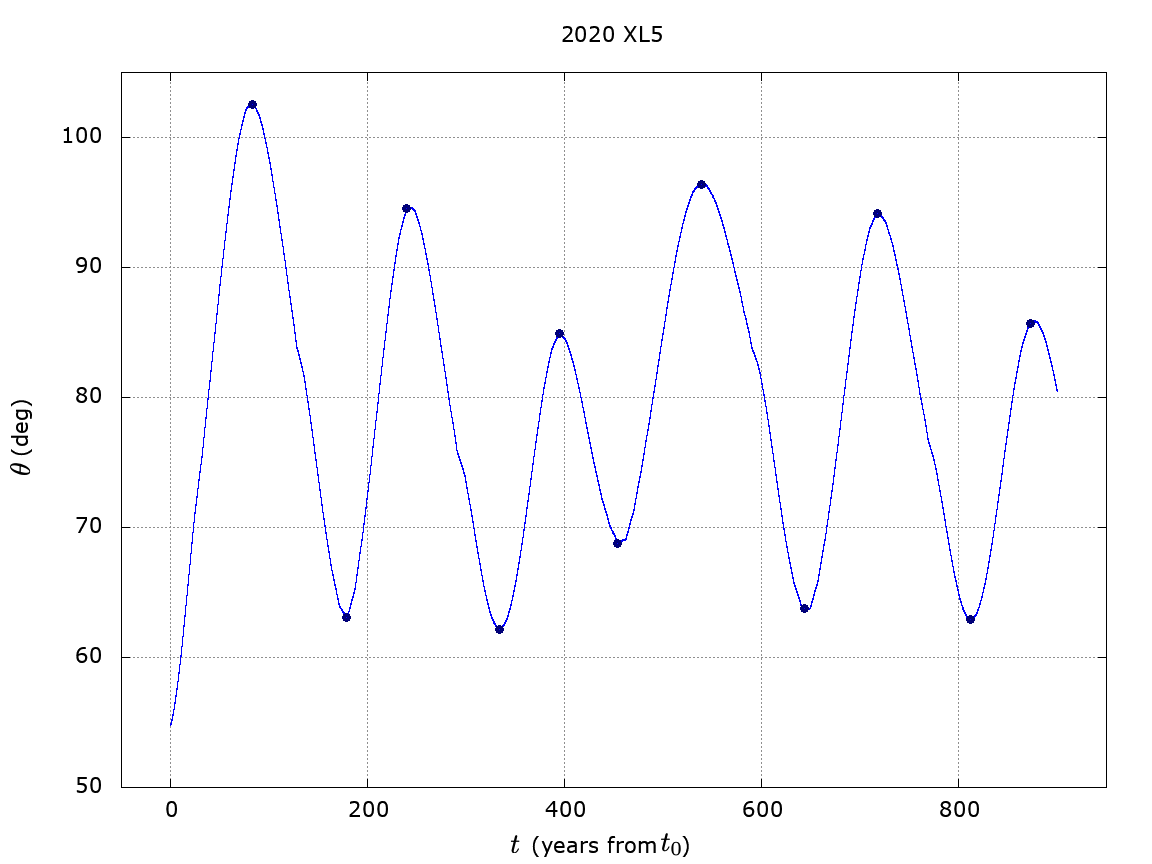}
      \includegraphics[scale = 0.19]{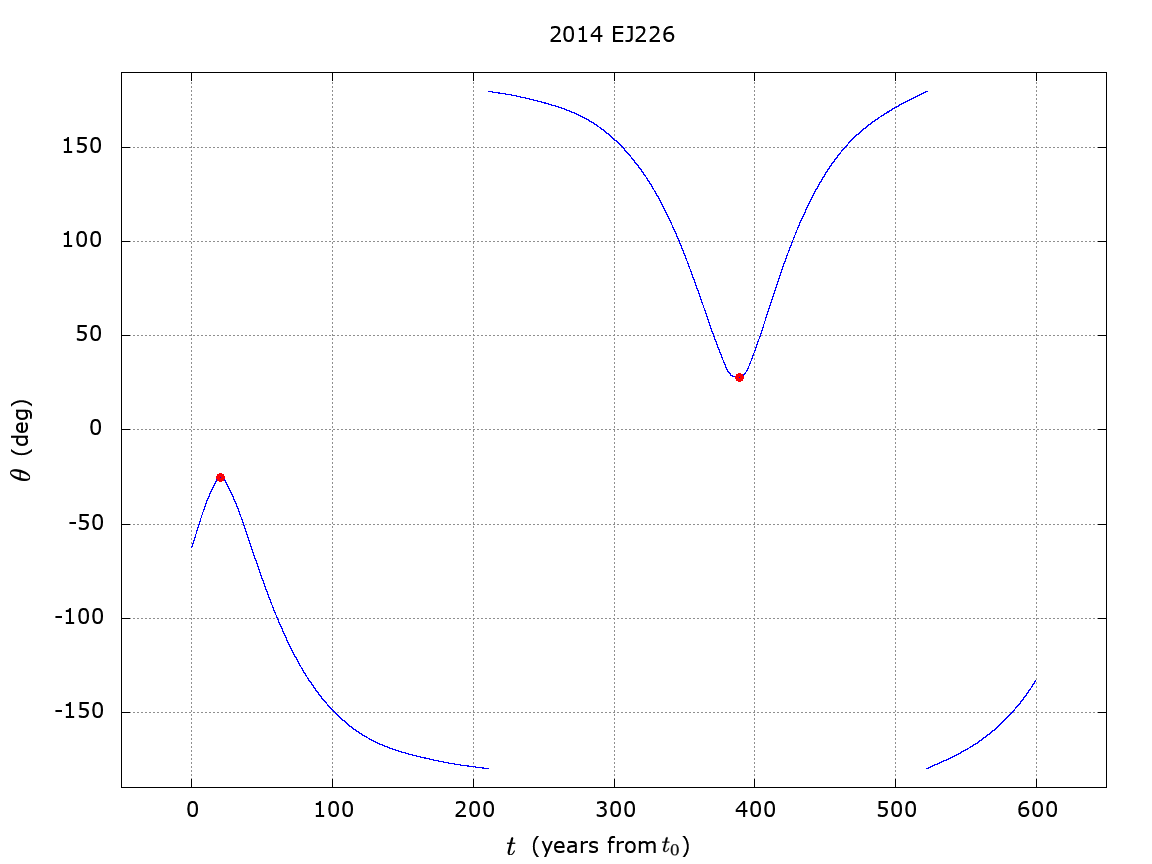}
        \includegraphics[scale = 0.19]{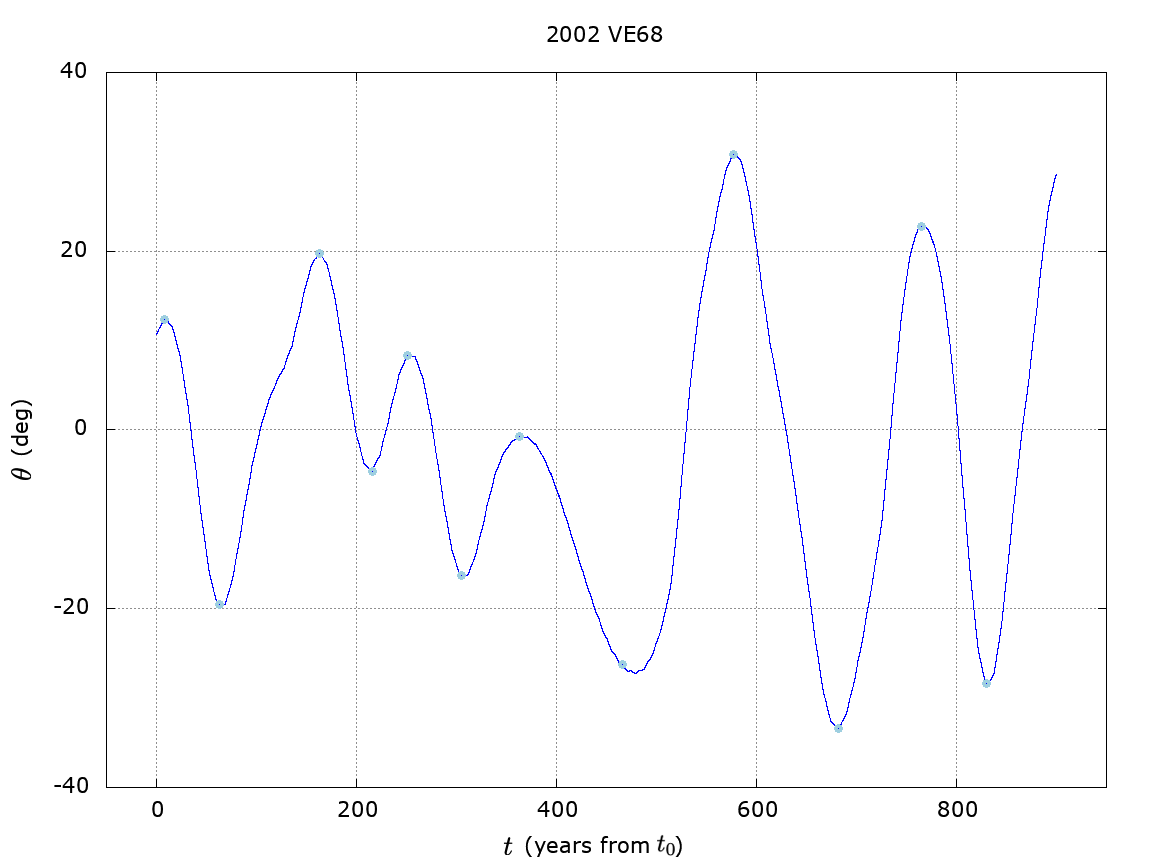}
    \caption{Upper: evolution of the angle $\theta$ versus time of three real asteroids in a regular co-orbital motion; from left to right, respectively, TP with Jupiter,  HS with Earth, QS with Jupiter. Bottom: evolution of the angle $\theta$ versus time of three real asteroids in co-orbital motion with non-regular oscillations; from left to right, respectively, TP with Earth, HS with Jupiter, QS with Venus. }
    \label{fig:QSHSTP_real}
\end{figure*}

%In Fig.~\ref{fig:QSHSTP}, the trajectories of asteroids in the three different regimes QS, HS, TP are represented in the synodic frame. 
In the upper panels of Fig.~\ref{fig:QSHSTP_real}, the graphs of the evolution of the time series $(t,\theta)$ of the three real examples of asteroids in the different regimes TP, HS, QS are plotted. In these cases, the evolution appears very regular, while in bottom panels, three less regular cases are reported for comparison.

It is important to underline that the analysis done in the current work, and described in the next Sections, takes specifically into account the time evolution of the resonant angle $\theta$. Subsequently, we will exploit the time series $(t,\theta)$ in order to recognize the different kinds of co-orbital regime as shown in Fig.~\ref{fig:QSHSTP_real}. 

\begin{figure*}
    \centering
    \includegraphics[scale = 0.19]{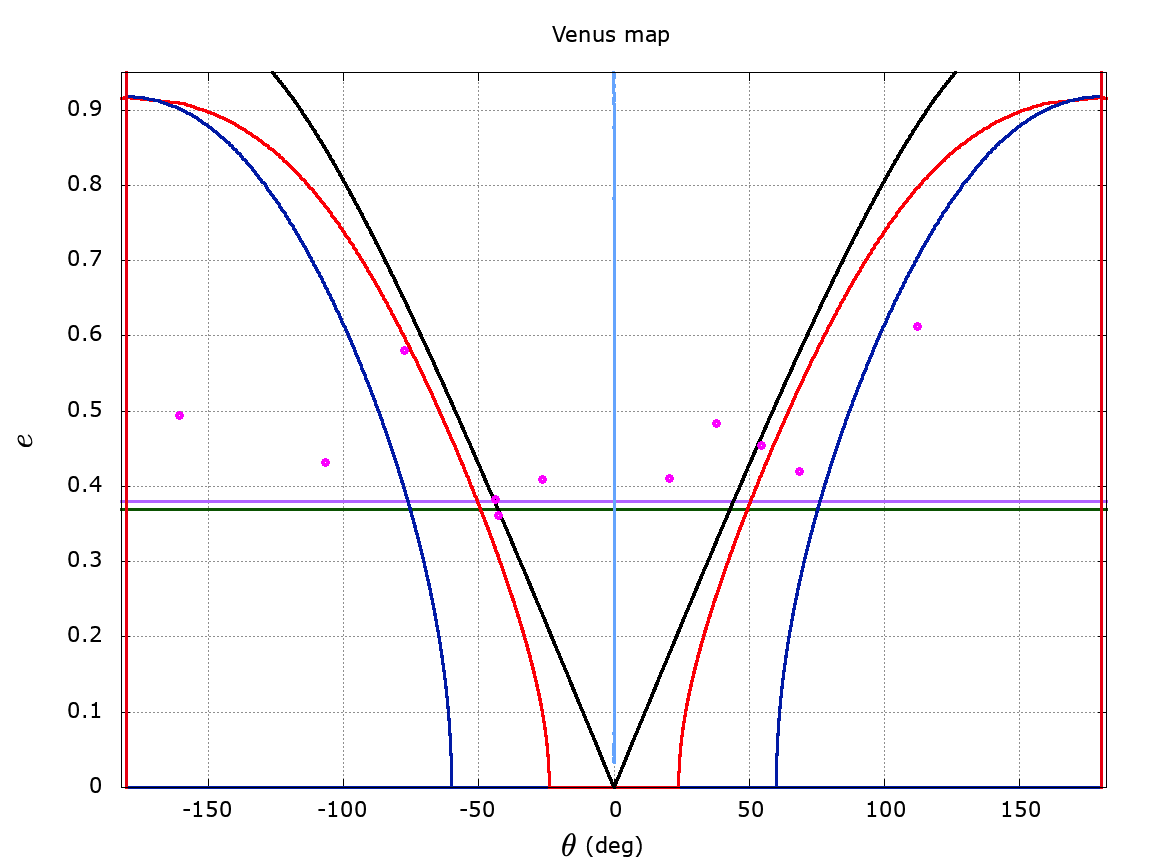}
      \includegraphics[scale = 0.19]{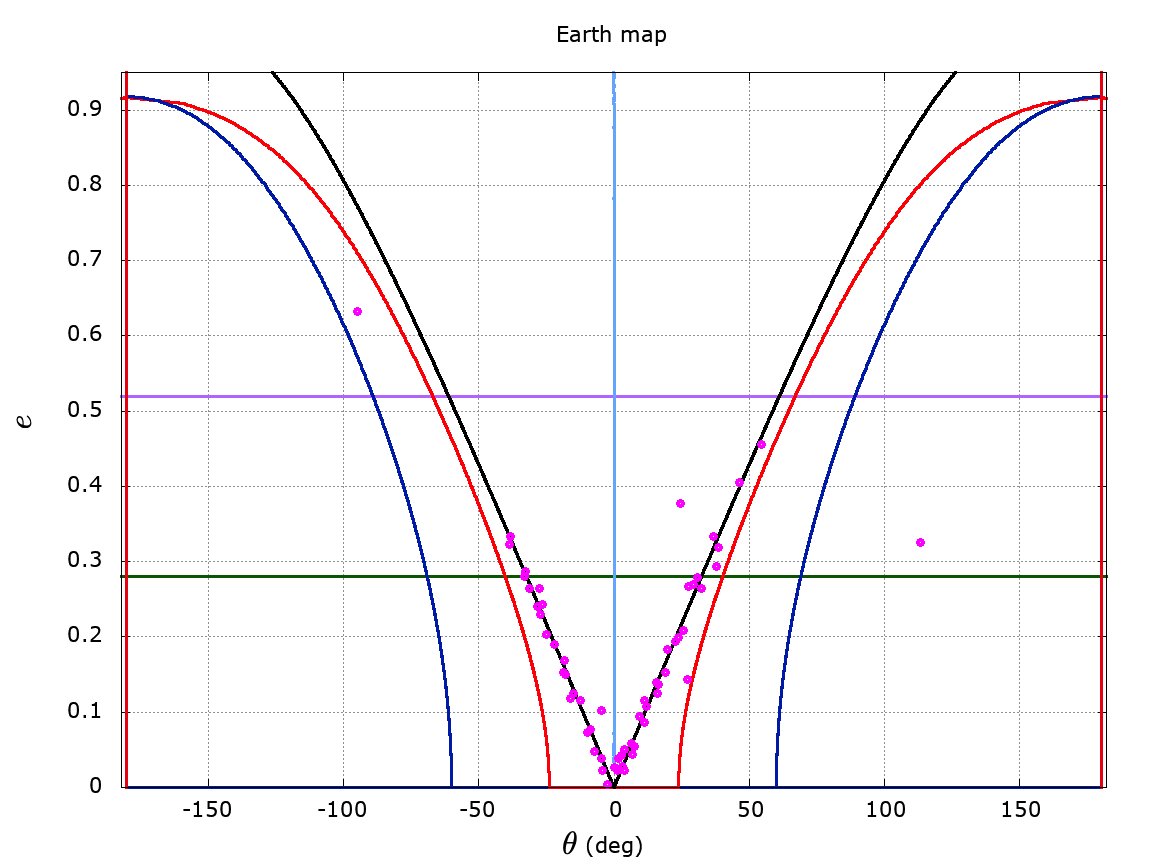}
        \includegraphics[scale = 0.19]{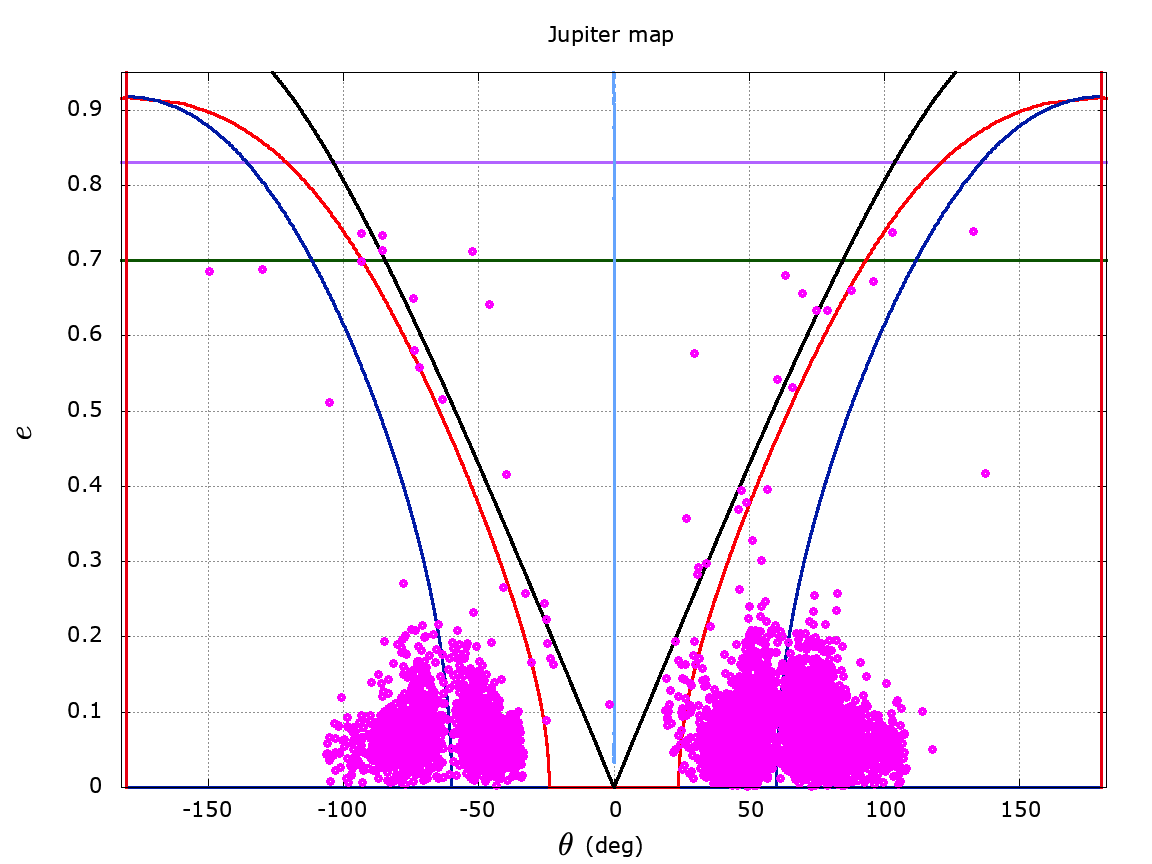}
    \caption{The $(\theta,e)$-maps for the three planets; from left to right, respectively, Venus, Earth and Jupiter.  The points in magenta represent the distribution of co-orbital asteroids in the $(\theta,e)$-map at a reference date, while the two horizontal lines stand for the eccentricities of an object in co-orbital motion with the considered planet $P$ when it crosses the orbit of the inner and the outer planet (respectively in green and purple) with respect to $P$. The figures are already used in \citet{DiRuzza2023}.}
    \label{fig:planet_map}
\end{figure*}

In \citet{DiRuzza2023}, co-orbital asteroids of Venus, Earth and Jupiter have been analyzed to show a practical application of the $(\theta,e)$-map just explained. After a suitable filtering on the asteroid orbital elements in order to fulfill the resonance condition and the quasi-coplanar configuration at a given epoch, the ephemerides of asteroids have been computed by means of JPL HORIZONS API service \citep{NASAHor} for an interval of time of about 900 years. The real data have been compared with the theoretical model and a very good correspondence has been found. Asteroids in quasi-coplanar co-orbital motion with Venus, Earth and Jupiter have been cataloged according to their co-orbital dynamics and their representation can be seen in Fig.~\ref{fig:planet_map}. A very refined analysis has been done checking {\it by hands} if the time series $(t,\theta)$ of each asteroid (as represented in Fig.~\ref{fig:QSHSTP_real}) was in agreement with its position in the $(\theta,e)$-map (Fig.~\ref{fig:planet_map}). The results presented in \citet{DiRuzza2023} are very promising for TP, HS and QS motion: under given assumptions, data of real observations fit very well with theory. The analyzed series comprised also transitions (TR) between different co-orbital regimes as well as the compound (CP) motion (a particular combination between QS and HS dynamics)\footnote{We refer to \citet{namouni99,namouni_etal99} for more details about the appearance of these kinds of motion.}. In this case, the map was not able to accurately catch the behaviour, as expected, since TR and CP are proper of the three-dimensional model, not of the planar one.

At this point, an automatic tool capable of distinguishing the different co-orbital regimes becomes essential in order to improve our study. Indeed, in the future we aim to extend the analysis for a longer time span (order of thousands of years or more), to consider the spatial problem including asteroids with very high inclination and to understand better and classify TR and CP motions. All these information would be desirable to create a complete catalogue of asteroids in co-orbital motion with all the planets in the solar system. 

For these reasons, a ML approach in this problem is highly recommended in order to deal with a huge number of very long time series that can exhibit very rich dynamical behaviors. The aim of the present and coming works is to become able to manage any kind of real data, for short, medium and long timescales also when transitions between different co-orbital motions occur or when new kinds of motion appear, as, for example, the compound motions. In what follows, we will consider only TP, HS and QS orbits since the foundations of the work are the results obtained in \citet{DiRuzza2023}. In particular, we will classify co-orbitals motions belonging to the four classes  QS, HS, TPL4 (a tadpole around the equilibrium position $L_4$) and TPL5 (a tadpole around the equilibrium position $L_5$).

%{\color{red} [Andrea: io aggiungerei figure per CP e TR, inoltre stresserei maggiormente questa parte che da la motivazione al lavoro. Una nota a margine pero': parlare di CP e TR senza averli analizzati nei dati forse e' rischioso. Occorre far capire bene cosa analizziamo noi, rispetto a quelli che potrebbero essere obiettivi futuri. Non vorrei che i revisori si aspettassero un lavoro su CP e TR. Nella sezione successiva sui dati infatti si dice esplicitamente che vengono esclusi i casi compound e transition.]}

\section{Data}
\label{Sec_data}

%\textcolor{magenta}{Io farei proprio come una descrizione della Tab.1, quindi partirei col dire consa intendiamo con il dataset Real e poi gli altri. Scriverei anche le dimensioni di tutti i dataset anche nel testo per dare un'idea di cosa intendiamo con 'pochi' e 'tanti' dati }

 Let us underline that our final goal is to be able to recognize, through the use of ML, co-orbital dynamics of real asteroids for short, medium and long timescales also when transitions between different co-orbital motions occur or when new kinds of motion appear, as, for example, the compound motions.

The data described in this section are the basis to outline the work done by the ML algorithms. As mentioned before, the information used in this work is the time evolution of the angle $\theta$, computed considering three different sources of data, as summarized in Tab.~\ref{tab:number_of_cases}.

In general, training a ML algorithm requires large amounts of data in order to provide accurate predictions. In our case, obtaining numerous time series of real asteroids with regular trends and clearly attributable to a single class (QS, HS, TPL4, TPL5) is not straightforward as real cases may present some complex behaviors, sometimes making labeling difficult and unclear. In particular, a high number of asteroids among those considered can escape from the given resonance or experience a co-orbital transitions. 

We start our work by using the time series of asteroids reported in Table~3,~4,~5 of the paper \citet{DiRuzza2023}. Looking at those tables, it is evident that most of the asteroids exhibit motions with different co-orbital dynamics and, as previously stated, these cases must be excluded so that, as shown in Tab.~\ref{tab:number_of_cases}, the real cases dataset used in the current work turns out to be composed by only 50 series, that is an absolutely insufficient number for a training set.

To overcome this issue, a dataset containing simulated data of ideal cases is introduced. This kind of data can be produced by using suitable model and initial conditions (as depicted in the following) in order to get the four desired classes. It is possible to obtain as many cases as we need and we produced a total number of 1999 time series of ideal cases. This dataset allows us to train the ML models with a consistent number of cases with well-known labels (i.e., motion clearly attributable to a single class), leaving the real cases dataset for testing purposes.

On the other hand, to have more data to evaluate the performance of our pipeline, we decided to increase the number of cases that can be used. To this aim, we generated time series deviating from the ideal ones by perturbing the model used to generate ideal cases. This process only partially enlarges the number of cases to be used; in fact, by adding perturbations, the time series become more similar to real cases and most of them must be eliminated because escapes from the resonance or transitions between different co-orbital regimes appear. For this reason, the number of perturbed cases can not be as large as the ideal ones. As reported in the last row of Tab.~\ref{tab:number_of_cases}, the total number of produced perturbed series is 347. 

A detailed description of how the data are obtained is provided below. 

\begin{enumerate}
\item[1.] Real ephemerides are obtained from the JPL HORIZONS system \citep{NASAHor}, following the approach adopted in \citet{DiRuzza2023}. In this case, from the database analyzed in \citet{DiRuzza2023}, we have selected 50 asteroids that exhibit a regular tadpole, horseshoe, quasi-satellite behavior, that is, we excluded the compound motions and transitions.
In this case, the simulated data cover an interval of time equal at most to 900 years. We refer to these data as {\it real data}.
\item[2.] Ideal cases of TP, HS, QS motions are generated by propagating the equations of motion of the Circular Restricted Three-Body Problem (CR3BP) with initial conditions obtained from the $(\theta, e)$-map in the corresponding orbital domain (see Fig.~\ref{fig:Map}). In this case, the initial condition in the synodic reference system is computed starting from the heliocentric orbital elements $(a,e,i,\omega,\Omega,M)$ in the inertial system, by assuming the initial semi-major axis $a$ equal to 1, the eccentricity $e$ given by the map, the initial inclination  $i$, the longitude of the ascending node $\Omega$ and the mean anomaly $M$ equal to 0 and the argument of pericenter $\omega$ equal to $\theta$. In this case, the simulated data cover an interval of time equal to 3000 years. We refer to these data as {\it ideal simulated data} and we produced a total number of 1999 time series of such cases. 
\item[3.] Perturbed cases from the ideal cases are computed by propagation of initial conditions obtained from the $(\theta, e)$-map, considering a dynamical model that accounts for Sun, Moon and the planets from Mercury to Mars. The propagation is performed by means of REBOUND \citet{ReinLiu2012}, taking the initial states for the massive bodies from \citet{NASAHor} assuming as initial epoch $t_0=JD \, \, 2305537.5$. The initial orbital elements for the asteroids are taken as above, except that now the argument of pericenter is set as $\omega=\theta+\lambda_{Earth}$, where the mean longitude of the Earth $\lambda_{Earth}$ is given by $\lambda_{Earth}=\omega_{Earth}+\Omega_{Earth}+M_{Earth}$ with $\, \omega_{Earth},\, \Omega_{Earth}, \, M_{Earth}$ being, respectively, the argument of pericenter, the longitude of the ascending node and the mean anomaly of the Earth at $t_0$. Also in this case, the simulated data cover an interval of time equal to 3000 years. We refer to these data as {\it perturbed simulated data} and we produced a total number of 347 time series for this dataset. They present variations to the ideal cases that resemble the behavior of real objects, although no further perturbations have been added otherwise the motion more frequently escapes from the resonance. However, we consider this dataset to test algorithms trained on ideal simulated data.
\end{enumerate}
We note that data produced as described in point 2. and 3. above could be also interpreted as a good test of the results obtained in the previous paper \citet{DiRuzza2023}. Indeed, we have chosen initial conditions $(\theta,e)$ in the $(\theta,e)$-map and propagated them in order to obtain the desired kind of co-orbital motion.

\begin{table}
\begin{center}
\caption{Summary of the data available.}\label{tab:number_of_cases}
\label{Tab_Data}
\begin{tabular}{|c|c c c c|c|} 
\hline
\textbf{Series} & HS & QS & TPL4 & TPL5 & Total \\ 
\hline
Real & 14 & 15 & 11 & 10 & 50\\ 
\hline
Ideal Simulated & 668 & 528 & 581 & 222 & 1999 \\
\hline
Perturbed Simulated & 61 & 54 & 147 & 85 & 347\\
\hline
\end{tabular}
\end{center}
\end{table}

%\subsection{Data strategy}
%\label{data_strategy}

 %In Sec.~\ref{sec:results}, we will describe how these kinds of data are used in the chosen ML techniques and we will show the robustness of the results obtained {\color{red} [Andrea: non mi torna il riferimento alla sezione risultati. Forse andrebbe fatto riferimento alla sezione precedente dove si parla del training degli algoritmi di ML. Non mi torna tanto neppure l'inglese "these kinds of data are used in the chosen ML techniques". Dipende cosa volete dire: dati utilizzati per addestrare/testare algoritmi ML? Non capisco bene cosa volete dire.]}.

\section{Data analysis workflow}
\label{pipeline}

\begin{figure*}
    \centering
    \includegraphics[width=0.8\textwidth]{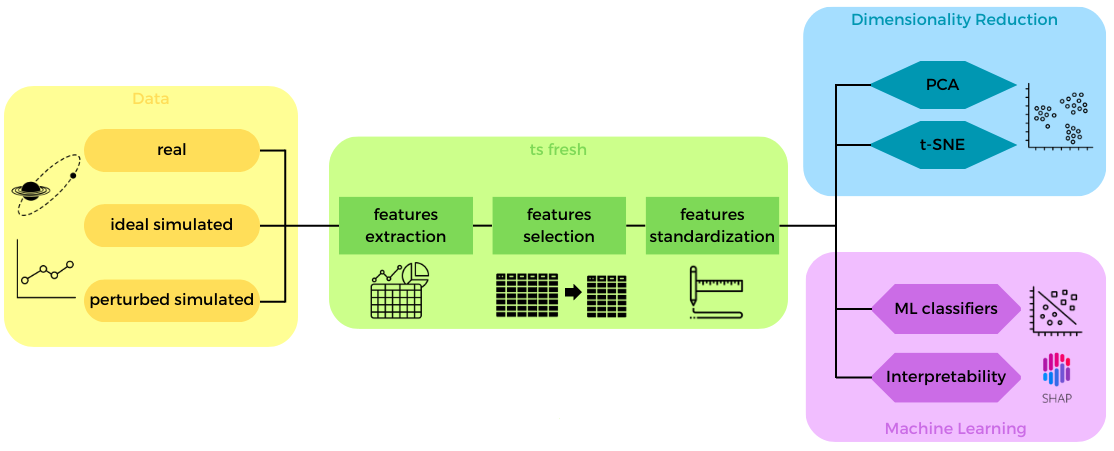}
    \caption{Data Analysis Workflow. The first step is the time series preparation, followed by the \textit{tsfresh} python package block where features are extracted and possibly selected and standardized. The final step regards the Machine Learning analysis performed using Dimensionality reduction algorithms (PCA and t-SNE) and classification algorithms (SVM, Random Forest and XGBoost).}
    \label{fig:workflow}
\end{figure*}

%The purpose of this study is to provide a Machine Learning based data analysis workflow aiming to the classification of time series coming from the co-orbitals motions belonging to the four classes  QS, HS, TPL4 (a tadpole around the equilibrium position $L_4$), TPL5 (a tadpole around the equilibrium position $L_5$), introduced above.

As shown in Fig.~\ref{fig:workflow}, our data analysis workflow can be conceptually divided in three macro blocks.
The first step consists in preparing and labelling the data described in Sec.~\ref{Sec_data}, i.e., the output of the propagation of orbital elements of the asteroids. The data are collected in .out format files: each file is associated with a single asteroid and it contains 7 columns corresponding, respectively, to time (in Julian date), elapsed time in years (starting from $t_0$), semi-major axis $a$, eccentricity $e$, inclination $i$, resonant angle $\theta$ and associated action $u$. The filenames contain acronyms useful to recognize the name of the asteroid, the kind of co-orbital motion, the planet that the asteroid is in resonance with and the kind of propagation used to get the data (points 1., 2., 3. in Sec.~\ref{Sec_data}). In this way, files can be easily shared if required. It is important to stress that in this work we focus only on the time evolution of the variable angle $\theta$, but the other information can turn out to be useful for future analysis.

These tabular data are passed to the next block, where the \textit{tsfresh} python package \citep[e.g.,][]{christ2018time} provides a systematic time series feature extraction thanks to the combination of established algorithms from statistics, time series analysis, signal processing and non-linear dynamics. 

Before giving the extracted features to the Machine Learning classification algorithms, two additional steps can be applied: selection and standardization. Selection can be performed thanks to \textit{tsfresh}, which represents a robust feature selection algorithm \citep[e.g.,][]{li2017feature}, while standardization can be obtained by any kind of library such as Scikit Learn pre-processing functions \citep[e.g.,][]{scikit-learn}. 

The final classification step (last two blocks in Fig.~\ref{fig:workflow}) is performed in two parallel branches, with two classes of ML algorithms involved, namely, Dimensionality Reduction and classification algorithms.

Before moving into a deeper explanation of all the details regarding the steps involved in the data analysis workflow, it is worth noting how our approach based on features extraction and standard Machine Learning algorithms is very well suited for our case where we have two constraints: data numerosity and physical interpretability. Both these constraints encourage an approach based on Machine Learning algorithms where the requirement on the number of data to train the algorithm is less tight with respect to Deep Learning. At the same time, thanks to the features extraction, a time series of any length can be converted into a finite number of features, all of them holding a physical meaning. This physical meaning is deeply important, because not only at the end of the whole data analysis workflow it is possible to identify the most important features responsible for a good time series classification (Feature Importance), but in addition we can look at the discriminating features between the different classes of signals, recovering a physical understanding of such processes.

\subsection{Features extraction and selection: the \textit{tsfresh} open-source package} 

In order to train a ML model, features need to be extracted from the data. In our case a total of 789 features are extracted from each time series representing the time evolution of the angle $\theta(t)$ by the Python package \textit{tsfresh} \citep[e.g.,][]{christ2018time}. For a detailed description of the meaning of each feature please refer to \citet{tsfresh-github}.

After feature extraction, usually, it is worth to introduce a step of \textit{Feature Selection}. This step can be performed in different ways or not performed at all. However, in general, it has been demonstrated \citep[e.g.,][]{guyon2003introduction} that Feature Selection can improve ML performances. Therefore, we decided to implement such step in our workflow using a built-in function of \textit{tsfresh}, which provides a feature selection method based on Mann-Whitney Test. In our case, this step reduces the number of features to 239.

\subsection{Features standardization}
Again, pre-processing data is an essential step to achieve good classification performance, with the importance of data standardization (or normalization) for improving the performance of ML algorithms described in many studies as stated in \citet{singh2020investigating}. In our study, features are standardized using the Scikit Learn function StandardScaler \citep[e.g.,][]{scikit-learn}.

\subsection{Dimensionality Reduction}
\label{dim_red}

The process of transforming data from a high-dimensional space into a low-dimensional space with the goal of keeping the low-dimensional representation as close as possible to the inherent dimension of the original data is known as \textit{ Dimensionality Reduction}. There exist many different ML algorithms able to perform such transformation on data. In this work, we focus on two of them, namely, \textit{Principal Components Analysis} (PCA) \citep[e.g.,][]{cozzolino2019interpreting} and \textit{t-distributed Stochastic Neighbor Embedding} (t-SNE) \citep[e.g.,][]{van2008visualizing,pmlr-v75-arora18a,kobak2019art}. PCA and t-SNE operate in two different ways: PCA is a linear method that seeks to preserve as much variance as possible and the global structure of the data, while t-SNE is a non-linear optimized technique that concentrates on preserving local similarities between data points. Additionally, PCA uses a well-known transformation making it a  deterministic technique. On the other hand, t-SNE is a stochastic optimized method, which tend to preserve points which are close to each other. However, the method doesn’t construct an explicit function that maps high dimensional points to a low dimensional space, but it just optimizes low dimensional positions of the data points directly. Since it does not define a data transformation function, the method cannot be applied to newer data, but a newer optimization must run.

Both algorithms are Dimensionality Reduction techniques particularly well suited for the visualization of high-dimensional datasets as in this case, where, after the feature selection step, the number of features is still above 200. The utility of such kind of algorithms is twofold: on the one hand they can be used as unsupervised learning methods which allow to visualize the data distribution in two dimension, providing a deep insight on whether and, in case, how the data can be divided in the higher dimensional space. Moreover, they usually can give an idea of how the classifiers will perform. Indeed, well clustered data visualized by Dimensionality Reduction methods are usually well classified by ML algorithms, whereas the contrary is not necessarily true, meaning there could be data with a low degree of clustering where the classification algorithms still perform very well.

\subsection{ML classification}
\label{sec:ML_class}

We use three ML algorithms: Support Vector Machine (SVM) \citep[e.g.,][]{cervantes2020comprehensive}, Random Forest (RF) \citep[e.g.,][]{biau2016random} and XGBoost (XGB) \citep[e.g.,][]{chen2016xgboost}. We evaluate the performances of these algorithms with different combinations of training and test sets, as reported here:
\begin{enumerate}
    \item trained on real data and tested on real data;
    \item trained on ideal simulated data and tested on real data;
    \item trained on ideal simulated data and tested on perturbed simulated data;
    \item trained on ideal simulated data and tested on real and perturbed simulated data.
\end{enumerate}

\subsubsection{Cross-Validation}
\label{CV}

When evaluating the performances of a ML model, it is highly important to validate its stability. This step is called \textit{validation} and it consists in making sure that the model has learned the right patterns of the data and it is not picking up too much noise. In other words, it evaluates the model's ability to generalize on unseen data.

In Machine Learning, the most used validation technique is \textit{Cross-Validation} (CV). It consists in splitting the dataset into multiple subsets, usually called \lq\lq folds\rq\rq, then training the model on some of the folds and evaluating it on the remaining fold. This process is repeated multiple times, each time changing the remaining fold. The result is the mean score of all the performed tests. This allows to train and test the model on different data partitions, providing a robust and unbiased estimate of a model's performance.

There are many types of Cross-Validation; for this work we use a technique named \textit{k-folds Cross-Validation} \citep[e.g.,][]{fushiki2011estimation}, where the dataset is divided in $k$ folds and $k-1$ folds are used as training set and the remaining one as test set.

\subsubsection{Hyperparameters Tuning}
\label{tuning}

When dealing with a ML model, one of the main aspects of designing the structure is a step called \textit{Hyperparameters Tuning}, which consists in finding the best combinations of hyperparameters' models in order to achieve the best performance. Unfortunately, there are no rules or formulas to calculate these parameters, and an approach based on an extensive exploration of the hyperparameters' space along with some experience is the only way to find them, making hyperparameters tuning a computationally long and tedious process. In Python, many techniques have been developed to automate the tuning of hyperparameters and in this work we apply two of them: \textit{GridSearchCV} and \textit{RandomizedSearchCV}. Both these techniques make use of \textit{k-fold Cross-Validation}.

\subsubsection{SHAP: features interpretability}
\label{shap}
Machine Learning models are frequently considered \lq\lq black boxes\rq\rq, which make their interpretation challenging. In order to understand the main features that affect the output of the model, we can leverage on Explainable Machine Learning techniques that can unravel some of these aspects \citep[e.g.,][]{roscher2020explainable}. One very promising technique is the SHapley Additive exPlanations, more commonly known as SHAP \citep[e.g.,][]{NIPS2017_7062,lundberg2018explainable,lundberg2020local2global,van2022tractability, mitchell2022gputreeshap}. It is based on Shapley values, which use game theory to assign credit for a model’s prediction to each feature or feature value, increasing the transparency and the interpretability of Machine Learning models \citep[e.g.,][]{molnar2022}. 
In particular SHAP is known for its "Consistency" property. SHAP values do not change when the model changes unless the contribution of a feature changes. This means that even when the model architecture or parameters change, SHAP values still offer a coherent interpretation of the behaviour of the model.

In our case, SHAP is applied to the ML models used for time series classification.

\section{Results}
\label{sec:results}

The results are presented in the following,  according to the considered techniques.

\subsection{Unsupervised ML: PCA and t-SNE}
\begin{figure*}
  \centering
  \vspace{-10ex}
  \subfloat[]{\includegraphics[width=\columnwidth]{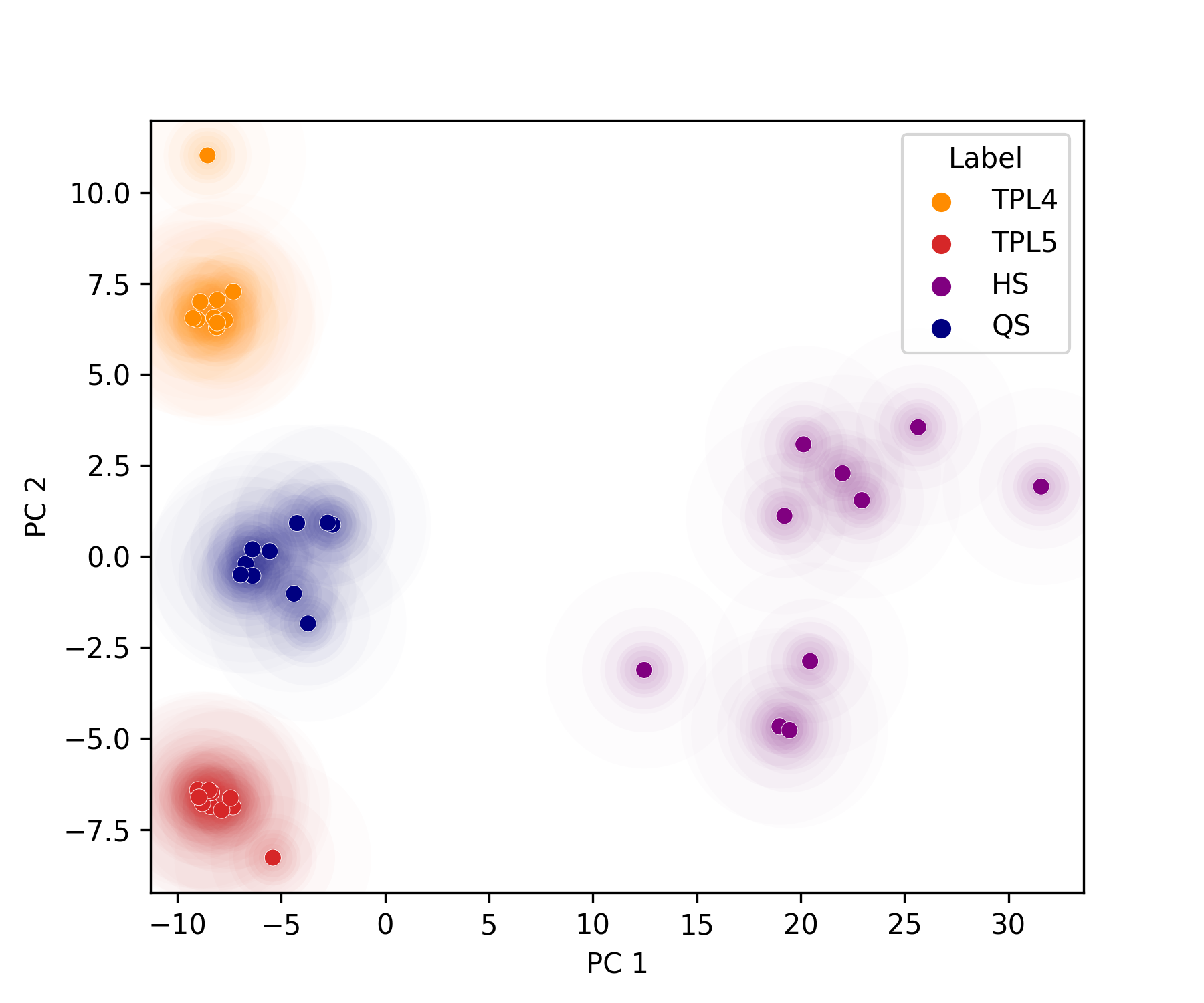}\label{fig: pca_real}}
  \hfil
  \subfloat[]{\includegraphics[width=\columnwidth]{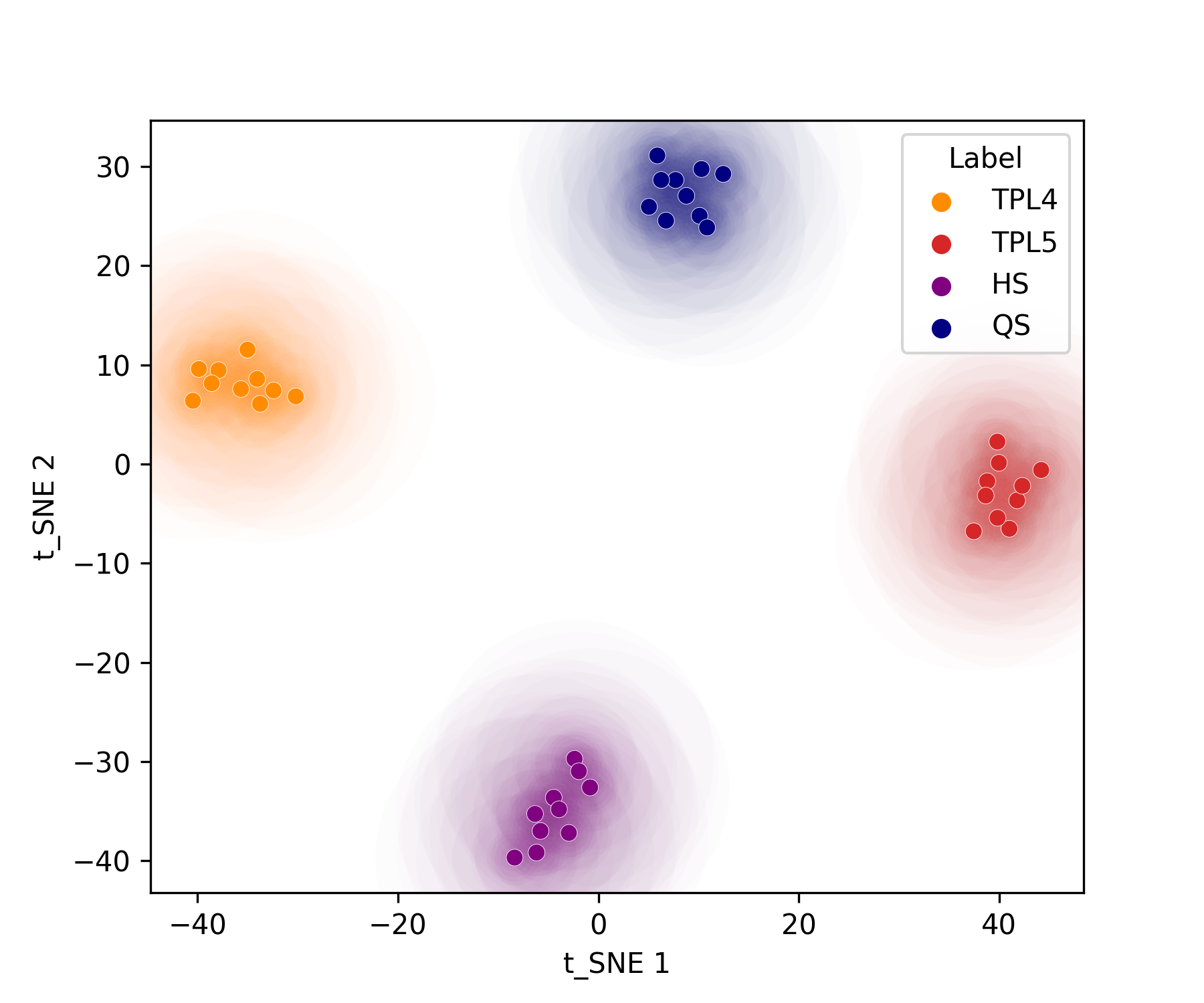}}
  \vspace{-3ex}
  \subfloat[]
  {\includegraphics[width=\columnwidth]{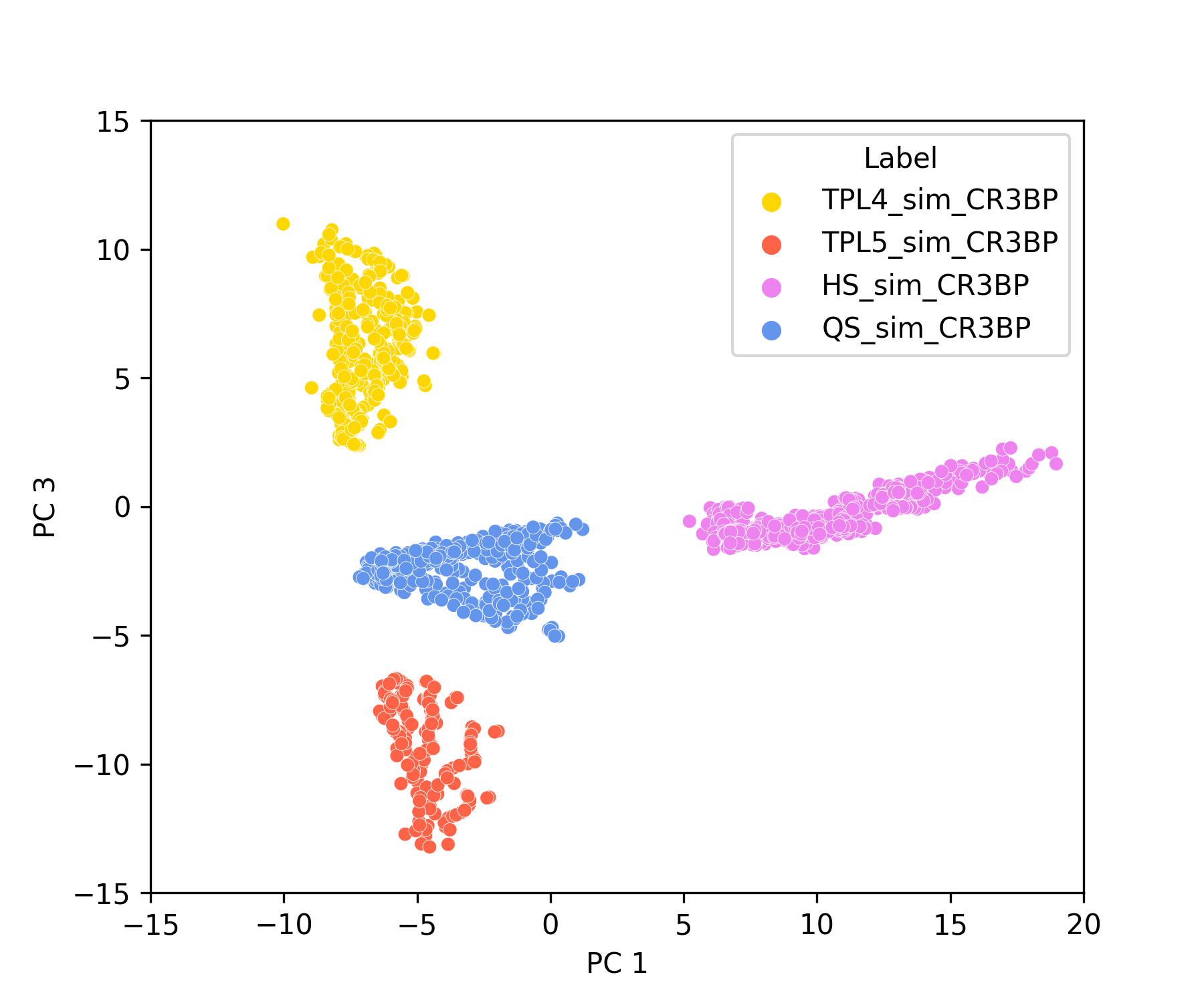}}
  \hfil
  \subfloat[]
  {\includegraphics[width=\columnwidth]{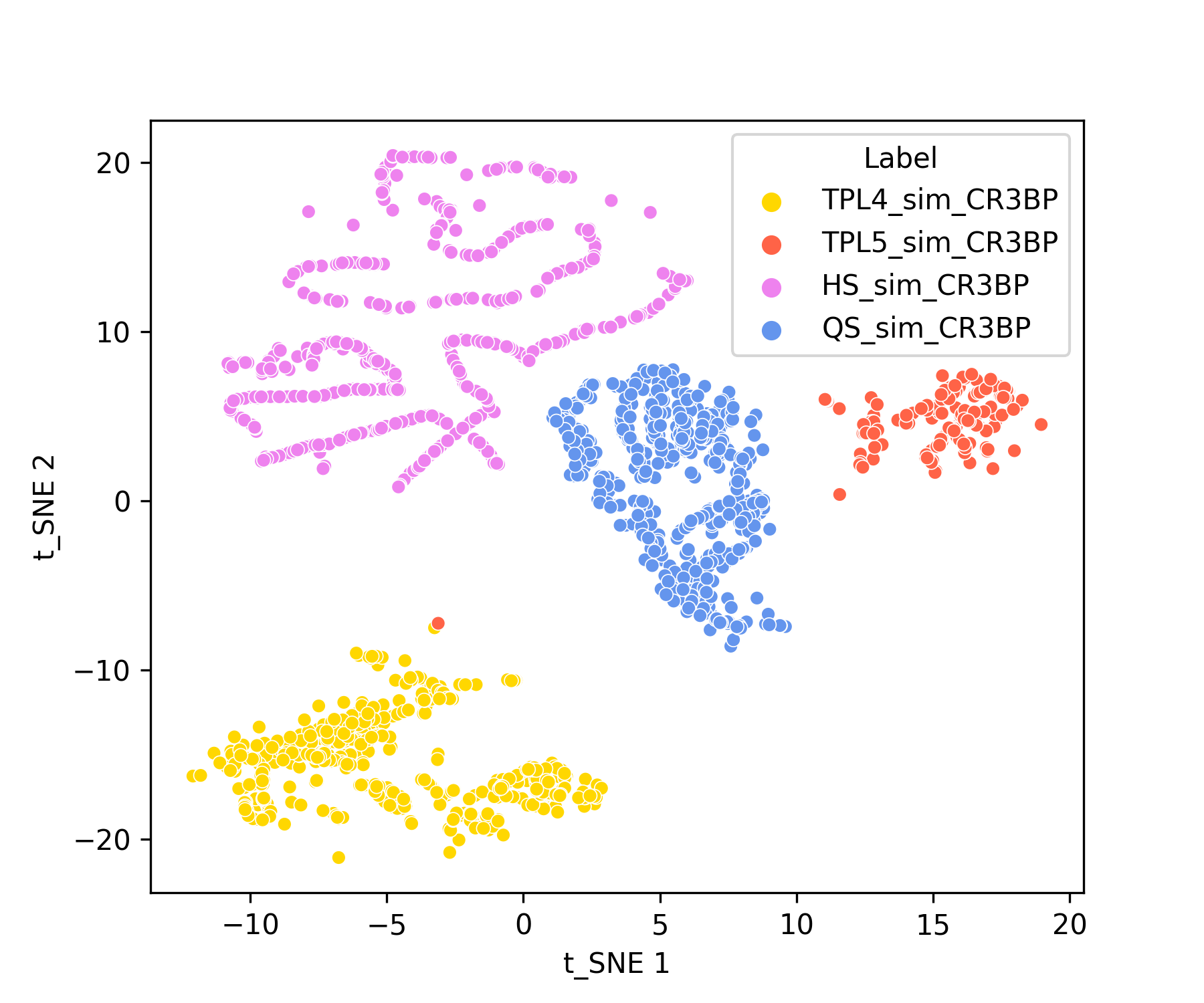}}
  \vspace{-3ex}
  \subfloat[]
  {\includegraphics[width=\columnwidth]{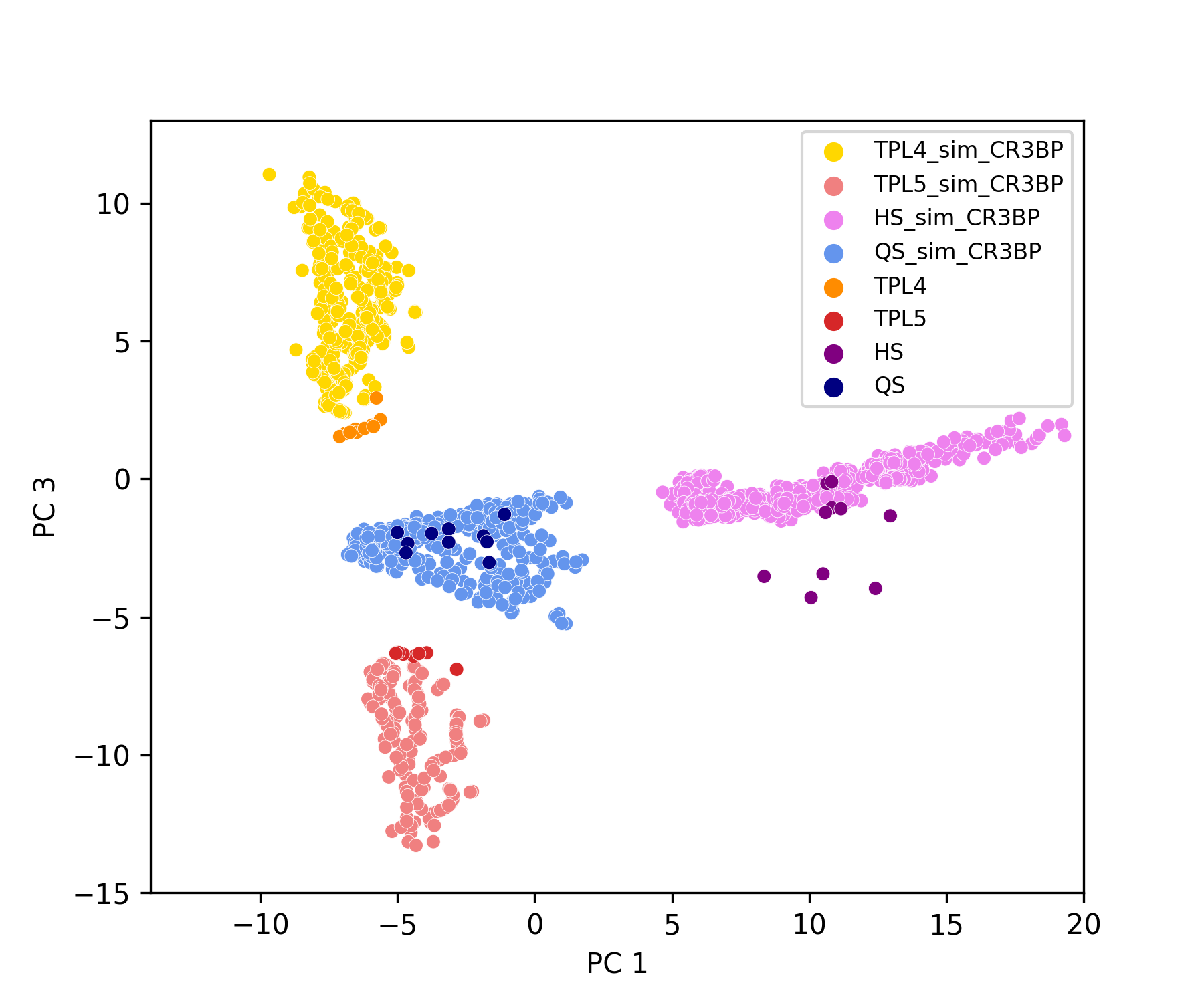}}
  \hfil
  \subfloat[]
  {\includegraphics[width=\columnwidth]{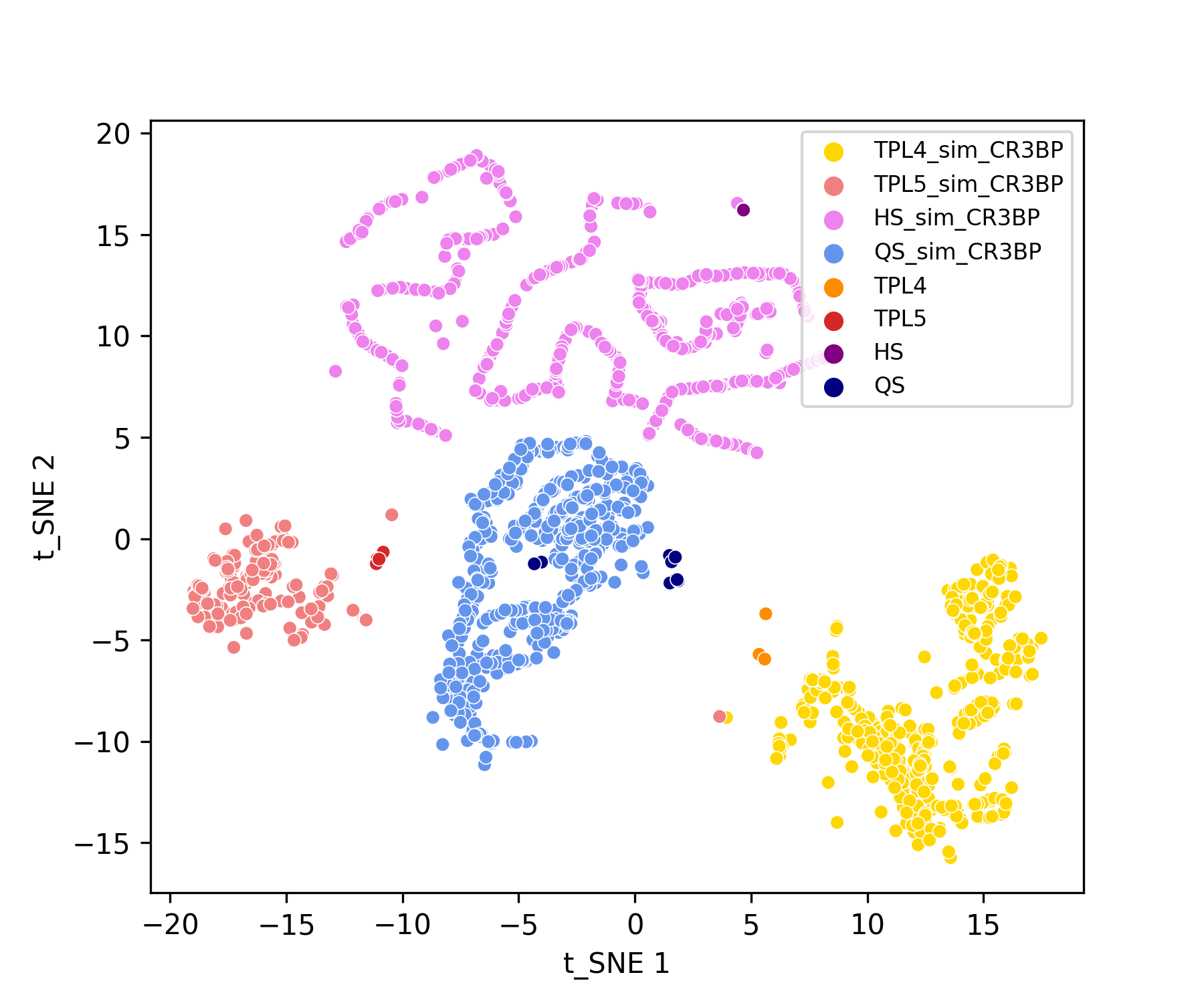}}
  \caption{PCA and t-SNE of selected and standardized features extracted from: real data (a) and (b); ideal simulated data (c) and (d); overlapping between ideal simulated and real data clusters (e) and (f). In this last case it is worth to note as the orange points representing the real TPL4 cases overlap the yellow points representing the simulated TPL4 cases; the red points representing the real TPL5 cases overlap the light-red points representing the simulated TPL5 cases; the purple points representing the real HS cases overlap the violet points representing the simulated HS cases; the blue points representing the real QS cases overlap the light-blue points representing the simulated QS cases.}
  \label{fig: pca_tsne}
\end{figure*}

As stated in Sec. \ref{dim_red}, Dimensionality Reduction techniques can be used to discover whether a high dimensional dataset presents separate clusters when projected in lower dimensional space (e.g., bi-dimensional). Therefore, the first step of our analysis has been to perform PCA and t-SNE on the features extracted from the real time series (real data) to see if they would cluster into four separated groups corresponding to four classes: QS, HS, TPL4, TPL5 (described  in Sec. \ref{coorbital}). PCA and t-SNE visualizations show four well separated clusters, as can be appreciated in Fig.~\ref{fig: pca_tsne} (a), (b), respectively,  where real data are considered.

Next, we performed PCA and t-SNE on the ideal simulated data to determine whether the trend of clustering in the four groups was also present in this dataset. As it can be appreciated in Fig.~\ref{fig: pca_tsne} (c), (d), clusters are still well visible.

Finally, given the positive results of the previous tests, we have applied the Dimensionality Reduction techniques on a dataset containing both the real and ideal simulated data expecting an overlap between the real and simulated clusters for each class.  The encouraging results of this analysis are reported in Fig.~\ref{fig: pca_tsne} (e), (f). It is worth to observe that in these plots, PCA and t-SNE show the overlapping between real and simulated data clusters. In particular, the orange points representing the real TPL4 cases overlap the yellow points representing the simulated TPL4 cases; the red points representing the real TPL5 cases overlap the light-red points representing the simulated TPL5 cases; the purple points representing the real HS cases overlap the violet points representing the simulated HS cases; finally, the blue points representing the real QS cases overlap the light-blue points representing the simulated QS cases. This overlapping between clusters of real and simulated data in the reduced space confirms that the features extracted from these two datasets are similar and meaningful. In particular, these results confirm our expectations that both datasets are extracted from the same data distribution, making them suitable for the deeper machine learning analysis shown hereafter.

\subsection{Supervised ML}

While Dimensionality Reduction techniques allow to visualize high-dimensional data and eventual clusters within them, supervised ML algorithms provide an actual classification of the data. In our case, six classification metrics are considered to evaluate the supervised ML algorithms performances: \textit{Accuracy, Balanced Accuracy, ROC AUC, Recall, Precision, f1}.
A full description of the metrics can be found in \citet{Sklearn_metrics}

It is worth to note how some ML algorithms do not require features normalization, such as Random Forest, while for some others, such as Support Vector Machine, the normalization step strongly improves the classification performances \citep[e.g.,][]{singh2020investigating,ozsahin2022impact}. This peculiarity can be ascribed to the intrinsic differences in the working principles at the basis of each algorithm.

As was already noted, another crucial step that is typically (but not always) necessary to enhance classification performances is features selection. Our data shows that this is not the case; the outcomes are unaffected by the pre-processing stage. It should be highlighted, nevertheless, that this step generally needs to be preserved in the data analysis workflow. This is not the case for our data, results not being affected by this pre-processing step. However, it should be noted that in general such step must be kept in the data analysis workflow, evaluating its importance case by case. Concerning our work, the results reported in this section are then relative to datasets containing all the extracted features.

\subsubsection{Test results}

The classification performances of the three used supervised ML algorithms (SVM, RF and XGB, see Sec.~\ref{sec:ML_class}) are reported in Tab.~\ref{tab: test_results} for four different combinations of training and test sets. Although the motivations behind the chosen approach have already been partially described above, we remark the following observations. First of all, the real cases dataset is limited, therefore  it is impossible to give a clear answer regarding the generalization capability of our models to unseen data when trained and tested on real data. For this particular reason we introduced the ideal and perturbed simulated datasets, where the ideal one is intended for training purposes leaving the perturbed one to testing ones. 

The hypothesis regarding the use of the ideal simulated as training set is confirmed by the fact that the classifiers trained in this way classify correctly the real series with an accuracy that reach 98\%.  Lastly, classifiers trained on ideal simulated data and tested on perturbed simulated data obtain an accuracy of 100\% for all algorithms, while a slightly lesser accuracy is achieved testing on real and perturbed data.

\begin{table*}
\centering
\caption{Machine Learning multi-class classifiers results obtained with different combinations of training and test sets divided by algorithm.
Because this is a multi-class classification, AUC, Recall, Precision and f1 are averaged. In the Average AUC the acronym "ovo" stands for One-vs-one and it computes the average AUC of all possible pairwise combinations of classes.}
\label{tab: test_results}
\begin{tabular}{|c c c c c c c c|}
\hline
Training set & Test set & Accuracy (\%) & Balanced Acc. (\%) & "ovo" Average AUC & Average Recall & Average Precision & Average f1 \\
\hline
\multicolumn{8}{|c|}{\textbf{Support Vector Machine}}\\
\hline
Real & Real & 100 & 100 & 1.0 & 1.0 & 1.0 & 1.0\\
Ideal & Real & 98.0 & 98.3 & 0.995 & 0.980 & 0.981 & 0.980\\
Ideal & Perturbed & 100 & 100 & 1.0 & 1.0 & 1.0 & 1.0\\
Ideal & Real+Perturbed & 99.7 & 99.7 & 0.999 & 0.997 & 0.998 & 0.997\\
\hline
\multicolumn{8}{|c|}
{\textbf{Random Forest}}\\
\hline
Real & Real & 100 & 100 & 1.0 & 1.0 & 1.0 & 1.0\\
Ideal & Real & 98.0 & 98.3 & 0.998 & 0.980 & 0.981 & 0.980\\
Ideal & Perturbed & 100 & 100 & 1.0 & 1.0 & 1.0 & 1.0\\
Ideal & Real+Perturbed & 99.5 & 99.2 & 1.0 & 0.995 & 0.995 & 0.995\\
\hline
\multicolumn{8}{|c|}{\textbf{XGBoost}}\\
\hline
Real & Real & 100 & 100 & 1.0 & 1.0 & 1.0 & 1.0\\
Ideal & Real & 98.0 & 97.7 & 1.0 & 0.980 & 0.981 & 0.980\\
Ideal & Perturbed & 100 & 100 & 1.0 & 1.0 & 1.0 & 1.0\\
Ideal & Real+Perturbed & 99.7 & 99.8 & 1.0 & 0.997 & 0.998 & 0.997\\
\hline
\end{tabular}
\end{table*}

\begin{figure*}
  \centering
    \vspace{-3ex}
    \subfloat[SVM]{
    \includegraphics[scale=0.36]{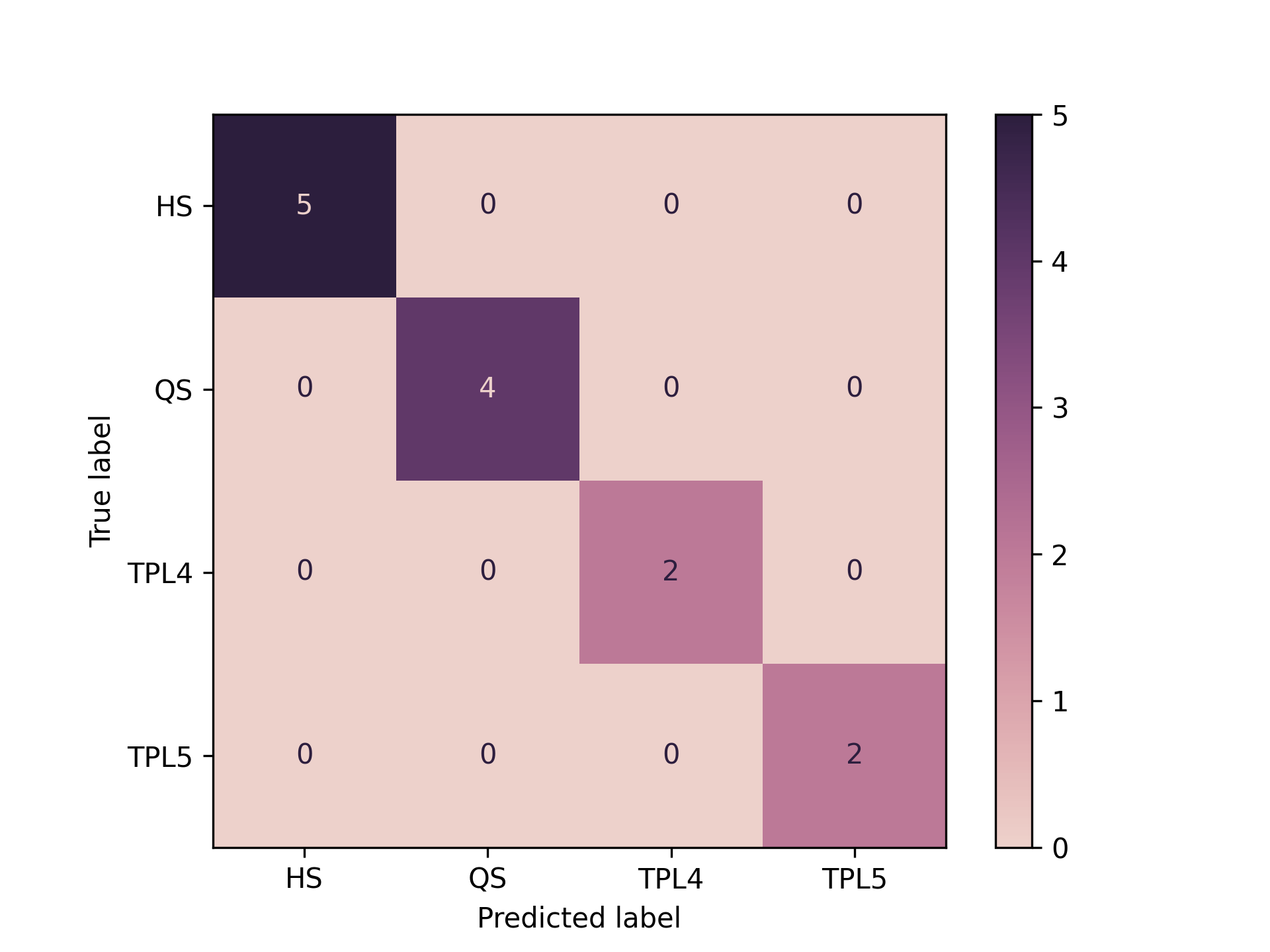}}
    \subfloat[RF]{
    \includegraphics[scale=0.36]{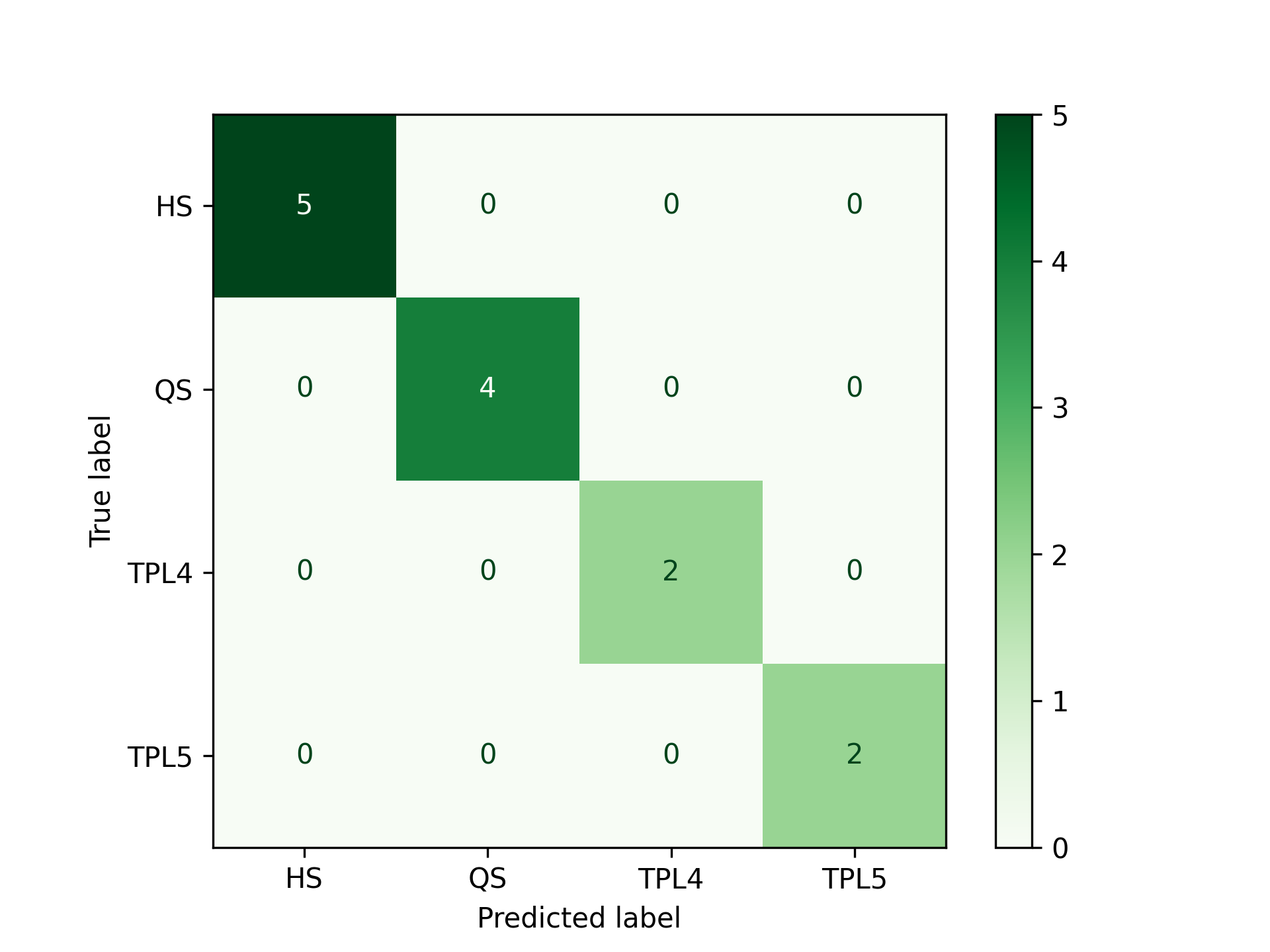}}
    \subfloat[XGB]{
    \includegraphics[scale=0.36]{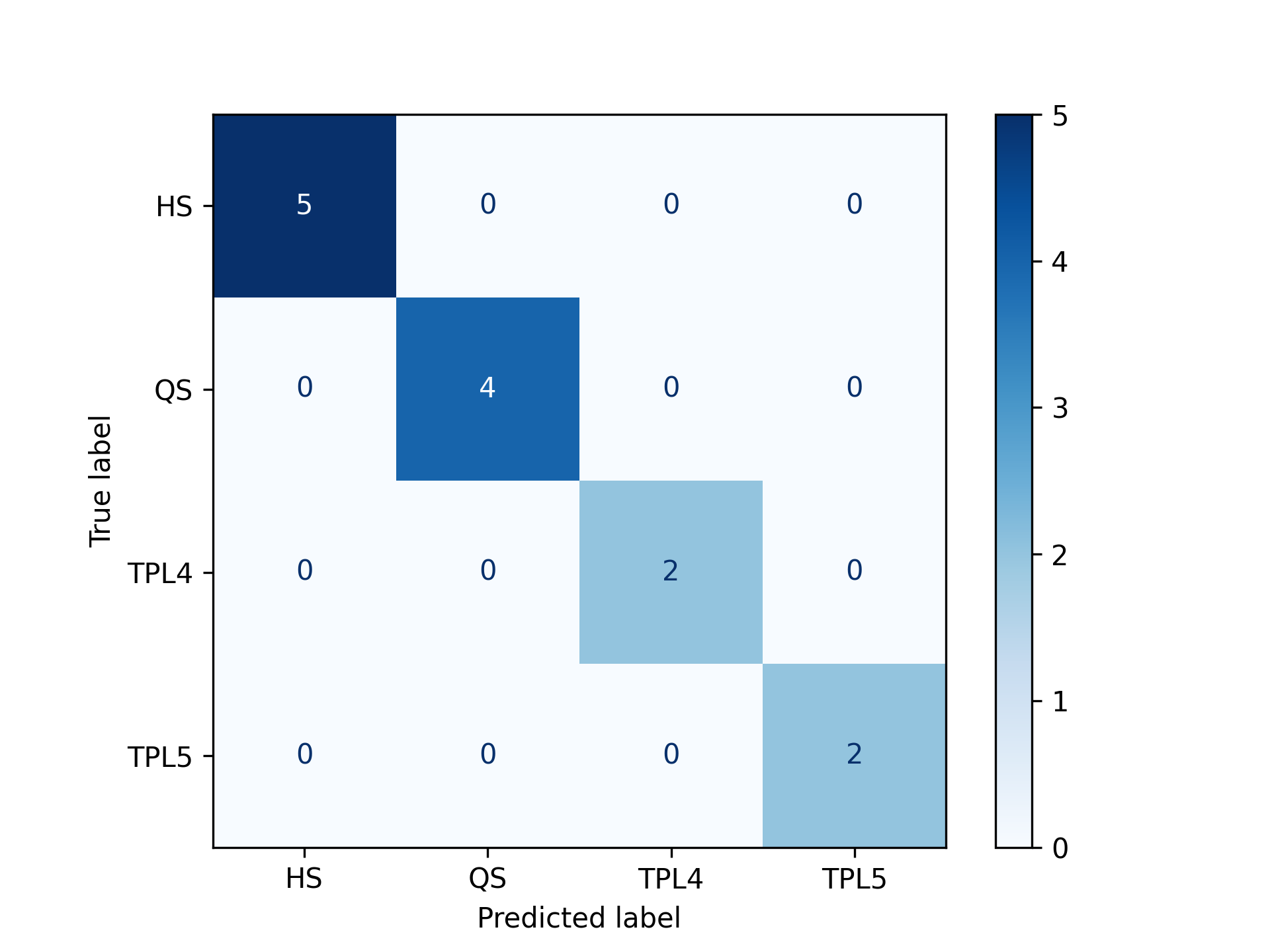}}
    
    \vspace{-3ex}
    \subfloat[SVM]{
    \includegraphics[scale=0.36]{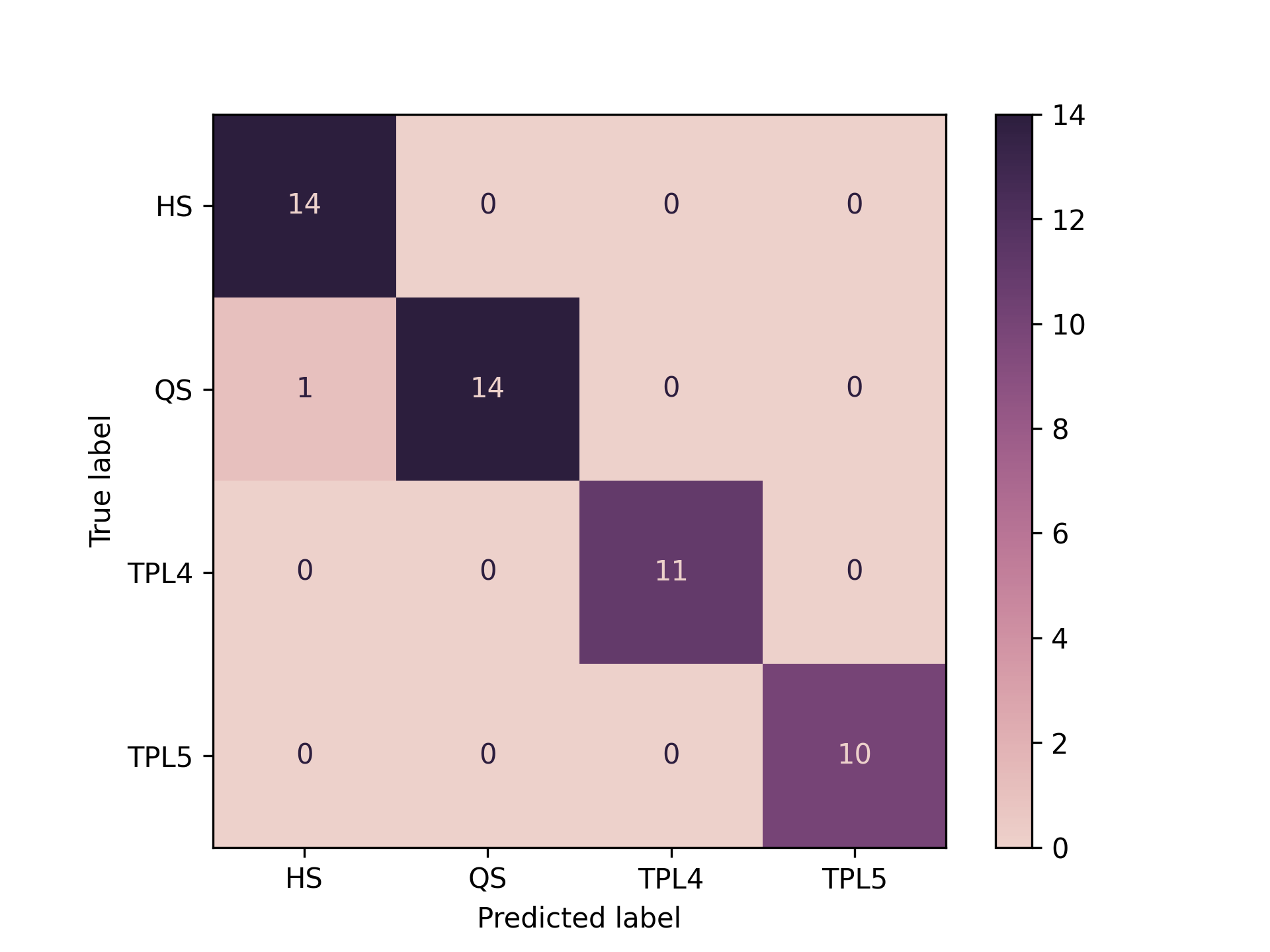}}
    \subfloat[RF]{
    \includegraphics[scale=0.36]{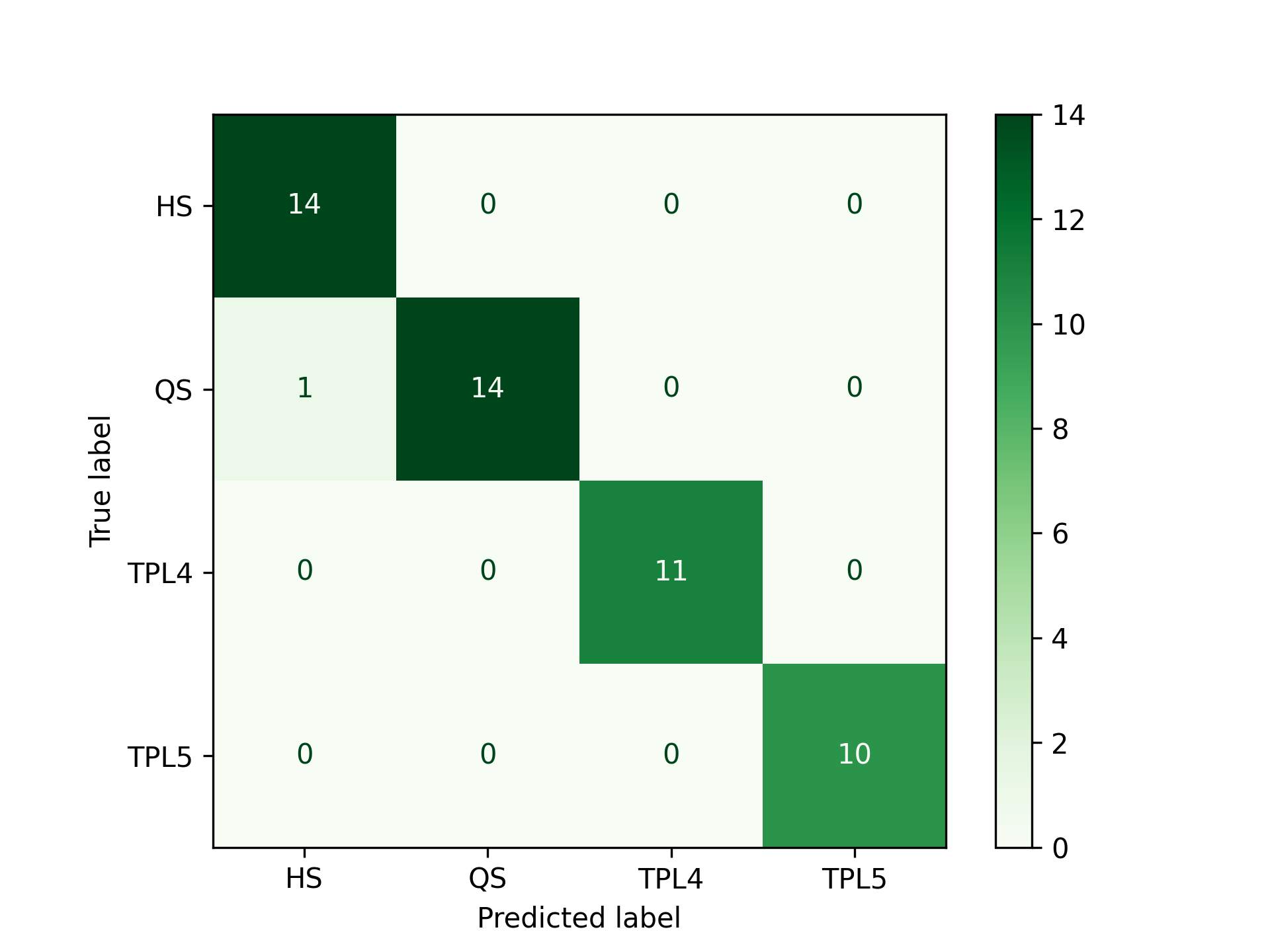}}
    \subfloat[XGB]{
    \includegraphics[scale=0.36]{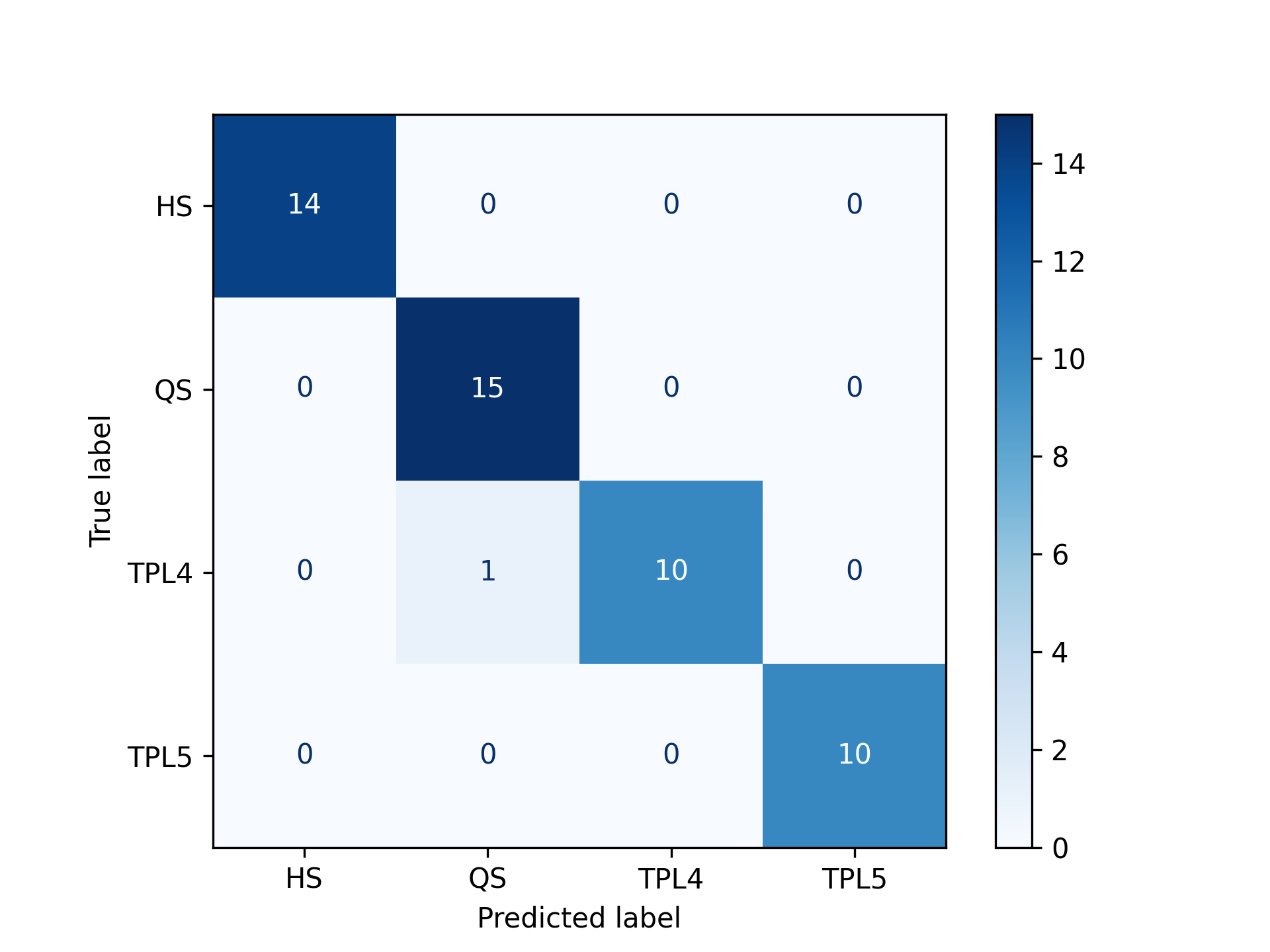}}
    
    \vspace{-3ex}   
    \subfloat[SVM]{
    \includegraphics[scale=0.36]{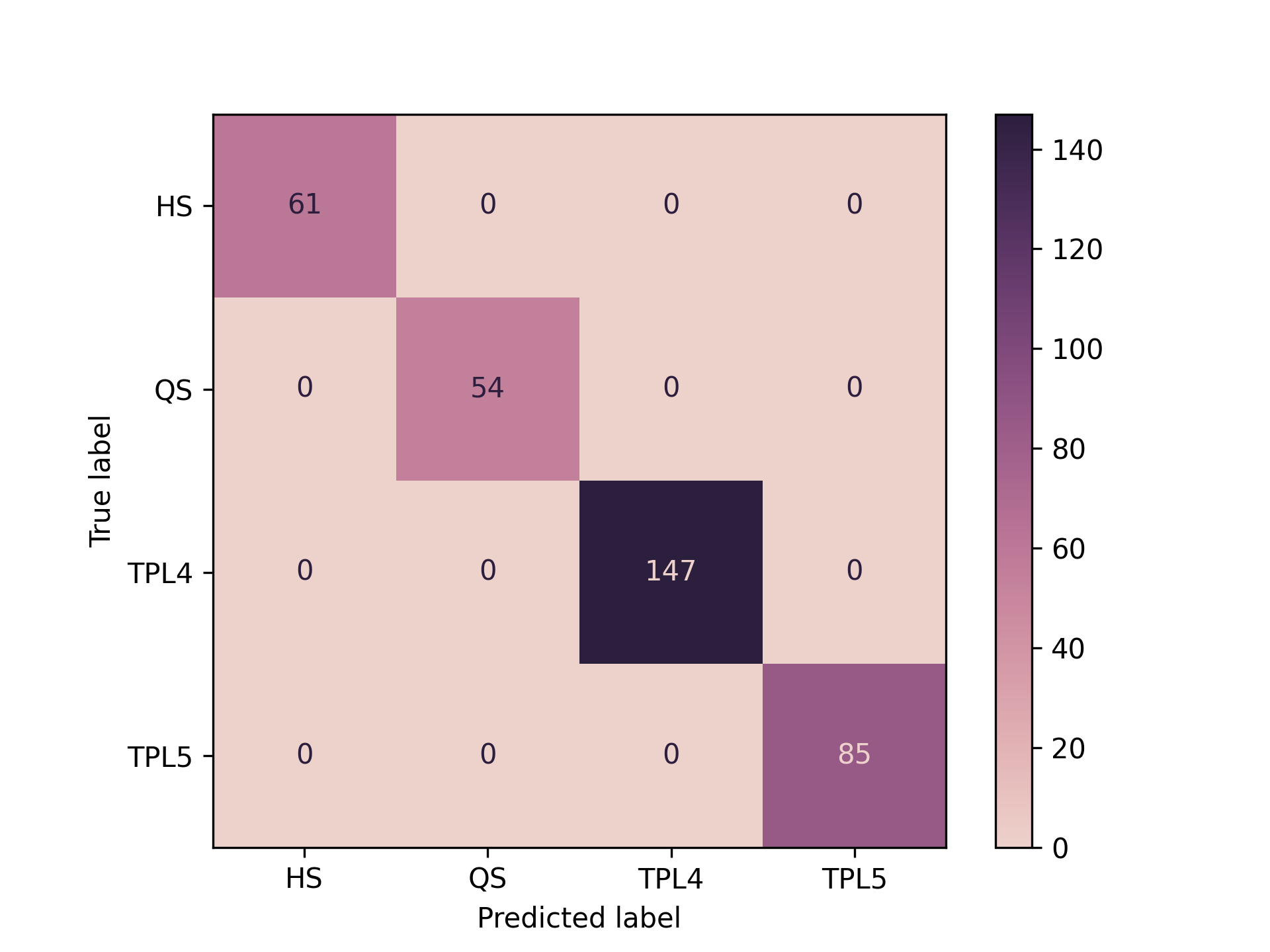}}
    \subfloat[RF]{
    \includegraphics[scale=0.36]{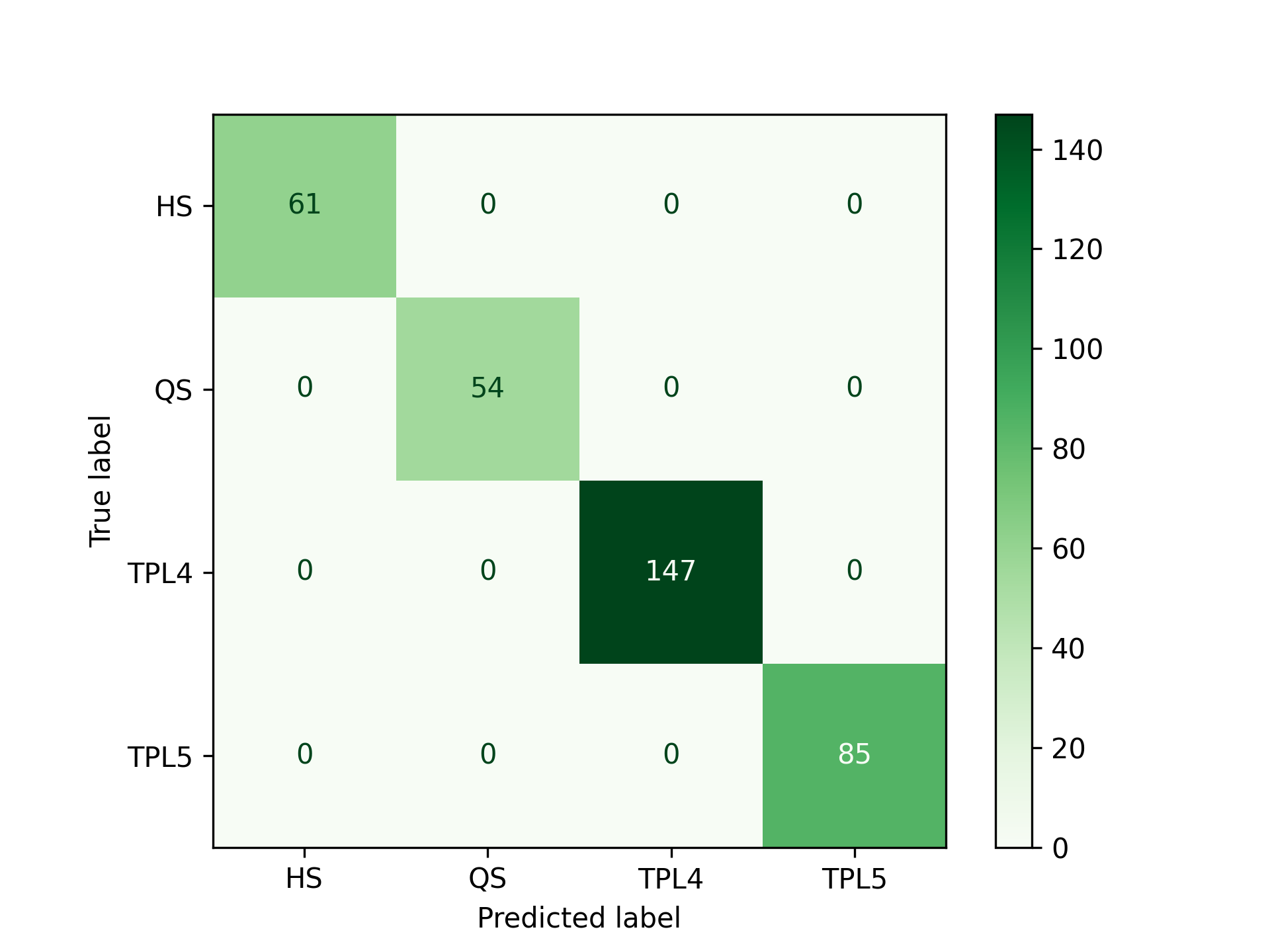}}
    \subfloat[XGB]{
    \includegraphics[scale=0.36]{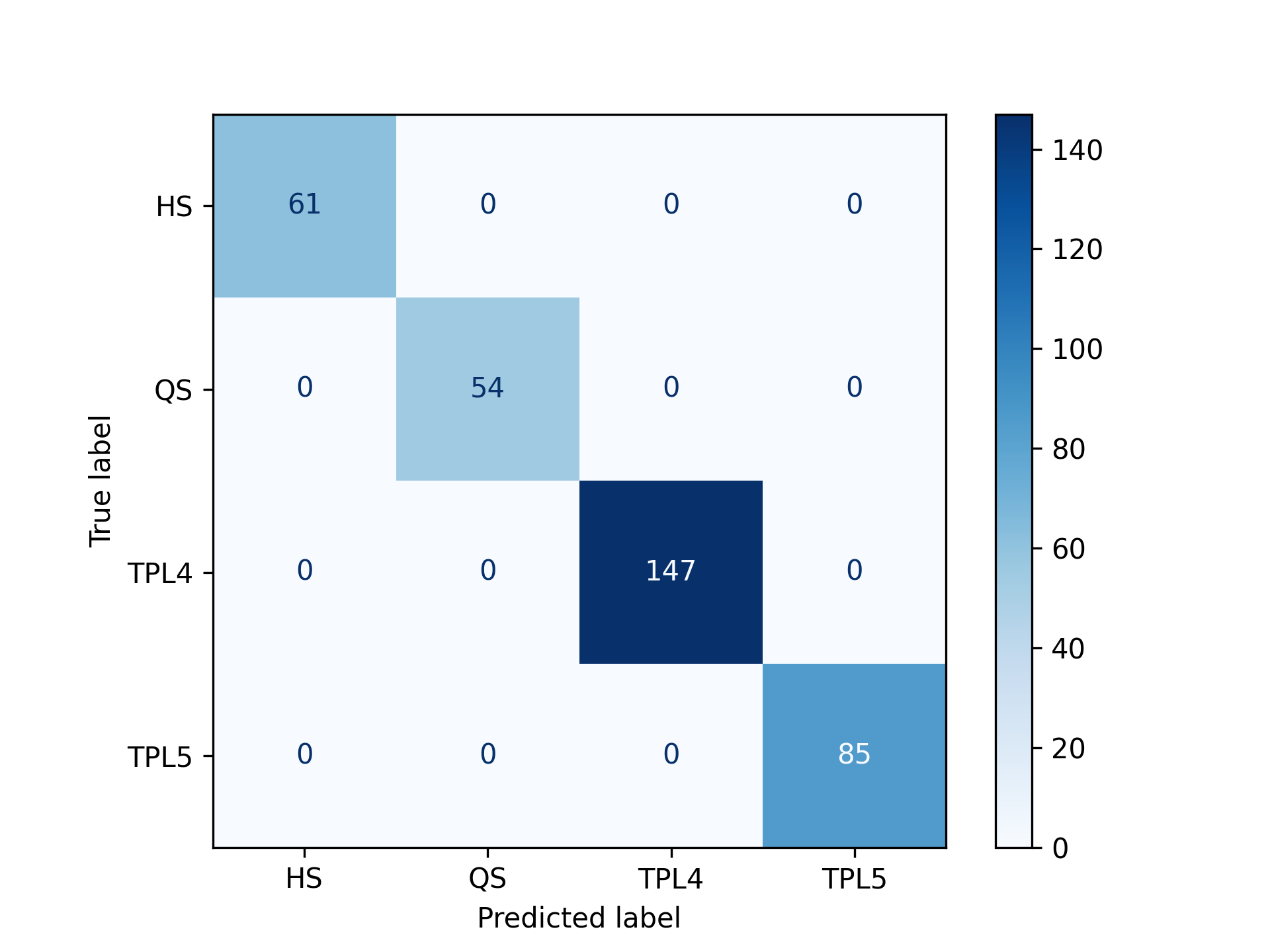}}
    
    \subfloat[SVM]{
    \includegraphics[scale=0.36]{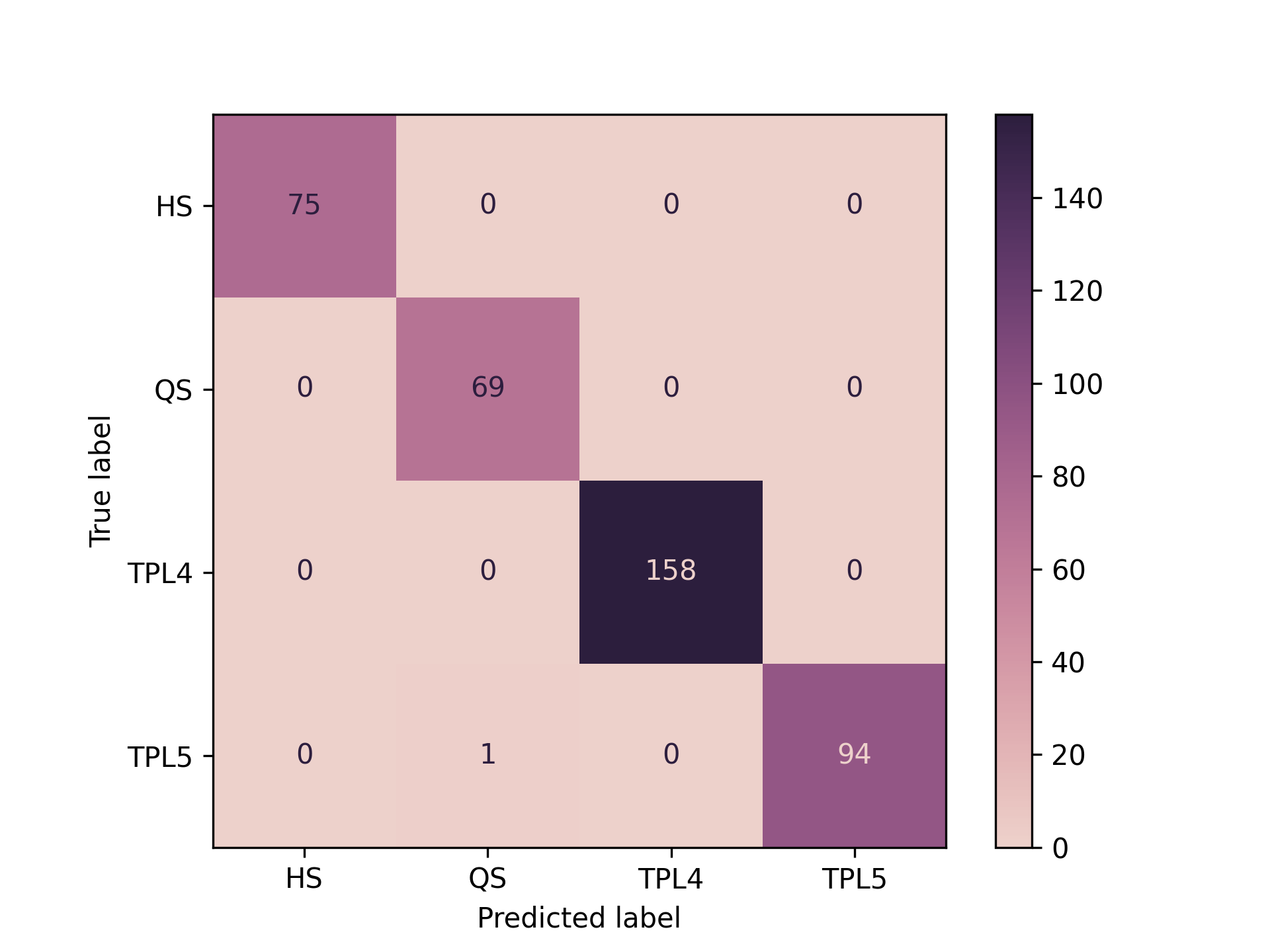}}
    \subfloat[RF]{
    \includegraphics[scale=0.36]{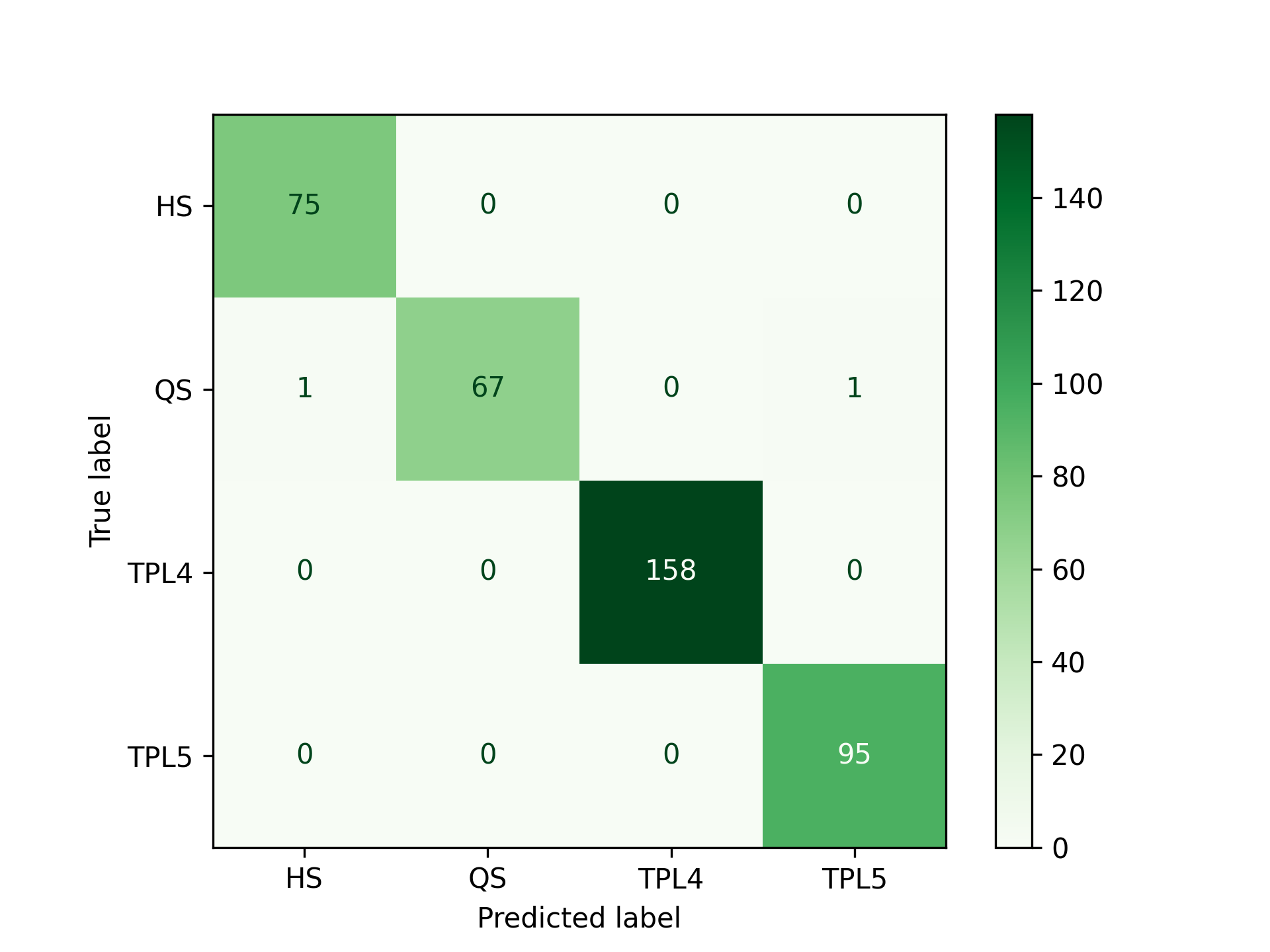}}
    \subfloat[XGB]{
    \includegraphics[scale=0.36]{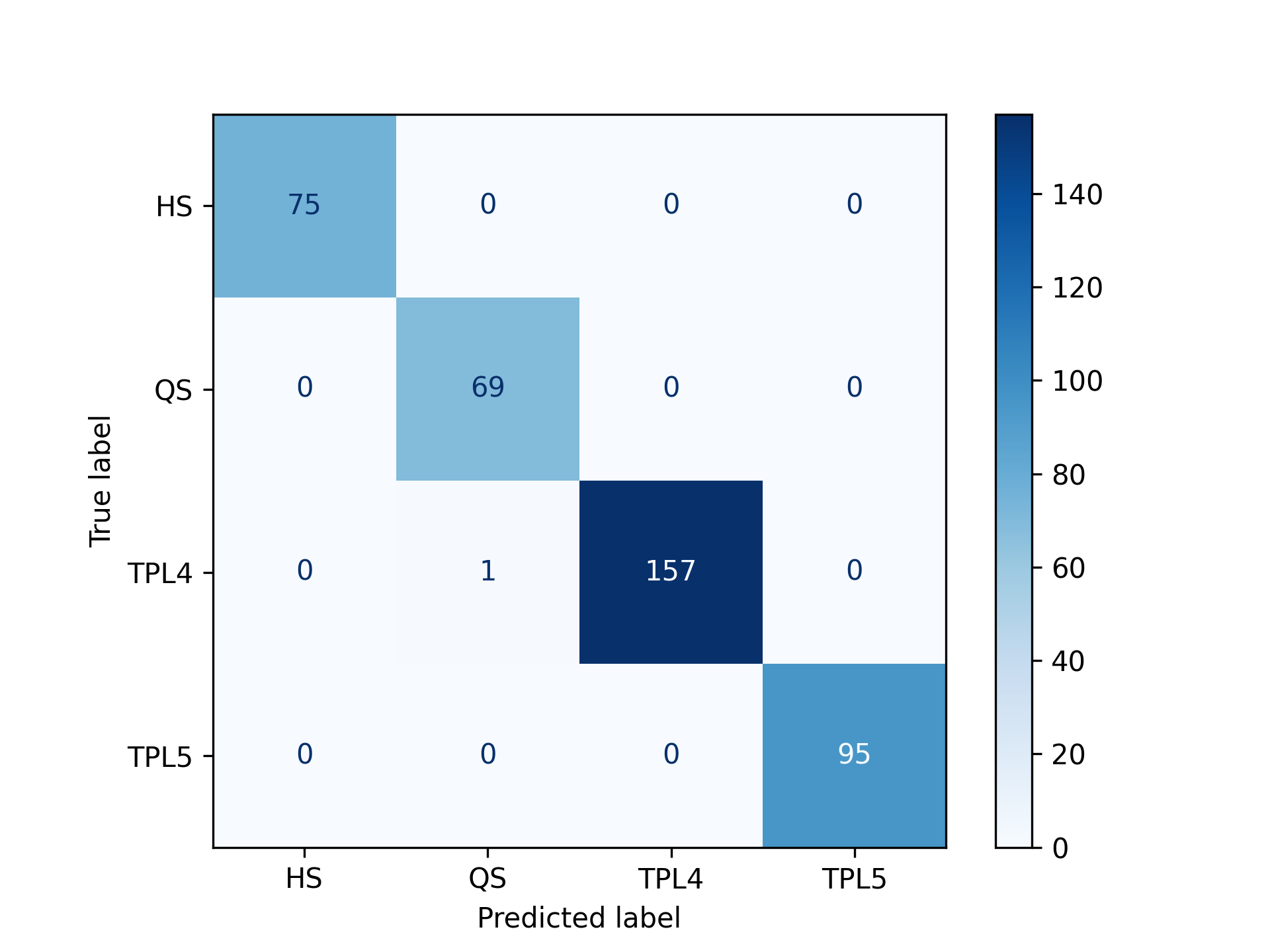}}
  \caption{Confusion matrix for SMV (a), RF (b) and XGB (c) algorithms when trained and tested on real data. Confusion matrix for SMV (d), RF (e) and XGB (f) algorithms when trained on ideal simulated data and tested on real data. Confusion matrix for SMV (g), RF (h) and XGB (i) algorithms when trained on ideal simulated data and tested on perturbed simulated data. Confusion matrix for SMV (j), RF (k) and XGB (l) algorithms when trained on ideal simulated data and tested on real and perturbed simulated data.}
    \label{fig:conf_matrix}
\end{figure*}

All classification results are reported in Fig.~\ref{fig:conf_matrix}, where confusion matrices for each performed test are presented.
A \textit{Confusion Matrix} is a type of visualization particularly well suited for evaluating the performance of a ML algorithm. The rows of the matrix represent the actual labels of the test set while the columns represent the labels predicted by the algorithm. Accordingly, the corrected predictions can be found along the diagonal of the matrix and the wrong ones outside of it.

In Tab.~\ref{tab:hyp_tun} they are reported all the selected hyperparameters for each performed test divided by algorithm.

\begin{table*}
\centering
\caption{Machine Learning selected hyperparameters. A full description of their meaning can be found, for instance, in
\citet{SklearnSVM, SklearnRF, XGBclassifier}.}
\label{tab:hyp_tun}
\begin{tabular}{|c c c c c|}
\cline{2-5}
  \multicolumn{1}{|c|}{} &
  \multicolumn{4}{|c|}{ Training set - Test set} \\
Algorithm Hyperparameters & Real-Real & Ideal-Real & Ideal-Pert. & Ideal-Real+Pert. \\
\hline
\multicolumn{5}{|c|}{\textbf{Support Vector Machine}}\\
\hline
C & 0.0001 & 1 & 0.001 & 1\\
gamma & 0.0001 & 0.001 & 0.1 & 0.001\\
kernel & linear & linear & linear & linear\\
\hline
\multicolumn{5}{c}{\textbf{Random Forest}}\\
\hline
n° estimators & 190 & 100 & 300 & 300\\
\hline
\multicolumn{5}{c}{\textbf{XGBoost}}\\
\hline
colsample bytree & 0.668 & 0.668 & 0.668 & 0.668\\
learning rate & 0.0765 & 0.0765 & 0.0765 & 0.0765\\
max depth & 5 & 5 & 5 & 5\\
min child weight & 1 & 1 & 1 & 1\\
n° estimators & 70 & 70 & 70 & 70\\
subsample & 0.409 & 0.409 & 0.409 & 0.409\\
\hline
\end{tabular}
\end{table*}

\subsubsection{Cross-Validated results}
As introduced in Sec.~\ref{CV} Cross-Validation is a crucial step to evaluate the model's ability to generalize on unseen data and it provides a more accurate evaluation of the model's performance.

Results obtained with a 5-fold Cross-Validation are reported in Tab.~\ref{tab: cv_results}, where we test on different combinations of the three datasets described in Sec.~\ref{Sec_data}.

The mean accuracy relative to the real cases dataset is quite high, but as already mentioned in the previous paragraph this may be due to the very limited dimensions of the dataset. In fact, this case is the one with the highest CV error score ($4\%$) appearing on the table. Adding the ideal simulated dataset, not only increases the mean accuracy (up to 99.9\% for XGB) but it also decreases the CV error score by an order of magnitude (0.09\% for XGB).

The third row of Tab.~\ref{tab: cv_results} is relative to the combination of the two simulated datasets, where we reach extremely high accuracy and quite low CV error score for all algorithms.

Finally, the algorithms' performances is cross-validated using all the available data. Although this is the case with the highest number of series and highest variability we still achieve remarkably good results with a mean accuracy that reaches 99.9 \% (for RF and XGB) and overall low CV error score.

It is important to note how in the current section we report extremely good results, sometimes reaching up to 100\% accuracy, but these high numbers should not mislead the reader. The main purpose of this work is to demonstrate that our approach based on features extraction and Machine Learning algorithms works. For this reason, we have considered about 2400 series with quite regular trends and belonging to only 4 possible classes. Increasing the number of series, the number of classes or the irregularity of the series trends may lead to a worsening of the performances.

In other words, in this work we establish that our approach perfectly works in the most basic settings and, considering the extremely satisfactory results obtained, we plan to extend our goal to a more complete analysis increasing the complexity of the data in future works.

\begin{table*}
\centering
\caption{Machine Learning multi-class classifiers results obtained in 5-fold Cross Validation. Training sets and test sets contain, respectively, 80\% and 20\% of the dataset. Standard deviation reported in parentheses. In the Average AUC the acronym "ovo" stands for One-vs-one and it computes the average
AUC of all possible pairwise combinations of classes} 
\label{tab: cv_results}
\begin{tabular}{|c c c c c c c c c|}
\hline
Dataset & Train & Test & Accuracy (\%) & Balanced Acc. (\%) & "ovo" AUC & Precision & Recall & f1 \\
\hline
\multicolumn{9}{|c|}{\textbf{Support Vector Machine}}\\
\hline
Real & 40 & 10 &98.0($\pm 4.0$) & 98.3($\pm3.3 $) & 0.994($\pm0.011 $) &0.987($\pm 0.027 $)&0.980($\pm0.040 $) &0.980($\pm0.040 $)\\
Real+Ideal & 1639 & 410 &99.3($\pm1.3 $) &99.4($\pm1.0 $) &0.999($\pm0.001 $) &0.993($\pm0.012 $) &0.993($\pm 0.013$) &0.993($\pm0.014 $)\\
Ideal+Pert. & 1877 & 469 &99.95($\pm0.09 $) &99.97($\pm0.07 $) &0.999($\pm0.001 $) &0.999($\pm0.001 $) &0.999($\pm0.001 $) &0.999($\pm0.001 $)\\
Real+Ideal+Pert. & 1917 & 179 &99.42($\pm1.17 $) &99.53($\pm0.94 $) &0.999($\pm0.001 $) &0.994($\pm0.010 $) &0.994($\pm0.010 $) &0.994($\pm0.010 $)\\
\hline
\multicolumn{9}{|c|}{\textbf{Random Forest}}\\
\hline
Real & 40 & 10 & 98.0($\pm4.0 $) &98.3($\pm3.3 $) &0.995($\pm0.009 $) &0.985($\pm0.030 $) &0.980($\pm0.040 $) &0.979($\pm0.041 $)\\
Real+Ideal & 1639 & 410 &99.9($\pm 0.2$) &99.9($\pm0.2 $) &1.0($\pm0.0 $) &0.999($\pm 0.002$) &0.999($\pm0.002 $) &0.999($\pm0.002 $)\\
Ideal+Pert. & 1877 & 469 &100.0($\pm0.0 $) &100.0($\pm0.0 $) &1.0($\pm0.0 $) &1.0($\pm0.0$) &1.0($\pm0.0 $) &1.0($\pm0.0 $)\\
Real+Ideal+Pert. & 1917 & 179 &99.92($\pm 0.17$) &99.92($\pm0.17 $) &1.0($\pm0.0 $) &0.999($\pm0.002 $) &0.999($\pm0.002 $) &0.999($\pm0.002 $)\\
\hline
\multicolumn{9}{|c|}{\textbf{XGBoost}}\\
\hline
Real & 40 & 10 &98.0($\pm4.0 $) &98.3($\pm3.3 $) &1.0($\pm0.0 $) &0.985($\pm0.030 $) &0.980($\pm 0.040$) &0.979($\pm 0.041$)\\
Real+Ideal & 1639 & 410 &99.95($\pm0.10 $) &99.96($\pm0.08 $) &1.0($\pm0.0 $) &0.999($\pm0.001 $) &0.999($\pm0.001 $) &0.999($\pm0.001 $)\\
Ideal+Pert. & 1877 & 469 &100.0($\pm0.0 $) &100.0($\pm0.0 $) &1.0($\pm0.0 $) &1.0($\pm0.0$) &1.0($\pm0.0 $) &1.0($\pm0.0 $)\\
Real+Ideal+Pert. & 1917 & 179 &99.96($\pm0.08 $) &99.97($\pm0.07 $) &1.0($\pm0.0 $) &0.999($\pm0.001$) &0.999($\pm0.001 $) &0.999($\pm0.001 $)\\
\hline
\end{tabular}
\end{table*}

\subsubsection{Features Importance}
Features Importance is one of the key points when using a Machine Learning algorithm for an application, where the interpretation and/or explanation of the results are as much important as finding good classification/regression results. The term \textit{Features Importance} relates to methods for scoring each input feature given to the model based on how useful they are when predicting a target variable; the scores indicate what we call \lq\lq importance\rq\rq \, of each feature. A higher score indicates that the particular feature will have a greater impact on the model. 
There are many ways to assign scores to the features; in our case we have used two different approaches: one based on a function provided by the algorithm library \citep[e.g.,][]{FeatImpRF,SklearnSVM,FeatImpXGB} and the other based on Shapley Values calculated by the SHAP package.

\begin{figure}
    \centering
     \includegraphics[scale=0.8]{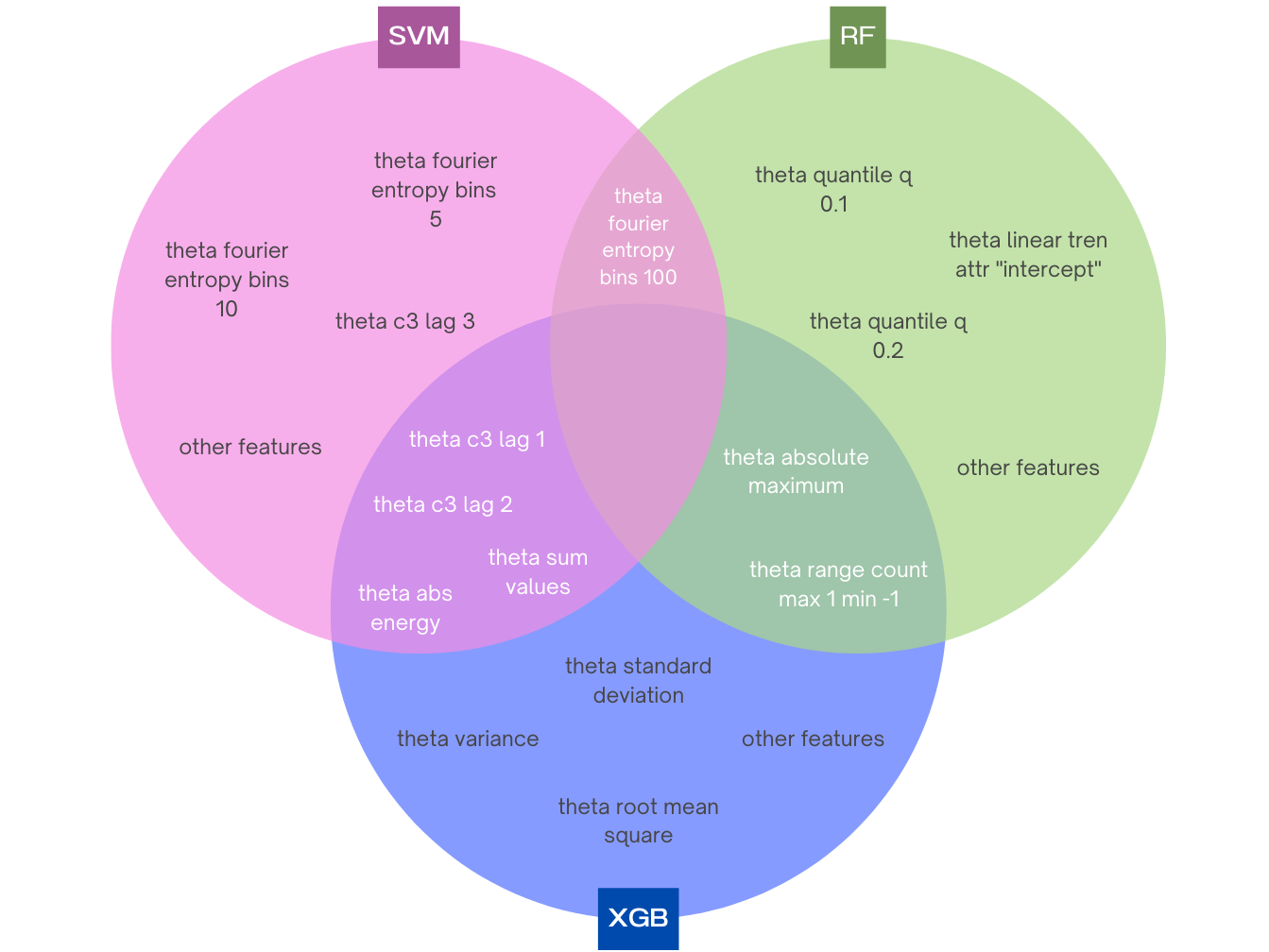}
    \caption{Common important features of the three supervised ML algorithms ranked by SHAP and Feature Importance tools.}
    \label{fig: common_feat}
\end{figure}

It is important to keep in mind that each algorithm has a tendency to weight features in a different way, even though some of them may be the same across all algorithms. %The signature of the signals we are attempting to classify is represented by this pool of features. 
In our case, it appears that there are no features common to all three algorithms, although
we can find some common ones when comparing the algorithms two at a time. These common features are reported in Fig.~\ref{fig: common_feat}.

\begin{figure*}
  \centering
  \subfloat[Scikit Learn Feature Importances for SVM]{\includegraphics[width=\columnwidth]{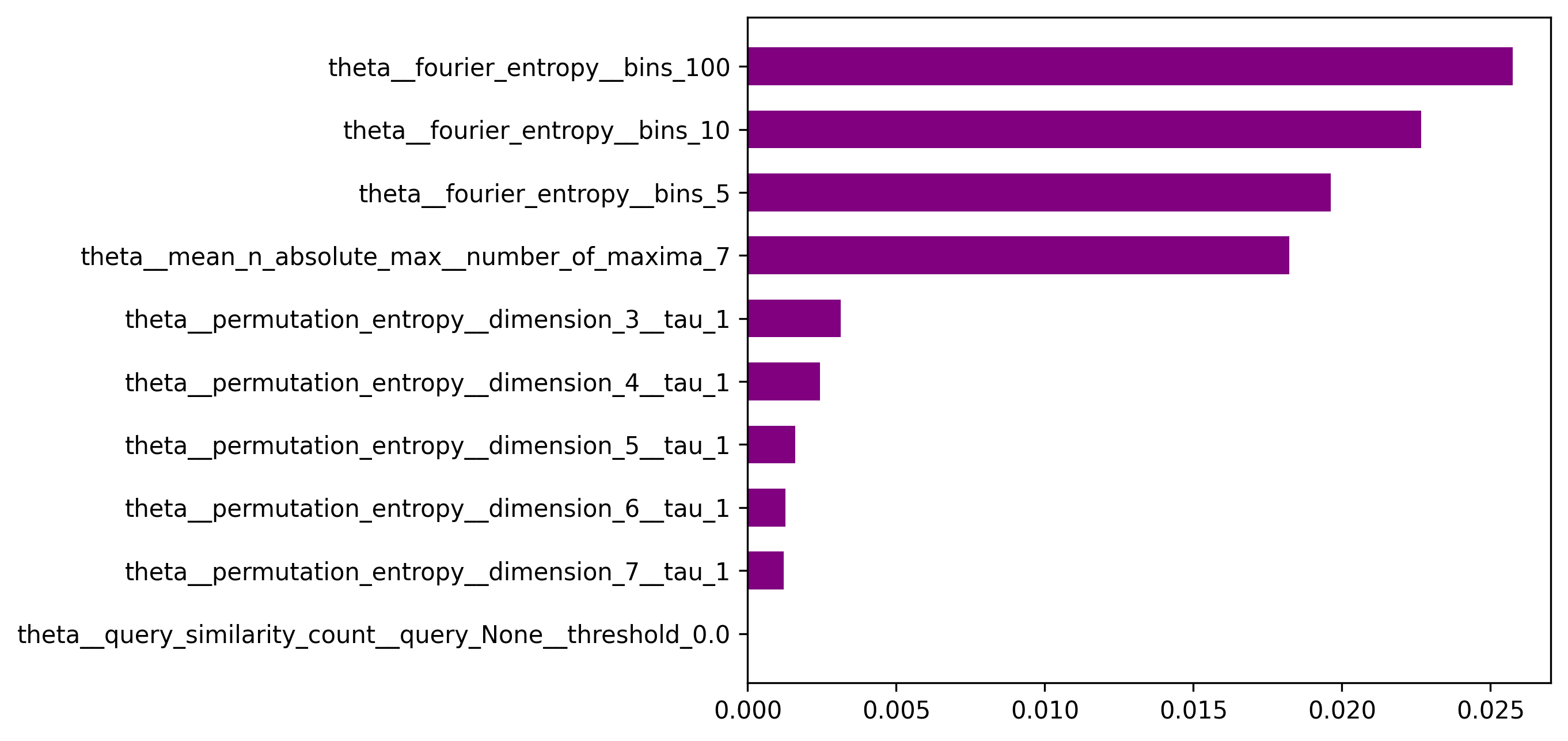}}
  \subfloat[SHAP summary plot for SVM]{\includegraphics[width=\columnwidth]{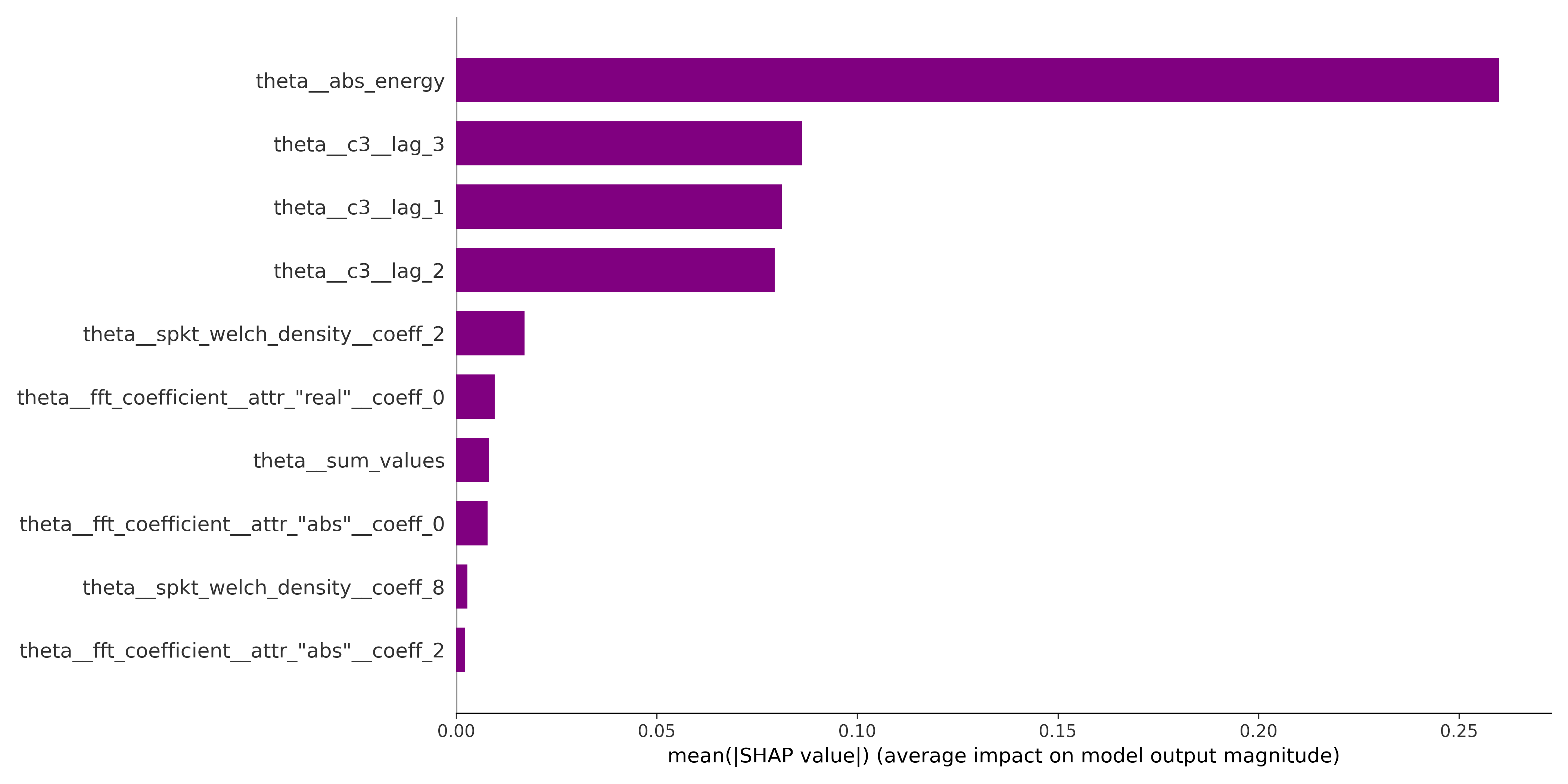}}
  \hfill
  \subfloat[Scikit Learn Feature Importances for RF]
  {\includegraphics[width=\columnwidth]{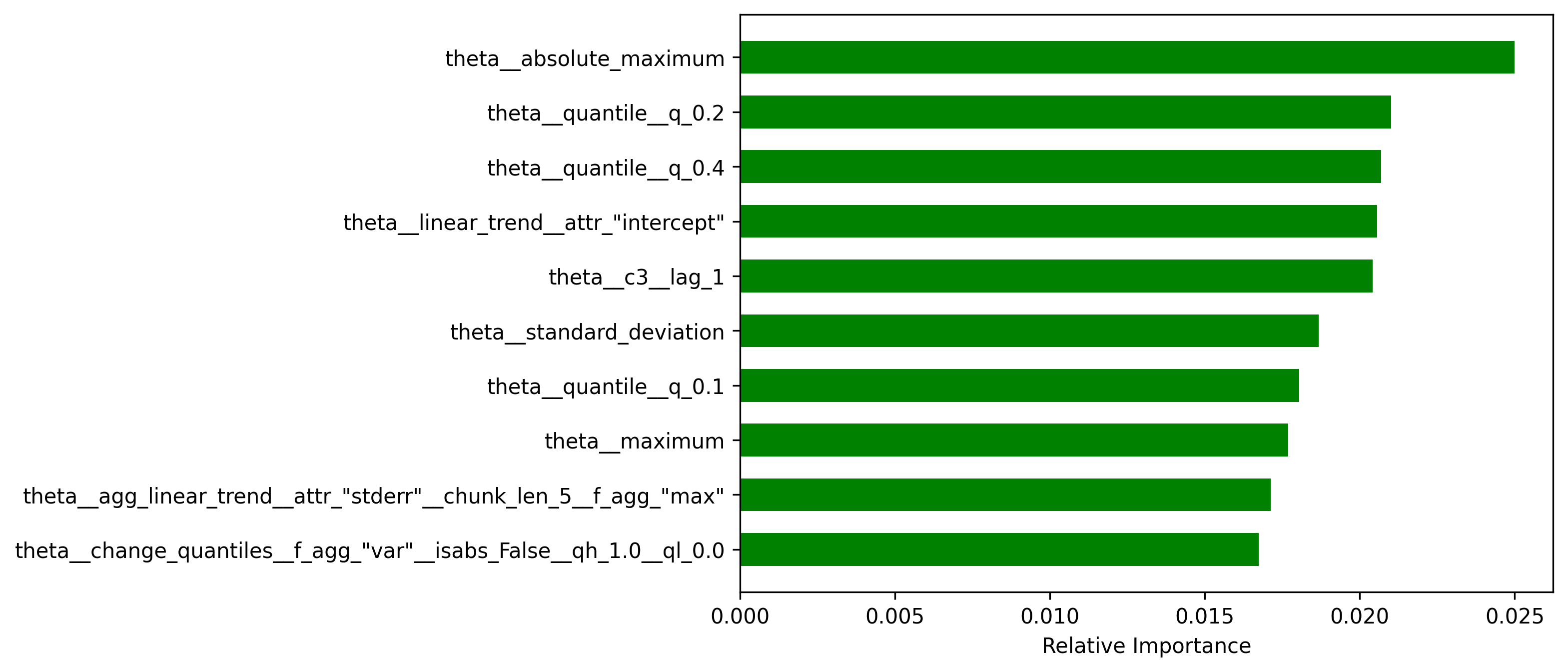}}
  \subfloat[SHAP summary plot for RF]
  {\includegraphics[width=\columnwidth]{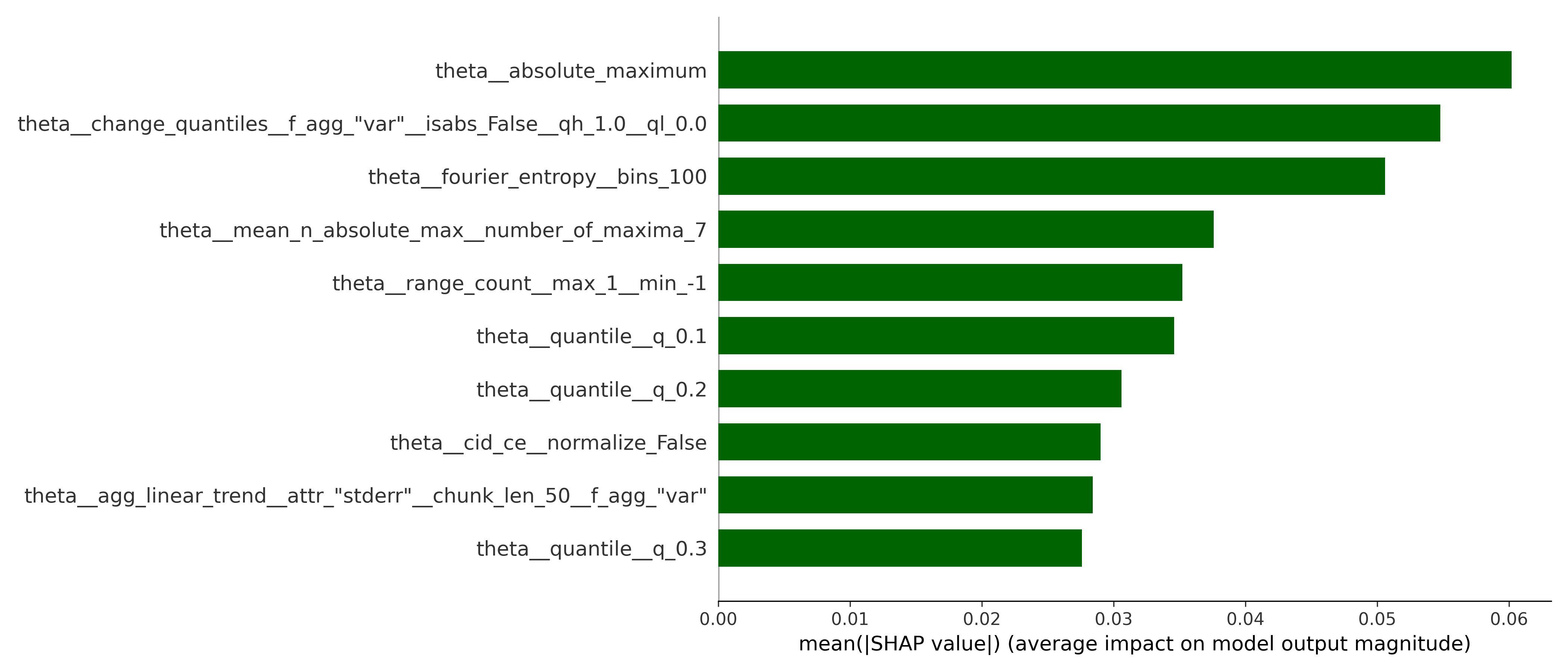}}
  \hfill
  \subfloat[XGBoost Feature Importances for XGB]
  {\includegraphics[width=\columnwidth]{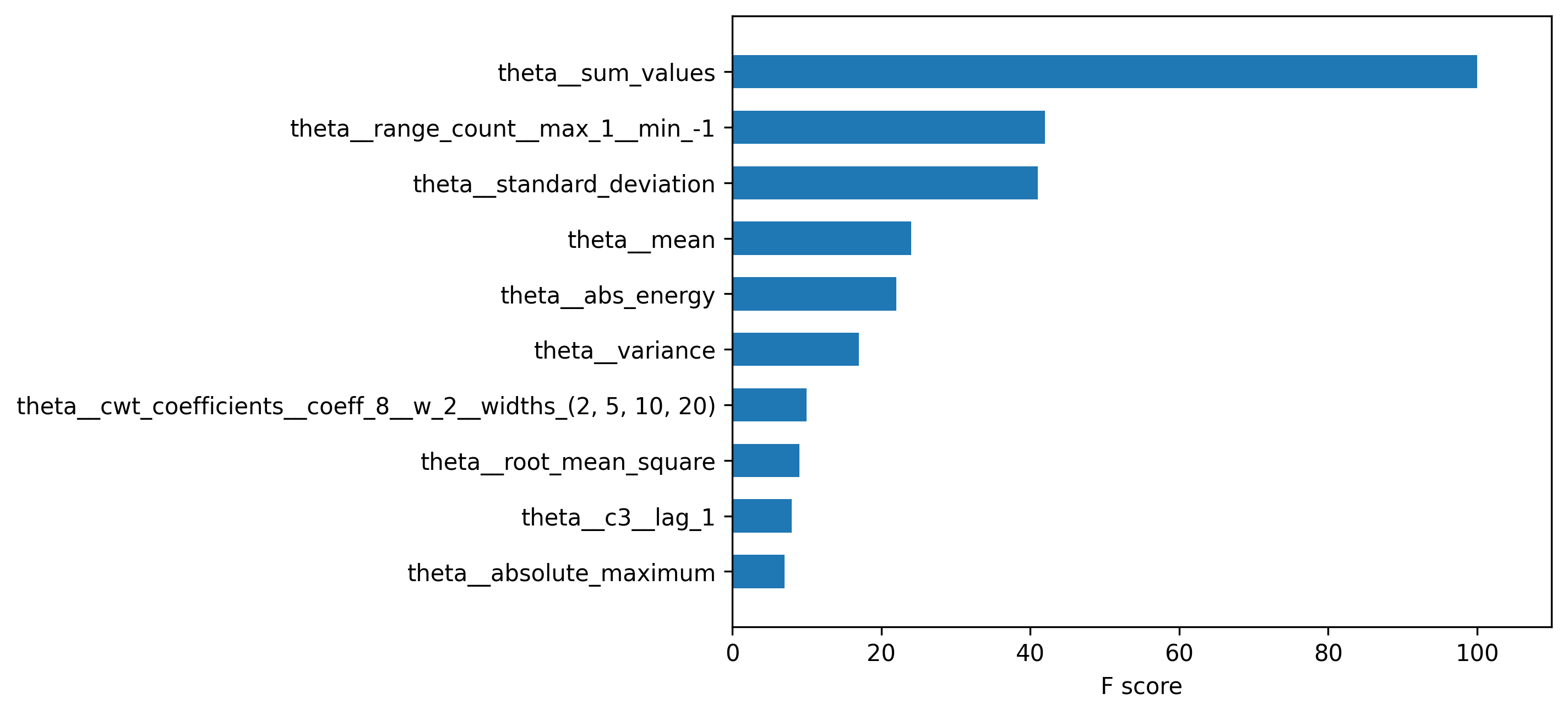}}
  \subfloat[SHAP summary plot for XGB]
  {\includegraphics[width=\columnwidth]{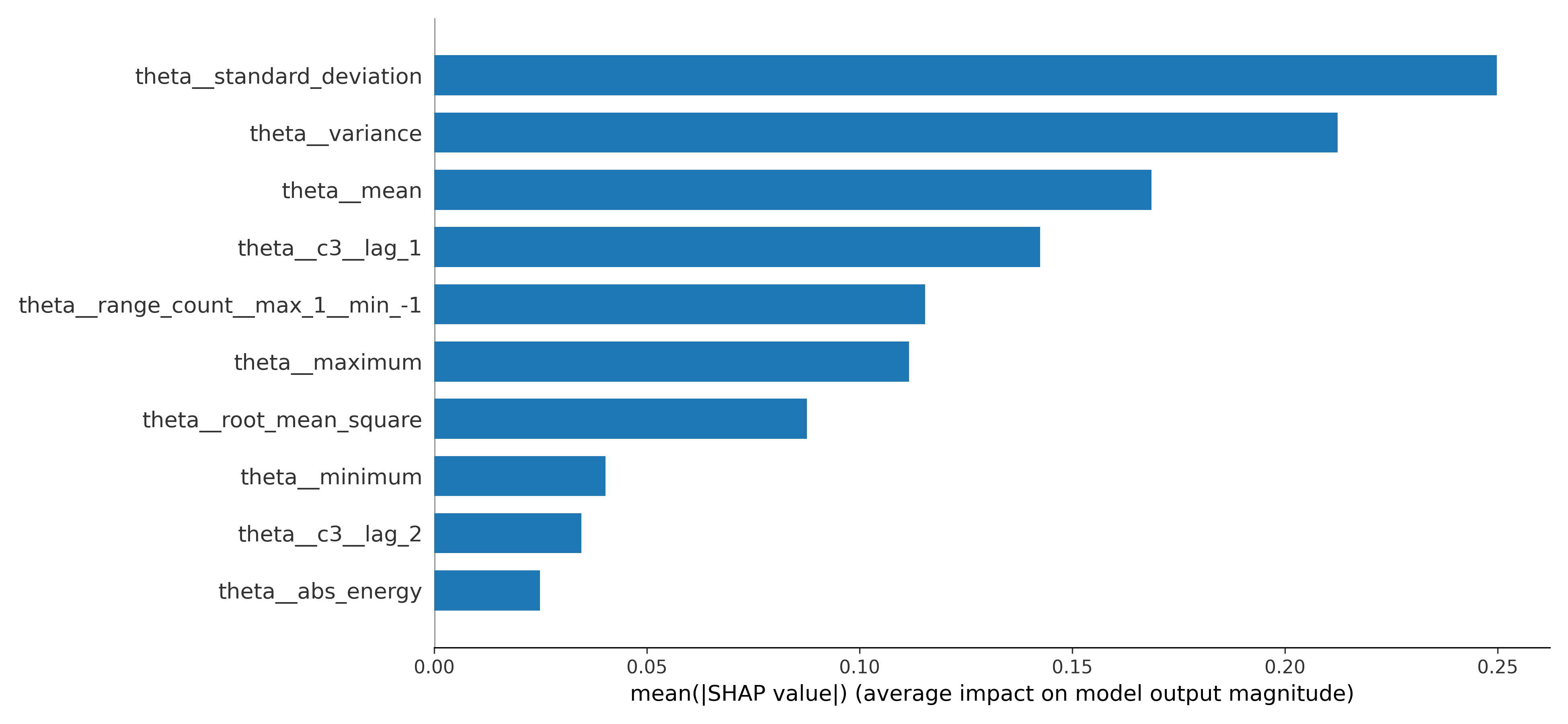}}
  \caption{Feature Importances for the three different Machine Learning Algorithms, evaluated with Scikit Learn packages and SHAP.}
  \label{fig: feat_imp}
\end{figure*}

\begin{figure*}
  \centering
  \subfloat[SHAP summary plot for XGB with color division based on importance for each class]{\includegraphics[width=\columnwidth]{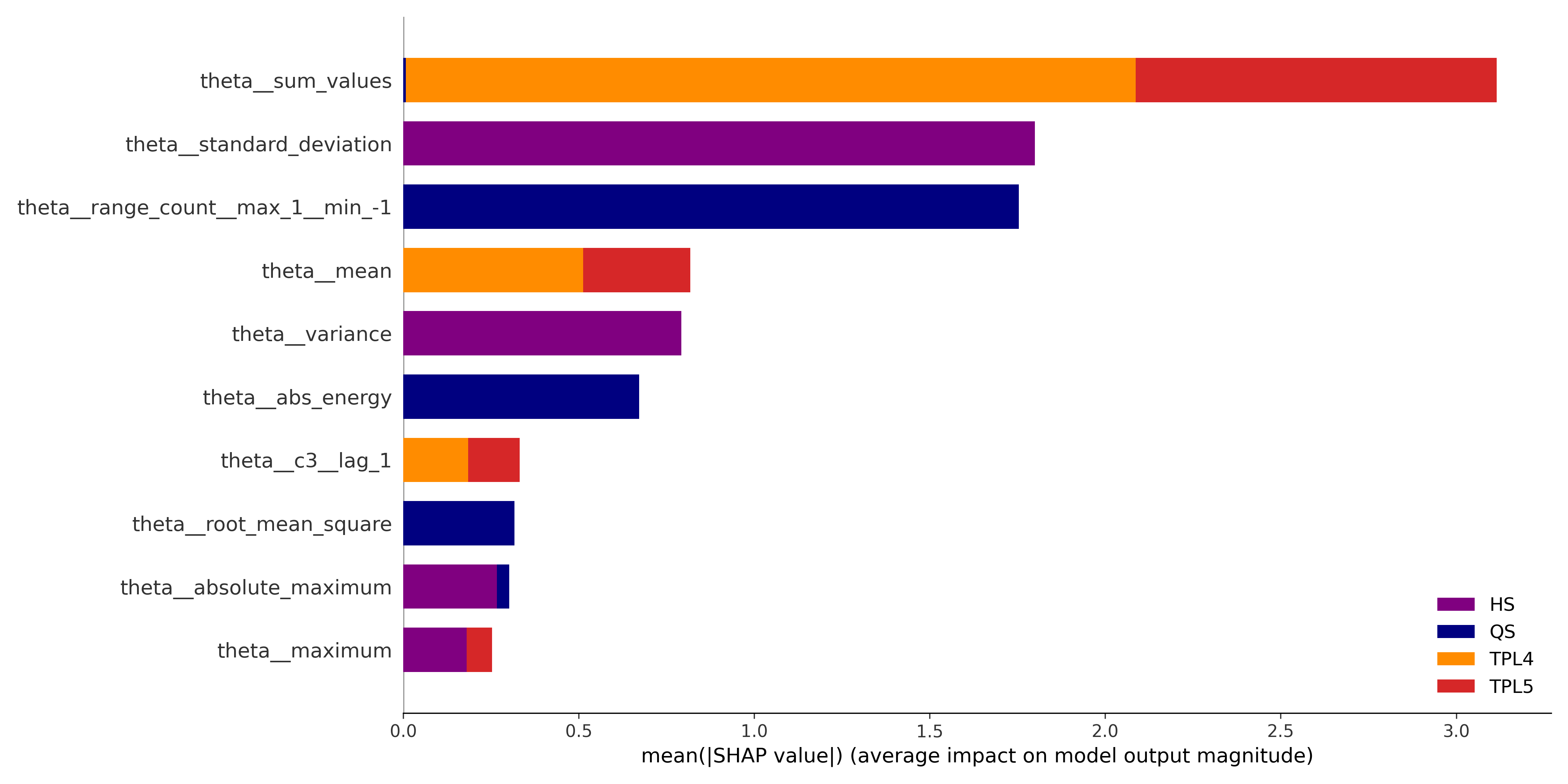}}
  \subfloat[SHAP beeswarm plot for XGB]{\includegraphics[width=\columnwidth]{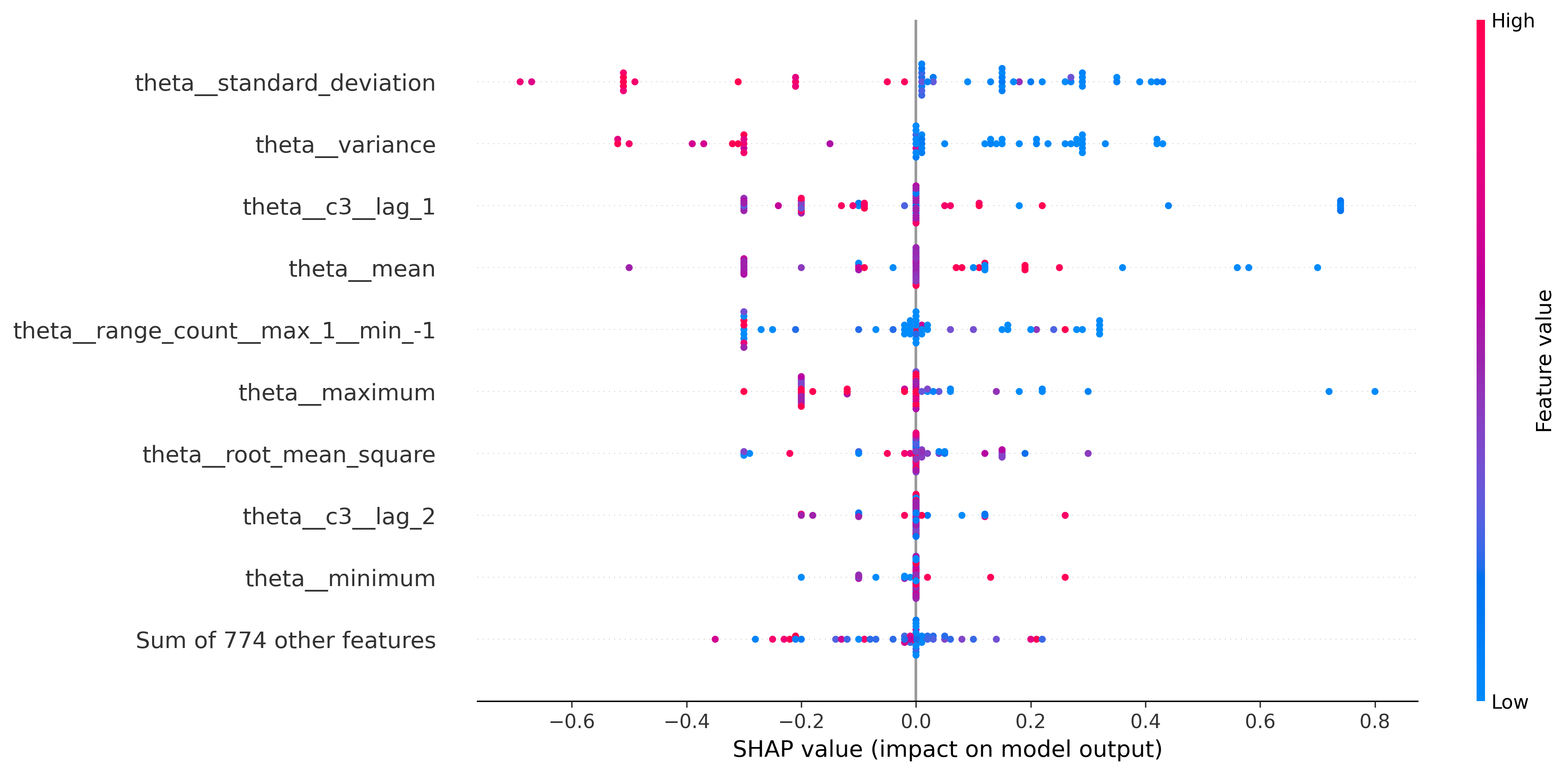}}
  \caption{SHAP results for XGBoost algorithm. On the left the summary plot while in the right the beeswarm plot.}
  \label{fig: shap_xgb}
\end{figure*}

Let us recall that, in this work, we have used three different classification algorithms: Random Forest, Support Vector Machines and XGBoost. Our results, reported in Fig.~\ref{fig: feat_imp} (a), (b), (c), (d) show that, for RF and SVM,  most features are quite difficult to interpret, while the features ranking provided by XGBoost (Fig.~\ref{fig: feat_imp} (e), (f)) propose a more straightforward and interpretable explanation of the model. For XGBoost in particular, the two approaches for Features Importance point out two similar pools of features, where 7 out of 10 are the same. In addition, as shown in Fig.~\ref{fig: feat_imp} (e) and (f), both approaches rank in the top positions features whose physical meaning is quite easy to deduct from their name, such as \textit{theta sum values, theta standard deviation, theta mean} and \textit{theta variance}. Additionally, for XGBoost in Fig.~\ref{fig: shap_xgb}, two other SHAP plots are shown: a \textit{summary plot} where each feature's bar has a division into colors based on importance for each class and a \textit{beeswarm plot}. A beeswarm plot is a  data visualization tool used to display a summary of how the top features impact the model’s output. Each point in the scatterplot represents a data point from the dataset, the vertical line represents the baseline value, which may be the model's average prediction or the expected value of the output. The position of the point in relation to the vertical line reveals whether a feature makes a positive (increasing the prediction) or negative (decreasing the prediction) contribution to the prediction and this position is determined by the Shapley value of the data point. What is important to understand is that the farther a point is from the vertical line, the higher its impact will be on the output of the model, regardless of whether it is on the left or on the right side of the plot. For a more detailed explanation of the plot please refer to \citet{beeswarm}.

\section{Future perspective: time series with transition between trends, an approach based on sliding windows}
\label{sec:future}

We are aware that the general case of time series observed could comprise different kinds of motion (such as the ones described and used in this work) due to transitions.
In order to move towards this more complex real scenario, we have begun to work to identify regions in the time series where the kind of motion is of the same type. This capability would allow our data analysis pipeline to deal with any kind of scenario. As first approach, we have decided to leverage on standard packages for time series data analysis in the case of segmentation of non-stationary signals \citep[e.g.,][]{truong2020selective} and anomaly detection \citep[e.g.,][]{gensler2018performing}.%However, there are a lot of issues we have to handle before we can address this problem in every real-world circumstance. In this work, we have decided to perform a first attempt, leveraging on standard packages for time series data analysis in the case of segmentation of non-stationary signals \citep[e.g.,][]{truong2020selective} and anomaly detection \citep[e.g.,][]{gensler2018performing}. 
We have performed some preliminary tests and some results are reported in this section and in the figure below. Our aim here is  to give a possible direction for the next works.

The results show that it is possible to arrange a semi-automatic division of the time series in the different trends, looking for example at the average over a fixed window length (in this case made of 8500 points) sliding over the $|\theta(t)|$ signal. The signal's mean of a window is compared to the mean of the following window; if the difference between those two values exceeds a certain threshold (empirically determined), a transition is detected.

\begin{figure*}
    \centering
    \includegraphics[scale=0.35]
    {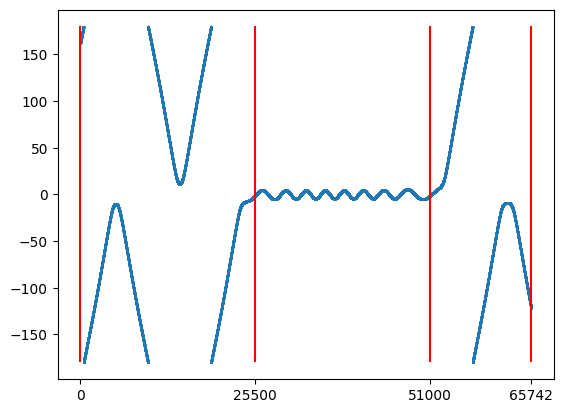}
    \includegraphics[scale=0.35]{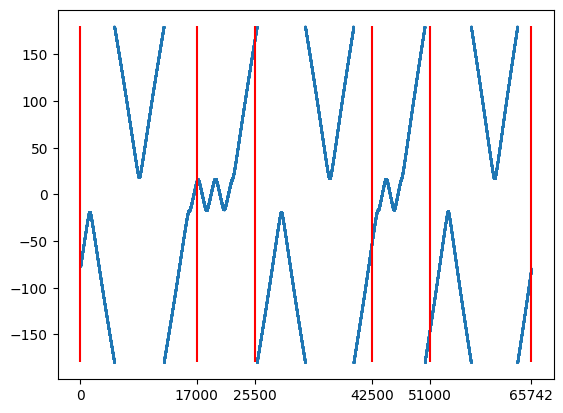}
    \includegraphics[scale=0.35]{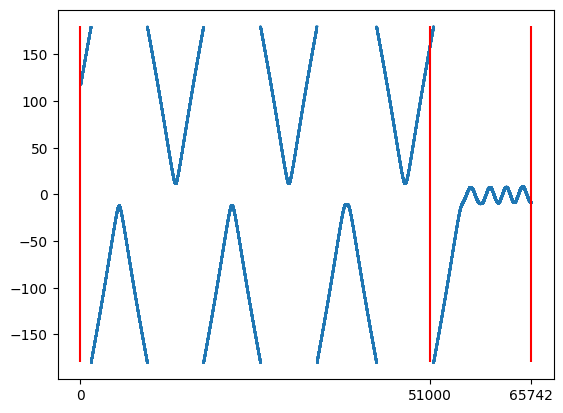}
    \caption{Three cases of real time series data: evolution of the resonant angle $\theta$ versus time; in the three cases several transitions between QS and HS regimes occur. }
    \label{finestre}
\end{figure*}

However, despite the results can be useful and sometimes impressive (see Fig.~\ref{finestre}), we have to investigate further how to generalize the definition of the time windows. This will be left to a future work.

\section{Conclusions}
\label{sec:conclusions}
%Concluding, in this work we faced the problem of time series classification of asteroids in co-orbital motion, leveraging on the action-angle variable $\theta(t)$ and restricting to four well-known cases: Tadpole (L4 and L5), Horseshoe and Quasi-satellite. Three different datasets were collected: one containing real cases, and other two for ideal simulated and perturbated ones.

This work deals with the problem of classification of asteroids in co-orbital motion with a given planet using a Machine Learning approach. The main parameter analysed to determine the type of co-orbital motion is a suitable angle $\theta$, that is defined following the assumption of the Planar Circular Restricted Three-Body Problem (PCR3BP) and its averaged approximation. The time evolution of $\theta$ allows to identify if the asteroid is in Tadpole motion, distinguishing between TPL4 (around the equilibrium point $L_4$) and TPL5 (around the equilibrium point $L_5$), Horseshoe (HS) motion or Quasi-satellite (QS) motion. We produce three different kinds of datasets called real, ideal simulated and perturbed simulated in order to apply Machine Learning algorithms. 
The datasets are formed by time series of the angle $\theta$, that consist in its evolution in time for short and medium timescale (about 900 years for real asteroid cases and 3000 years for simulated cases).

The Python package {\it tsfresh} is applied to such time series, extracting meaningfully features, which are selected and, if needed, standardized.
Then, a Machine Learning pipeline based on algorithms for Dimensionality Reduction and Classification, is built, with the features extracted as input.
The results show the power of such approach, with very well evident clusters in Dimensionality Reduction visualization plot and classification accuracy above 99\%.
This paper aims to define a methodological approach to such kind of data, serving as a backbone model for further studies, where more and more complex cases are faced.

\section*{Acknowledgements}

Authors express their gratitude to Tiago Azevedo and Pietro Lió from the University of Cambridge (UK), Michela Baccini, Chiara Marzi, Fabrizio Argenti and Simone Marinai from the University of Florence (Italy), Stefano Diciotti from the University of Bologna (Italy), Alessandro Mecocci from the University of Siena (Italy), for fruitful discussion and advices on data analysis and to Lona Ceccherini for her support.

%%%%%%%%%%%%%%%%%%%%%%%%%%%%%%%%%%%%%%%%%%%%%%%%%%
\section*{Data Availability}

 The data underlying this article will be shared on request
to the corresponding author.

%The inclusion of a Data Availability Statement is a requirement for articles published in MNRAS. Data Availability Statements provide a standardised format for readers to understand the availability of data underlying the research results described in the article. The statement may refer to original data generated in the course of the study or to third-party data analysed in the article. The statement should describe and provide means of access, where possible, by linking to the data or providing the required accession numbers for the relevant databases or DOIs.

%%%%%%%%%%%%%%%%%%%% REFERENCES %%%%%%%%%%%%%%%%%%

% The best way to enter references is to use BibTeX:
\clearpage
\bibliographystyle{mnras}
\bibliography{references} % if your bibtex file is called example.bib

\begin{thebibliography}{}
\makeatletter
\relax
\def\mn@urlcharsother{\let\do\@makeother \do\$\do\&\do\#\do\^\do\_\do\%\do\~}
\def\mn@doi{\begingroup\mn@urlcharsother \@ifnextchar [ {\mn@doi@}
  {\mn@doi@[]}}
\def\mn@doi@[#1]#2{\def\@tempa{#1}\ifx\@tempa\@empty \href
  {http://dx.doi.org/#2} {doi:#2}\else \href {http://dx.doi.org/#2} {#1}\fi
  \endgroup}
\def\mn@eprint#1#2{\mn@eprint@#1:#2::\@nil}
\def\mn@eprint@arXiv#1{\href {http://arxiv.org/abs/#1} {{\tt arXiv:#1}}}
\def\mn@eprint@dblp#1{\href {http://dblp.uni-trier.de/rec/bibtex/#1.xml}
  {dblp:#1}}
\def\mn@eprint@#1:#2:#3:#4\@nil{\def\@tempa {#1}\def\@tempb {#2}\def\@tempc
  {#3}\ifx \@tempc \@empty \let \@tempc \@tempb \let \@tempb \@tempa \fi \ifx
  \@tempb \@empty \def\@tempb {arXiv}\fi \@ifundefined
  {mn@eprint@\@tempb}{\@tempb:\@tempc}{\expandafter \expandafter \csname
  mn@eprint@\@tempb\endcsname \expandafter{\@tempc}}}

\bibitem[\protect\citeauthoryear{Arora, Hu  \& Kothari}{Arora
  et~al.}{2018}]{pmlr-v75-arora18a}
Arora S.,  Hu W.,   Kothari P.~K.,  2018, in Bubeck S.,  Perchet V.,   Rigollet
  P.,  eds,  Proceedings of Machine Learning Research Vol. 75, Proceedings of
  the 31st Conference On Learning Theory. PMLR, pp 1455--1462, \url
  {https://proceedings.mlr.press/v75/arora18a.html}

\bibitem[\protect\citeauthoryear{Ball \& Brunner}{Ball \&
  Brunner}{2010}]{ball2010data}
Ball N.~M.,  Brunner R.~J.,  2010, \mn@doi [International Journal of Modern
  Physics D] {https://doi.org/10.1142/S0218271810017160}, 19, 1049

\bibitem[\protect\citeauthoryear{Biau \& Scornet}{Biau \&
  Scornet}{2016}]{biau2016random}
Biau G.,  Scornet E.,  2016, Test, 25, 197

\bibitem[\protect\citeauthoryear{Carruba, Aljbaae  \& Lucchini}{Carruba
  et~al.}{2019}]{carruba_etal2019}
Carruba V.,  Aljbaae S.,   Lucchini A.,  2019, \mn@doi [Monthly Notices of the
  Royal Astronomical Society] {10.1093/mnras/stz1795}, 488, 1377

\bibitem[\protect\citeauthoryear{Carruba, Aljbaae, Domingos, Lucchini  \&
  Furlaneto}{Carruba et~al.}{2020}]{carruba_etal2020}
Carruba V.,  Aljbaae S.,  Domingos R.~C.,  Lucchini A.,   Furlaneto P.,  2020,
  \mn@doi [Monthly Notices of the Royal Astronomical Society]
  {10.1093/mnras/staa1463}, 496, 540

\bibitem[\protect\citeauthoryear{Carruba, Aljbaae, Domingos, Huaman  \&
  Barletta}{Carruba et~al.}{2022}]{carruba_etal2022}
Carruba V.,  Aljbaae S.,  Domingos R.~C.,  Huaman M.,   Barletta W.,  2022,
  \mn@doi [Celestial Mechanics and Dynamical Astronomy]
  {10.1007/s10569-022-10088-2}, 134, 36

\bibitem[\protect\citeauthoryear{Celletti, Gales, Rodriguez-Fernandez  \&
  Vasile}{Celletti et~al.}{2022}]{celletti_etal2022}
Celletti A.,  Gales C.,  Rodriguez-Fernandez V.,   Vasile M.,  2022, \mn@doi
  [Scientific Reports] {10.1038/s41598-022-05696-9}, 12, 1890

\bibitem[\protect\citeauthoryear{Cervantes, Garcia-Lamont,
  Rodr{\'\i}guez-Mazahua  \& Lopez}{Cervantes
  et~al.}{2020}]{cervantes2020comprehensive}
Cervantes J.,  Garcia-Lamont F.,  Rodr{\'\i}guez-Mazahua L.,   Lopez A.,  2020,
  Neurocomputing, 408, 189

\bibitem[\protect\citeauthoryear{Chen \& Guestrin}{Chen \&
  Guestrin}{2016}]{chen2016xgboost}
Chen T.,  Guestrin C.,  2016, in Proceedings of the 22nd acm sigkdd
  international conference on knowledge discovery and data mining. pp 785--794,
  \mn@doi{https://doi.org/10.1145/2939672.2939785}

\bibitem[\protect\citeauthoryear{{Chen} et~al.,}{{Chen}
  et~al.}{2018}]{chen_etal2018}
{Chen} Y.-T.,  et~al., 2018, \mn@doi [Publications of the Astronomical Society
  of Japan] {10.1093/pasj/psx145}, \href
  {https://ui.adsabs.harvard.edu/abs/2018PASJ...70S..38C} {70, S38}

\bibitem[\protect\citeauthoryear{Christ, Braun, Neuffer  \& Kempa-Liehr}{Christ
  et~al.}{2018}]{christ2018time}
Christ M.,  Braun N.,  Neuffer J.,   Kempa-Liehr A.~W.,  2018, Neurocomputing,
  307, 72

\bibitem[\protect\citeauthoryear{{Christ et al.}}{{Christ et
  al.}}{2023}]{tsfresh-github}
{Christ et al.} 2023, tsfresh github documentation, \url
  {https://tsfresh.readthedocs.io/en/latest/}

\bibitem[\protect\citeauthoryear{Connor \& van Leeuwen}{Connor \& van
  Leeuwen}{2018}]{Connor_etal2018}
Connor L.,  van Leeuwen J.,  2018, \mn@doi [The Astronomical Journal]
  {10.3847/1538-3881/aae649}, 156, 256

\bibitem[\protect\citeauthoryear{Cozzolino, Power  \& Chapman}{Cozzolino
  et~al.}{2019}]{cozzolino2019interpreting}
Cozzolino D.,  Power A.,   Chapman J.,  2019, Food Analytical Methods, 12, 2469

\bibitem[\protect\citeauthoryear{Di~Ruzza, Pousse  \& Alessi}{Di~Ruzza
  et~al.}{2023}]{DiRuzza2023}
Di~Ruzza S.,  Pousse A.,   Alessi E.~M.,  2023, \mn@doi [Icarus]
  {10.1016/j.icarus.2022.115330}, 390, 115330

\bibitem[\protect\citeauthoryear{Erasmus, Mommert, Trilling, Sickafoose, van
  Gend  \& Hora}{Erasmus et~al.}{2017}]{Erasmus_etal2017}
Erasmus N.,  Mommert M.,  Trilling D.~E.,  Sickafoose A.~A.,  van Gend C.,
  Hora J.~L.,  2017, \mn@doi [The Astronomical Journal]
  {10.3847/1538-3881/aa88be}, 154, 162

\bibitem[\protect\citeauthoryear{Erasmus, McNeill, Mommert, Trilling,
  Sickafoose  \& van Gend}{Erasmus et~al.}{2018}]{Erasmus_etal2018}
Erasmus N.,  McNeill A.,  Mommert M.,  Trilling D.~E.,  Sickafoose A.~A.,   van
  Gend C.,  2018, \mn@doi [The Astrophysical Journal Supplement Series]
  {10.3847/1538-4365/aac38f}, 237, 19

\bibitem[\protect\citeauthoryear{Farah et~al.,}{Farah
  et~al.}{2018}]{farah_etal2018}
Farah W.,  et~al., 2018, \mn@doi [Monthly Notices of the Royal Astronomical
  Society] {10.1093/mnras/sty1122}, 478, 1209

\bibitem[\protect\citeauthoryear{Fluke \& Jacobs}{Fluke \&
  Jacobs}{2020}]{fluke&jacobs2020}
Fluke C.~J.,  Jacobs C.,  2020, \mn@doi [WIREs Data Mining and Knowledge
  Discovery] {https://doi.org/10.1002/widm.1349}, 10, e1349

\bibitem[\protect\citeauthoryear{Francis, Hewett, Foltz  \& Chaffee}{Francis
  et~al.}{1992}]{francis_etal1992}
Francis P.,  Hewett P.~C.,  Foltz C.~B.,   Chaffee F.~H.,  1992, \mn@doi [The
  Astrophysical Journal] {10.1086/171870}, 398, 476

\bibitem[\protect\citeauthoryear{Fushiki}{Fushiki}{2011}]{fushiki2011estimation}
Fushiki T.,  2011, Statistics and Computing, 21, 137

\bibitem[\protect\citeauthoryear{Gensler \& Sick}{Gensler \&
  Sick}{2018}]{gensler2018performing}
Gensler A.,  Sick B.,  2018, Pattern Analysis and Applications, 21, 543

\bibitem[\protect\citeauthoryear{Goodfellow, Bengio  \& Courville}{Goodfellow
  et~al.}{2016}]{goodfellow2016deep}
Goodfellow I.,  Bengio Y.,   Courville A.,  2016, Deep learning.
MIT press

\bibitem[\protect\citeauthoryear{Guyon \& Elisseeff}{Guyon \&
  Elisseeff}{2003}]{guyon2003introduction}
Guyon I.,  Elisseeff A.,  2003, Journal of machine learning research, 3, 1157

\bibitem[\protect\citeauthoryear{Hastie, Tibshirani  \& Friedman}{Hastie
  et~al.}{2009a}]{hastie_etal2009}
Hastie T.,  Tibshirani R.,   Friedman J.,  2009a, The elements of statistical
  learning: data mining, inference and prediction, 2 edn.
Springer, \url {http://www-stat.stanford.edu/~tibs/ElemStatLearn/}

\bibitem[\protect\citeauthoryear{Hastie, Tibshirani, Friedman  \&
  Friedman}{Hastie et~al.}{2009b}]{hastie2009elements}
Hastie T.,  Tibshirani R.,  Friedman J.~H.,   Friedman J.~H.,  2009b, The
  elements of statistical learning: data mining, inference, and prediction.
 Vol. 2, Springer

\bibitem[\protect\citeauthoryear{Ivezi{\'c}, Connolly, VanderPlas  \&
  Gray}{Ivezi{\'c} et~al.}{2014}]{ivezic2014statistics}
Ivezi{\'c} {\v{Z}}.,  Connolly A.~J.,  VanderPlas J.~T.,   Gray A.,  2014, in ,
  Statistics, Data Mining, and Machine Learning in Astronomy.
Princeton University Press

\bibitem[\protect\citeauthoryear{Jacobs et~al.,}{Jacobs
  et~al.}{2019}]{jacobs_etal2017}
Jacobs C.,  et~al., 2019, \mn@doi [Monthly Notices of the Royal Astronomical
  Society] {10.1093/mnras/stz272}, 484, 5330

\bibitem[\protect\citeauthoryear{Jordan \& Mitchell}{Jordan \&
  Mitchell}{2015}]{jordan2015machine}
Jordan M.~I.,  Mitchell T.~M.,  2015, Science, 349, 255

\bibitem[\protect\citeauthoryear{Kamath}{Kamath}{2022}]{kamath_2022}
Kamath C.,  2022, \mn@doi [International Journal of Data Science and Analytics]
  {10.1007/s41060-022-00368-3}

\bibitem[\protect\citeauthoryear{{Knezevic} \& {Milani}}{{Knezevic} \&
  {Milani}}{1994}]{knezevic_milani1994}
{Knezevic} Z.,  {Milani} A.,  1994, in {Milani} A.,  {di Martino} M.,
  {Cellino} A.,  eds,  Vol. 160, Asteroids, Comets, Meteors 1993. p.~143

\bibitem[\protect\citeauthoryear{Knezevic, Lema{\^i}tre  \& Milani}{Knezevic
  et~al.}{2002}]{knezevic_etal2002}
Knezevic Z.,  Lema{\^i}tre A.,   Milani A.,  2002, The determination of
  Asteroid Proper Elements.
pp 603--612

\bibitem[\protect\citeauthoryear{Kobak \& Berens}{Kobak \&
  Berens}{2019}]{kobak2019art}
Kobak D.,  Berens P.,  2019, Nature communications, 10, 5416

\bibitem[\protect\citeauthoryear{Lanusse, Ma, Li, Collett, Li, Ravanbakhsh,
  Mandelbaum  \& PÃczos}{Lanusse et~al.}{2017}]{lanusse_etal2018}
Lanusse F.,  Ma Q.,  Li N.,  Collett T.~E.,  Li C.-L.,  Ravanbakhsh S.,
  Mandelbaum R.,   PÃczos B.,  2017, \mn@doi [Monthly Notices of the Royal
  Astronomical Society] {10.1093/mnras/stx1665}, 473, 3895

\bibitem[\protect\citeauthoryear{LeCun, Bengio  \& Hinton}{LeCun
  et~al.}{2015}]{lecun2015deep}
LeCun Y.,  Bengio Y.,   Hinton G.,  2015, nature, 521, 436

\bibitem[\protect\citeauthoryear{Li, Cheng, Wang, Morstatter, Trevino, Tang  \&
  Liu}{Li et~al.}{2017}]{li2017feature}
Li J.,  Cheng K.,  Wang S.,  Morstatter F.,  Trevino R.~P.,  Tang J.,   Liu H.,
   2017, ACM computing surveys (CSUR), 50, 1

\bibitem[\protect\citeauthoryear{Liu, Zhang, Hou, Mian, Wang, Zhang  \&
  Tang}{Liu et~al.}{2021}]{liu2021self}
Liu X.,  Zhang F.,  Hou Z.,  Mian L.,  Wang Z.,  Zhang J.,   Tang J.,  2021,
  IEEE transactions on knowledge and data engineering, 35, 857

\bibitem[\protect\citeauthoryear{Lundberg \& Lee}{Lundberg \&
  Lee}{2017}]{NIPS2017_7062}
Lundberg S.~M.,  Lee S.-I.,  2017, in Guyon I.,  Luxburg U.~V.,  Bengio S.,
  Wallach H.,  Fergus R.,  Vishwanathan S.,   Garnett R.,  eds, , Advances in
  Neural Information Processing Systems 30.
Curran Associates, Inc., pp 4765--4774, \url
  {http://papers.nips.cc/paper/7062-a-unified-approach-to-interpreting-model-predictions.pdf}

\bibitem[\protect\citeauthoryear{Lundberg et~al.,}{Lundberg
  et~al.}{2018}]{lundberg2018explainable}
Lundberg S.~M.,  et~al., 2018, Nature Biomedical Engineering, 2, 749

\bibitem[\protect\citeauthoryear{Lundberg et~al.,}{Lundberg
  et~al.}{2020}]{lundberg2020local2global}
Lundberg S.~M.,  et~al., 2020, Nature Machine Intelligence, 2, 2522

\bibitem[\protect\citeauthoryear{Mitchell, Frank  \& Holmes}{Mitchell
  et~al.}{2022}]{mitchell2022gputreeshap}
Mitchell R.,  Frank E.,   Holmes G.,  2022, GPUTreeShap: Massively Parallel
  Exact Calculation of SHAP Scores for Tree Ensembles (\mn@eprint {arXiv}
  {2010.13972})

\bibitem[\protect\citeauthoryear{Molnar}{Molnar}{2022}]{molnar2022}
Molnar C.,  2022, Interpretable Machine Learning, 2 edn.
\url {https://christophm.github.io/interpretable-ml-book}

\bibitem[\protect\citeauthoryear{{NASA}}{{NASA}}{2022}]{NASAHor}
{NASA} 2022, https://ssd-api.jpl.nasa.gov/doc/horizons.htm

\bibitem[\protect\citeauthoryear{Namouni}{Namouni}{1999}]{namouni99}
Namouni F.,  1999, \mn@doi [Icarus] {https://doi.org/10.1006/icar.1998.6032},
  137, 293

\bibitem[\protect\citeauthoryear{Namouni, Christou  \& Murray}{Namouni
  et~al.}{1999}]{namouni_etal99}
Namouni F.,  Christou A.~A.,   Murray C.~D.,  1999, \mn@doi [Phys. Rev. Lett.]
  {10.1103/PhysRevLett.83.2506}, 83, 2506

\bibitem[\protect\citeauthoryear{Ozsahin, Mustapha, Mubarak, Ameen  \&
  Uzun}{Ozsahin et~al.}{2022}]{ozsahin2022impact}
Ozsahin D.~U.,  Mustapha M.~T.,  Mubarak A.~S.,  Ameen Z.~S.,   Uzun B.,  2022,
  in 2022 International Conference on Artificial Intelligence in Everything
  (AIE). pp 87--94

\bibitem[\protect\citeauthoryear{{Pearson}, {Palafox}  \& {Griffith}}{{Pearson}
  et~al.}{2018}]{pearson_etal2018}
{Pearson} K.~A.,  {Palafox} L.,   {Griffith} C.~A.,  2018, \mn@doi [Monthly
  Notices of the Royal Astronomical Society] {10.1093/mnras/stx2761}, \href
  {https://ui.adsabs.harvard.edu/abs/2018MNRAS.474..478P} {474, 478}

\bibitem[\protect\citeauthoryear{Pedregosa et~al.,}{Pedregosa
  et~al.}{2011}]{scikit-learn}
Pedregosa F.,  et~al., 2011, Journal of Machine Learning Research, 12, 2825

\bibitem[\protect\citeauthoryear{Pourrahmani, Nayyeri  \& Cooray}{Pourrahmani
  et~al.}{2018}]{Pourrahmani_etal2018}
Pourrahmani M.,  Nayyeri H.,   Cooray A.,  2018, \mn@doi [The Astrophysical
  Journal] {10.3847/1538-4357/aaae6a}, 856, 68

\bibitem[\protect\citeauthoryear{Pousse \& Alessi}{Pousse \&
  Alessi}{2022}]{pousse&alessi2022}
Pousse A.,  Alessi E.~M.,  2022, \mn@doi [Nonlinear Dynamics]
  {10.1007/s11071-022-07229-5}, 108, 959

\bibitem[\protect\citeauthoryear{Rein \& Liu}{Rein \& Liu}{2012}]{ReinLiu2012}
Rein H.,  Liu S.~F.,  2012, \mn@doi [A\&A] {10.1051/0004-6361/201118085}, 537,
  A128

\bibitem[\protect\citeauthoryear{Roscher, Bohn, Duarte  \& Garcke}{Roscher
  et~al.}{2020}]{roscher2020explainable}
Roscher R.,  Bohn B.,  Duarte M.~F.,   Garcke J.,  2020, Ieee Access, 8, 42200

\bibitem[\protect\citeauthoryear{{SHAP}}{{SHAP}}{2023}]{beeswarm}
{SHAP} 2023, beeswarm plot,
  \url{https://shap.readthedocs.io/en/latest/example\_notebooks/api\_examples/plots/beeswarm.html}

\bibitem[\protect\citeauthoryear{{Scikit-Learn}}{{Scikit-Learn}}{2023d}]{FeatImpRF}
{Scikit-Learn} 2023d, Feature importances with a forest of trees,
  \url{https://scikit-learn.org/stable/auto\_examples/ensemble/plot\_forest\_importances.html}

\bibitem[\protect\citeauthoryear{{Scikit-Learn}}{{Scikit-Learn}}{2023a}]{Sklearn_metrics}
{Scikit-Learn} 2023a, Metrics and scoring: quantifying the quality of
  predictions,
  \url{https://scikit-learn.org/stable/modules/model\_evaluation.html}

\bibitem[\protect\citeauthoryear{{Scikit-Learn}}{{Scikit-Learn}}{2023c}]{SklearnRF}
{Scikit-Learn} 2023c, Random Forest Scikit-Learn,
  \url{https://scikit-learn.org/stable/modules/generated/sklearn.ensemble.RandomForestClassifier.html}

\bibitem[\protect\citeauthoryear{{Scikit-Learn}}{{Scikit-Learn}}{2023b}]{SklearnSVM}
{Scikit-Learn} 2023b, SVC Scikit-Learn,
  \url{https://scikit-learn.org/stable/modules/generated/sklearn.svm.SVC.html}

\bibitem[\protect\citeauthoryear{Shallue \& Vanderburg}{Shallue \&
  Vanderburg}{2017}]{Shallue_etal2018}
Shallue C.~J.,  Vanderburg A.~M.,  2017, The Astronomical Journal, 155

\bibitem[\protect\citeauthoryear{Singh \& Singh}{Singh \&
  Singh}{2020}]{singh2020investigating}
Singh D.,  Singh B.,  2020, Applied Soft Computing, 97, 105524

\bibitem[\protect\citeauthoryear{Singh, Gulati  \& Gupta}{Singh
  et~al.}{1998}]{singh_etal1998}
Singh H.~P.,  Gulati R.~K.,   Gupta R.,  1998, \mn@doi [Monthly Notices of the
  Royal Astronomical Society] {10.1046/j.1365-8711.1998.01255.x}, 295, 312

\bibitem[\protect\citeauthoryear{Smirnov}{Smirnov}{2023}]{SMIRNOV2023}
Smirnov E.,  2023, \mn@doi [Astronomy and Computing]
  {https://doi.org/10.1016/j.ascom.2023.100707}, 43, 100707

\bibitem[\protect\citeauthoryear{{Smirnov} \& {Markov}}{{Smirnov} \&
  {Markov}}{2017}]{Smirnov&Markov2017}
{Smirnov} E.~A.,  {Markov} A.~B.,  2017, \mn@doi [Monthly Notices of the Royal
  Astronomical Society] {10.1093/mnras/stx999}, \href
  {https://ui.adsabs.harvard.edu/abs/2017MNRAS.469.2024S} {469, 2024}

\bibitem[\protect\citeauthoryear{Smirnov \& Shevchenko}{Smirnov \&
  Shevchenko}{2013}]{Smirnov_etal2013}
Smirnov E.~A.,  Shevchenko I.~I.,  2013, \mn@doi [Icarus]
  {https://doi.org/10.1016/j.icarus.2012.10.034}, 222, 220

\bibitem[\protect\citeauthoryear{Smullen \& Volk}{Smullen \&
  Volk}{2020}]{smullen&volk2020}
Smullen R.~A.,  Volk K.,  2020, \mn@doi [Monthly Notices of the Royal
  Astronomical Society] {10.1093/mnras/staa1935}, 497, 1391

\bibitem[\protect\citeauthoryear{Truong, Oudre  \& Vayatis}{Truong
  et~al.}{2020}]{truong2020selective}
Truong C.,  Oudre L.,   Vayatis N.,  2020, Signal Processing, 167, 107299

\bibitem[\protect\citeauthoryear{Van~den Broeck, Lykov, Schleich  \&
  Suciu}{Van~den Broeck et~al.}{2022}]{van2022tractability}
Van~den Broeck G.,  Lykov A.,  Schleich M.,   Suciu D.,  2022, Journal of
  Artificial Intelligence Research, 74, 851

\bibitem[\protect\citeauthoryear{Van~der Maaten \& Hinton}{Van~der Maaten \&
  Hinton}{2008}]{van2008visualizing}
Van~der Maaten L.,  Hinton G.,  2008, Journal of machine learning research, 9

\bibitem[\protect\citeauthoryear{{Whitmore}}{{Whitmore}}{1984}]{whitmore1984}
{Whitmore} B.~C.,  1984, \mn@doi [Astrophysical Journal] {10.1086/161768},
  \href {https://ui.adsabs.harvard.edu/abs/1984ApJ...278...61W} {278, 61}

\bibitem[\protect\citeauthoryear{{xgboost}}{{xgboost}}{2023a}]{XGBclassifier}
{xgboost} 2023a, XGBclassifier,
  \url{https://xgboost.readthedocs.io/en/latest/python/python_api.html}

\bibitem[\protect\citeauthoryear{{xgboost}}{{xgboost}}{2023b}]{FeatImpXGB}
{xgboost} 2023b, xgboost.plot\_importance,
  \url{https://xgboost.readthedocs.io/en/stable/python/python\_api.html#module-xgboost.plotting}

\makeatother
\end{thebibliography}

% Alternatively you could enter them by hand, like this:
% This method is tedious and prone to error if you have lots of references
%\begin{thebibliography}{99}
%\bibitem[\protect\citeauthoryear{Author}{2012}]{Author2012}
%Author A.~N., 2013, Journal of Improbable Astronomy, 1, 1
%\bibitem[\protect\citeauthoryear{Others}{2013}]{Others2013}
%Others S., 2012, Journal of Interesting Stuff, 17, 198
%\end{thebibliography}

%%%%%%%%%%%%%%%%%%%%%%%%%%%%%%%%%%%%%%%%%%%%%%%%%%

%%%%%%%%%%%%%%%%% APPENDICES %%%%%%%%%%%%%%%%%%%%%

%\appendix

%\section{Some extra material}

%If you want to present additional material which would interrupt the flow of the main paper,
%it can be placed in an Appendix which appears after the list of references.

%%%%%%%%%%%%%%%%%%%%%%%%%%%%%%%%%%%%%%%%%%%%%%%%%%

% Don't change these lines
\bsp	% typesetting comment
\label{lastpage}
\end{document}